\documentclass[aps,prc,reprint,superscriptaddress,nofootinbib,floatfix,longbibliography]{revtex4-2}

\usepackage[T1]{fontenc}
\usepackage[utf8]{inputenc}
\usepackage[english]{babel}
\usepackage{textcomp}
\usepackage{bm}
\usepackage{amsmath,amssymb,amsfonts,amsthm}
\usepackage{latexsym}
\usepackage{slashed}
\usepackage{graphicx}
\usepackage{esint}
\usepackage{booktabs}
\usepackage{array}
\usepackage{multirow}
\usepackage{siunitx}
\usepackage{xcolor}
\usepackage{comment}

\makeatletter
\def\label#1{%
  \@bsphack
  \begingroup
    \UseHookWithArguments{label}{1}{#1}%
    \protected@write\@auxout{}%
      {\string\newlabel{#1}{{\@currentlabel}{\thepage}%
      {\@currentlabelname}{\@currentHref}{\@kernel@reserved@label@data}}}%
  \endgroup
  \@esphack
}
\let\ltx@label\label
\makeatother

\usepackage[colorlinks=true,allcolors=blue,unicode=true]{hyperref}

\makeatletter
\def\@bibdataout@aps#1{}
\def\@bibstyleout@aps#1{}
\makeatother

\hypersetup{
  pdftitle={Full Uncertainty Quantification of Sign-Problem-Free Quantum Monte Carlo Methods and Nuclear Lattice Effective Field Theory Benchmarks},
  pdfauthor={BNLU},
  bookmarksnumbered=false,
  bookmarksopen=false,
  breaklinks=true,
  pdfborder={0 0 0}
}


\makeatletter

\def\NAT@sort@cites#1{%
  \let\NAT@cite@list\@empty
  \@for\@citeb:=#1\do{%
    \expandafter\NAT@star@cite\@citeb\@@
  }%
  \@ifnum{\NAT@sort>\z@}{%
    \expandafter\NAT@sort@cites@\expandafter{\NAT@cite@list}%
  }{}%
}

\let\auto@bib\@empty
\let\auto@bib@innerbib\@empty

\let\write@bibliographystyle\relax

\makeatother
\begin{document}

\title{\texorpdfstring{Full Uncertainty Quantification of Sign-Problem-Free Quantum Monte Carlo Methods \\ and Nuclear Lattice Effective Field Theory Benchmarks}{Full Uncertainty Quantification of Sign-Problem-Free Quantum Monte Carlo Methods and Nuclear Lattice Effective Field Theory Benchmarks}}

\author{Zhong-Wang~Niu}
\email{niuzhongwang21@gscaep.ac.cn}
\affiliation{Graduate School of China Academy of Engineering Physics, Beijing 100193, China}

\author{Bing-Nan~Lu}
\email{bnlv@gscaep.ac.cn}
\affiliation{Graduate School of China Academy of Engineering Physics, Beijing 100193, China}

\author{Shuang~Zhang}
\email{shu.zhang@fz-juelich.de}
\affiliation{Institute for Advanced Simulation (IAS-4), Forschungszentrum J\"ulich, D-52425 J\"ulich, Germany}

\author{Yuan-Zhuo~Ma}
\email{mayu@frib.msu.edu}
\affiliation{Facility for Rare Isotope Beams, Michigan State University, East Lansing, Michigan 48824, USA}
\affiliation{Department of Physics and Astronomy, Michigan State University, East Lansing, Michigan 48824, USA}

\author{Serdar~Elhatisari}
\email{selhatisari@gmail.com}
\affiliation{Faculty of Engineering and Natural Sciences,
Gaziantep Islam Science and Technology University, Gaziantep 27010, T\"urkiye}

\author{Dean~Lee}
\email{leed@frib.msu.edu}
\affiliation{Facility for Rare Isotope Beams, Michigan State University, East Lansing, Michigan 48824, USA}

\author{Ulf-G.~Mei{\ss}ner}
\email{meissner@hiskp.uni-bonn.de}
\affiliation{Helmholtz-Institut f\"ur Strahlen- und Kernphysik, Bethe Center for Theoretical Physics and Cluster of Excellence - Color meets Flavor, Universit\"at Bonn,
D-53115 Bonn, Germany}
\affiliation{Institute for Advanced Simulation (IAS-4), Forschungszentrum J\"ulich, D-52425 J\"ulich, Germany}
\affiliation{Peng Huanwu Collaborative Center for Research and Education, International Institute for Interdisciplinary
and Frontiers, Beihang University, Beijing 100191, China}

\begin{abstract}
Sign-problem-free quantum Monte Carlo (QMC) methods provide one of the few polynomial-scaling routes to controlled, nonperturbative benchmarks of medium-mass and heavy nuclei.  We present a detailed uncertainty analysis of the recently developed sign-problem-free spin-orbit lattice action LAT-OPT1 and use it to benchmark nuclear lattice effective field theory (NLEFT).  We quantify statistical, imaginary-time, finite-volume, temporal-discretization, auxiliary-field-induced, and residual-sign uncertainties, finding that the cumulative many-body computational uncertainty in ground-state energies of doubly magic nuclei up to $^{100}$Sn is well below the percent level.  We also show that the spin-orbit interaction must be included nonperturbatively in the Euclidean-time projection: treating it only as a perturbative insertion misses shell-structure contributions of several to ten MeV in medium-mass nuclei.

In response to recent criticism of NLEFT benchmarks, we also revisit the relation between lattice transfer matrices, lattice Hamiltonians, Hartree--Fock variational bounds, finite-box and thermodynamic-limit calculations, and the continuum-limit behavior of regulated lattice interactions.  We identify several conceptual and technical errors in the analysis of Ref.~\cite{Rothman2026_NuLattice}.  These include (i) the comparison of inequivalent lattice transfer-matrix and lattice-Hamiltonian calculations, (ii) an inconsistent determination of correlation energies from comparisons of Hartree--Fock and full ground-state calculations with different boundary conditions, (iii) the attribution of nuclear saturation to lattice artifacts rather than to nonlocal smearing of interactions, a mechanism that can be demonstrated in continuous space, and (iv) an incorrect renormalization of short-range two-body interactions in the continuum limit.  When the same regulated lattice theory, renormalization prescription, and finite-volume boundary conditions are used consistently and analyzed properly, the reported discrepancies and concerns about the corresponding published NLEFT results are resolved.
\end{abstract}
\date{\today}

\maketitle

\section{Introduction}
\label{sec:Introduction}

Over the past few decades, modern nuclear \textit{ab initio} theory has witnessed rapid advancements by combining microscopic interactions, particularly those derived from chiral effective field theory (EFT)~\cite{Epelbaum2009_RMP81-1773,Hammer2020_RMP92-025004,Epelbaum2012PPNP67-343,Machleidt2014_EPJConf66-01011,Machleidt2024_PPNP137-104117,Machleidt2011PhysRept503-1,Epelbaum2020_FrontPhys8-98}, with systematically improvable many-body methods~\cite{Carlson2015_RMP87-1067,Barrett2013_PPNP69-131,Hagen2014_ReptProgPhys77-096302,Hergert2017_PhysScripta92-023002,Bogner2020_FrontinPhys8-379,Barbieri2004_PPNP52-377,Lee2009_PPNP63-117}.
Driven by the continuous innovation of diverse algorithmic approaches, this framework has achieved remarkable success in reproducing a wealth of observables crucial to both experimental nuclear physics and nuclear astrophysics.
Given these achievements, it is increasingly argued that quantum many-body methods have reached a level of maturity where the primary remaining challenges and uncertainties stem not from the many-body solvers themselves, but from the underlying nuclear forces~\cite{Ekstrom2023_FrontinPhys11-1129094, Maris2023_FrontinPhys11-1098262, Machleidt2023_FewBodySyst64-77}.
This optimistic perspective, however, can only be justified once the systematic errors inherent to the many-body algorithms have been thoroughly controlled and eliminated. While such control is accessible for light nuclei due to their modest computational demands, it remains formidably difficult for medium-mass and heavy systems.
In particular, although \textit{ab initio} calculations have recently been extended to heavy systems such as tin isotopes~\cite{Morris2018_PRL120-152503, Mougeot2021_NatPhys17-1099, Arthuis2020_PRL125-182501, Tichai2024_PLB851-138571, Niu2025_PRL135-222504, Hildenbrand2026_PRL136-062501} and $^{208}\text{Pb}$~\cite{Hu2022_NatPhys18-1196}, these simulations often lack a comprehensive quantification of their various systematic algorithmic uncertainties.

A major consequence of insufficient error quantification is the obfuscation of the intrinsic link between nuclear forces and nuclear structure within \textit{ab initio} frameworks. Because early calculations revealed that chiral interactions optimized solely to nucleon-nucleon (NN) scattering and few-body data frequently failed to reproduce the properties of medium-mass nuclei and infinite nuclear matter~\cite{Hagen2012_PRL109-032502, Otsuka2010_PRL105-032501, Hebeler2010_PRL105-161102, Hebeler2011_PRC83-031301, Binder2014_PLB736-119}, it became common practice to readjust three-nucleon (3N) low-energy constants (LECs), such as $c_D$ and $c_E$~\cite{Gazit2009_PRL103-102502, Ekstrom2015_PRC91-051301R, Ekstrom2018_PRC97-024332}, or to restrict simulations to specific resolution scales~\cite{Hebeler2011_PRC83-031301, Roth2012_PRL109-052501, Simonis2017_PRC96-014303}.
This readjustment procedure inadvertently absorbs the systematic errors of the many-body solvers into the interaction parameters, thereby masking the true origin of theoretical discrepancies.
This situation closely mirrors phenomenological mean-field models, where omitted many-body correlations are compensated for by replacing bare interactions with effective ones~\cite{Bender2003_RMP75-121, Meng2006_PPNP57-470, Robledo2019_JPG46-013001}.
Consequently, the deficiencies of the many-body solver are masked by the fine-tuning of the Hamiltonian, yielding superficially precise predictions through fortuitous error cancellations.
Particularly, this strategy compromises the core tenet of the \textit{ab initio} approach, which demands predictive power rooted strictly in few-body inputs, and poses severe risks when extrapolating to unexplored regions of the nuclear chart~\cite{Ekstrom2023_FrontinPhys11-1129094, Drischler2021_PPNP121-103888}.
To resolve these issues, it is paramount to rigorously quantify many-body uncertainties, thereby decoupling algorithm-induced uncertainties from the intrinsic limitations of the underlying nuclear forces.

Generally, systematic errors such as basis truncation errors or finite-volume effects can be mitigated via numerical extrapolation to the infinite-basis or infinite-volume limits. While such limits usually cannot be physically reached due to computational constraints, these extrapolation procedures provide a rigorous estimate of the associated uncertainties. In certain regimes, such as heavy nuclei under typical computational setups, these systematic uncertainties can be substantial.
Consequently, benchmarking different many-body methods or comparing results with experiment becomes unreliable without a systematic control of these errors.
A historical parallel can be found in lattice QCD, where early tensions among different collaborations, and between theory and experiment were successfully resolved through rigorous systematic error control.
Similarly, thoroughly examining internal systematic errors is an indispensable first step toward resolving current inconsistencies in modern nuclear \textit{ab initio} calculations.

Unfortunately, the accurate quantification of systematic errors within modern many-body algorithms is generally an exceptionally daunting task, particularly in the heavy-mass realm.
While certain uncertainties, such as basis truncation, can be straightforwardly estimated by comparing results calculated using different model space parameters, the fundamental challenge stems primarily from the systematic approximations inherent to various many-body solvers.
For instance, coupled-cluster (CC) theory~\cite{Hagen2014_ReptProgPhys77-096302} and the in-medium similarity renormalization group (IMSRG)~\cite{Hergert2016_PhysRept621-165} rely on truncating many-body operator expansions, whereas nuclear lattice effective field theory (NLEFT) employs perturbative expansions around non-perturbative actions designed to mitigate the sign problem.
Estimating the magnitude of these many-body or perturbative truncation errors by incorporating higher-order corrections demands a prohibitive increase in computational resources, often scaling with high power-law or exponential complexity. Consequently, the uncertainty quantification is frequently not rigorously performed in heavy-mass applications due to these severe computational bottlenecks.

Notably, there exist nuclear many-body models where systematic errors can be rigorously estimated and controlled.
For example, few-body systems can be solved exactly via direct diagonalization~\cite{Borasoy2006_NPA768-179, Epelbaum2009_EPJA41-125, Elhatisari2024_PLB859-139086, Wang2025_PRC112-025502, Zhang2025_PRD111-036002}, meaning that systematic errors originate exclusively from basis truncation or finite-volume effects, which scale as negative powers of the system size.
For medium-mass and heavy nuclei, sign-problem-free QMC methods represent the only polynomial-scaling frameworks with this capability.
By recasting the many-body problem into a high-dimensional integral evaluated through stochastic sampling, these methods yield a statistical uncertainty that scales as $\mathcal{O}(1/\sqrt{N})$ regardless of the system size.
However, for generic interactions, the sampling weights are not positive-definite, giving rise to the notorious sign problem where the signal-to-noise ratio decays exponentially with system size~\cite{Troyer2005_PRL94-170201}.
Fortunately, certain tailored models, such as those respecting Wigner SU(4) symmetry or time-reversal symmetry after an auxiliary-field transformation, are completely free from this bottleneck, allowing both statistical and systematic errors to scale with low power-law complexity that are straightforward to remove~\cite{ProgTheorPhys110-615, PRB71-155115, PRB91-241117, AnnuRevCondMattPhys10-337}.

NLEFT leverages precisely this advantage. To circumvent the sign problem in heavy-element simulations, modern NLEFT calculations typically construct a non-perturbative leading-order (LO) action that is free of, or exhibits only a mild, sign problem, while leaving complex higher-order chiral operators to be treated via advanced algorithms such as symmetry-sign extrapolation~\cite{Lahde2015_EPJA51-92}, perturbative quantum Monte Carlo~\cite{Lu2022_PRL128-242501}, wave function matching~\cite{Elhatisari2024_Nature630-59}, rank-one operator~\cite{Ma2024_PRL132-232502}, etc.
As an essential first step toward a comprehensive uncertainty quantification within NLEFT, this work focuses exclusively on these non-perturbative LO interactions.
Because the sign problem is absent or well-controlled at this level, all systematic errors can be rigorously quantified, thereby establishing a solid and indispensable foundation for the full error analysis of future NLEFT simulations with full chiral forces.

In NLEFT, nuclear interactions are formulated on a spatial lattice and solved via Euclidean-time auxiliary-field Monte Carlo methods~\cite{FrontPhys8-174, AnnRevNuclPartSci75-109}. This approach has been widely applied to calculate nuclear ground states~\cite{EPJA31-105, PhysRevLett.104.142501, EPJA45-335, PLB732-110, Lu2019_PLB797-134863, Lu2022_PRL128-242501, Elhatisari2024_Nature630-59} and excited states~\cite{PhysRevLett.112.102501, Nat.Comm.14-2777, PhysRevLett.132.062501,PRL134-162503}, explore intrinsic nuclear densities and clustering phenomena~\cite{PhysRevLett.106.192501, PhysRevLett.109.252501, PhysRevLett.110.112502, PRL119-222505, Zhang2025_PLB869-139839}, extract nucleus-nucleus scattering cross sections~\cite{PhysRevC.86.034003, Nature528-111}, simulate zero- and finite-temperature nuclear matter~\cite{Elhatisari2016_PRL117-132501, PhysRevLett.125.192502, PLB850-138463, Ma2024_PRL132-232502,Agar:2026nxr}, and investigate hypernuclear structures~\cite{PhysRevLett.115.185301,Frame2020_EPJA56-24, EPJA60-215}.

Most recent NLEFT studies initiate calculations with an LO interaction that respects Wigner SU(4) symmetry, either exactly or approximately, where the nuclear force is independent of spin and isospin~\cite{Wigner1937_PhysRev51-106, Lee2021_PRL127-062501}.
Consequently, spin-up and spin-down nucleons do not mix within the auxiliary-field formalism, yielding identical, real fermionic determinants that guarantee a positive-definite overall determinant for even-even nuclei.
This interaction can be smeared either locally or non-locally on the lattice to capture the finite, albeit short, range of the nuclear force.
Notably, the locality of this smearing has been identified as an essential element for binding nuclear clusters into finite nuclei~\cite{Elhatisari2016_PRL117-132501, PRL119-222505}.

Furthermore, a Wigner SU(4) three-body force was introduced in Ref.~\cite{Lu2019_PLB797-134863} to improve the description of light nuclei such as $^3\text{H}$ and $^4\text{He}$ and medium-mass nuclei up to calcium isotopes, where the predicted binding energies align well with experimental data below the $sd$-shell.
The major limitation of a purely Wigner-SU(4) action is that it cannot generate realistic spin-orbit shell splittings, and therefore cannot describe the observed non-harmonic-oscillator magic numbers in heavier nuclei.  The recently developed LAT-OPT1 action resolves this by incorporating a spin-orbit density into the same antiunitary-symmetric auxiliary-field structure that protects positivity of the Monte Carlo weight.  This is the key new physics ingredient highlighted in the present work: the spin-orbit force is not added as an uncontrolled post-processing correction, but is embedded in the nonperturbative propagation while preserving sign-problem-free sampling for even-even nuclei.
The resulting interaction serves as an exactly solvable, nonperturbative benchmark model that reaches the accuracy of popular phenomenological mean-field interactions for light and medium-mass nuclei up to $N, Z \le 28$~\cite{Niu2025_PRL135-222504}.

So far, the various sign-problem-free interactions have been successfully employed at the phenomenological level to investigate light-nucleus binding energies and charge radii, describe emergent geometry in $^{12}\text{C}$~\cite{Nat.Comm.14-2777}, examine clustering in hot nuclear matter~\cite{PhysRevLett.125.192502, PLB850-138463,Agar:2026nxr}, resolve Beryllium isotope spectra~\cite{PRL134-162503}, analyze $\alpha$-particle monopole transitions~\cite{PhysRevLett.132.062501}, conduct multi-channel calculations~\cite{Wang2025_arXiv2512.21942}, and study the $^5\text{He}$ resonant state~\cite{Niu2026_arXiv2603.25081}. Moreover, it provides a foundation for perturbative calculations using realistic chiral forces up to second order~\cite{Lu2022_PRL128-242501, Liu2025EPJA61-85, Shi2026_PLB874-140303, Wang2026_arXiv2604.20681}.
Furthermore, several recent NLEFT calculations non-perturbatively incorporate a regulated one-pion-exchange potential (OPEP) alongside the sign-problem-free interactions~\cite{Elhatisari2024_Nature630-59, Ma2024_PRL132-232502, Zhang2025_PLB869-139839, Ren2025_PRL135-152502, Song2026_PLB872-140086, Hildenbrand2026_PRL136-062501, Wang2025_PRC112-025502}.
While this does not strictly preclude the sign problem, the resulting sign oscillation remains remarkably weak and negligible in many physically relevant cases. Such a non-perturbative OPEP is indispensable for reproducing nucleon-nucleon scattering data and crucial single-particle properties.

It is worth noting that these LO interactions are not rigidly fixed by few-body inputs alone, but can be phenomenologically tailored to absorb essential elements of the realistic nuclear force.
Consequently, they can function either as self-contained, exactly solvable phenomenological forces or as optimized foundations for full chiral EFT expansions.
In the latter context, any adjustments made to the LO interaction can be systematically absorbed by a renormalization of the higher-order low-energy constants (LECs).
Nevertheless, a well-chosen LO interaction captures the dominant components of the nuclear force, thereby substantially accelerating the convergence of the chiral expansion.

This study has two intertwined goals.  First, we give a quantitative error budget for the recently developed sign-problem-free spin-orbit action~\cite{Niu2025_PRL135-222504}.  We show that the auxiliary-field implementation, induced operators, temporal discretization, finite-volume effects, and residual phase fluctuations are all quantitatively controlled.  Second, we use these controlled calculations to clarify a set of recent benchmarking misunderstandings concerning NLEFT.  Because different LO lattice actions share the same transfer-matrix and auxiliary-field logic, the conclusions about regulator consistency, variational bounds, and thermodynamic-limit extrapolation apply broadly beyond LAT-OPT1.  To make this point explicit, we supplement the LAT-OPT1 analysis with comparative calculations using Wigner-SU(4), OPEP-inclusive, transfer-matrix, and Hamiltonian formulations.

The remainder of this paper is organized as follows.  In Sec.~\ref{sec:Method}, we define the lattice Hamiltonian, the smeared density and spin-orbit operators, and the
auxiliary-field transformation used in the AFQMC projection.  We also review
the antiunitary symmetry that guarantees determinant positivity for even-even
nuclei and summarize the lattice actions used in the benchmark comparisons.
In Sec.~\ref{sec:Systematic}, we quantify the systematic uncertainties of the LAT-OPT1
spin-orbit action, including projection-time and finite-volume extrapolations,
finite-temporal-step effects, auxiliary-field-induced operators, the
nonperturbative role of the spin-orbit interaction, and residual phase
fluctuations in odd systems.  In Sec.~\ref{sec:NLEFT}, we use controlled calculations with several NLEFT actions to examine benchmark consistency.  We compare transfer-matrix and Hamiltonian formulations, analyze finite-nucleus
Hartree--Fock bounds, correlation energies, many-body correlations in
neutron matter, thermodynamic-limit extrapolations,
nuclear saturation, and the high-density behavior of regulated interactions.
In Sec.~\ref{sec:benchmarking_lessons}, we summarize the lessons learned regarding recent benchmark criticisms, emphasizing the need to compare the same regulated lattice theory, renormalization prescription, and finite-volume limit.  Finally, Sec.~\ref{sec:Summary} presents our summary and outlook.

\section{Method}
\label{sec:Method}

We work within the framework of NLEFT, where nucleons are defined on a three-dimensional cubic lattice and the many-body problem is solved by auxiliary-field quantum Monte Carlo (AFQMC). 
Ground-state observables are extracted through imaginary-time projection. For a trial state $|\Phi_T\rangle$ with nonvanishing overlap with the ground state, the projected expectation value of an operator $O$ is
\begin{equation}
\langle O \rangle =
\lim_{\tau\to\infty}
\frac{
\langle \Phi_T|e^{-\tau H/2} O e^{-\tau H/2}|\Phi_T\rangle
}{
\langle \Phi_T|e^{-\tau H}|\Phi_T\rangle
},
\label{eq:projected_expectation}
\end{equation}
where $H$ is the many-body Hamiltonian and $\tau$ is the total imaginary time. In practical calculations, the projection time is discretized as $\tau=L_t a_t$, with $L_t$ the number of temporal slices and $a_t$ the temporal step. The propagation is then discretized into a product of many transfer matrices $M = :e^{-a_t H}:$ and extrapolated to the $\tau\to\infty$ limit, where the colons denote the normal ordering for single particle operators.

The auxiliary-field transformation in AFQMC rewrites the many-body propagation as an integral over auxiliary fields. 
For a Slater-determinant trial state, the projected amplitude takes the schematic form
\begin{equation}
\langle \Phi_T|M^{L_t}|\Phi_T\rangle
=
\int Ds\, e^{-S_B[s]}\,\det Z(s),
\label{eq:AFQMC_weight}
\end{equation}
where $s$ denotes the auxiliary field, $S_B[s]$ is the corresponding bosonic action, and $Z(s)$ is the fermionic correlation matrix generated by the associated one-body propagation. The sign or phase of $\det Z(s)$ therefore determines whether the Monte Carlo sampling remains numerically tractable. The sign problem is absent only when this determinant is nonnegative for all relevant auxiliary-field configurations.

The central action studied in this work is LAT-OPT1, a minimal sign-problem-free lattice nuclear Hamiltonian that combines a Wigner--SU(4) central interaction with a spin-orbit density while retaining the antiunitary symmetry needed for positive Monte Carlo weights~\cite{Niu2025_PRL135-222504},
\begin{equation}
H=
\sum_{\mathbf n}
\left[
-\frac{\Psi^\dagger \nabla^2 \Psi}{2\,m_{N}}
+
:
\frac{C_2}{2}\,(\overline{\rho} + \frac{C_s}{2C_2}\overline{\rho}_s)^{\,2}
+\frac{C_3}{6}\,\overline{\rho}^{\,3}
:
\right].
\label{eq:H_LATOPT1}
\end{equation}
Here $\Psi(\mathbf n)$ and $\Psi^\dagger(\mathbf n)$ are the nucleon annihilation and creation operators at lattice site $\mathbf n$, $m_{N}$ is the nucleon mass, and implicit spin and isospin indices are summed. 
The square-completed form is central to the construction.  It allows the linear spin-orbit term to be generated by the same real auxiliary field that couples to the density, rather than by a complex auxiliary field that would reintroduce a severe sign problem.  The price is the accompanying $\mathcal{O}(\rho_s^2)$ operator, whose size is quantified below and shown to be a higher-order correction.
The overline denotes smeared operators.  
The nonlocally smeared single-particle operator is defined as
\begin{equation}
\overline{\Psi}(\mathbf n)
=
\Psi(\mathbf n)
+
s_{\rm NL}
\sum_{|\mathbf n'-\mathbf n|=1}
\Psi(\mathbf n').
\label{eq:nonlocal_smearing}
\end{equation}
Using these operators, we define the density
\begin{equation}
\rho(\mathbf n)=\overline{\Psi}^{\dagger}(\mathbf n)\,\overline{\Psi}(\mathbf n)
\end{equation}
and the spin-orbit density
\begin{equation}
\rho_s(\mathbf n)
=
\frac{i}{4}
\sum_{ijk}
\epsilon_{ijk}\,
\nabla_i
\left[
\overline{\Psi}^{\dagger}(\mathbf n)
\bigl(
\overrightarrow{\nabla}_j-\overleftarrow{\nabla}_j
\bigr)
\sigma_k
\overline{\Psi}(\mathbf n)
\right],
\label{eq:rho_s_def}
\end{equation}
where $\epsilon_{ijk}$ is the Levi-Civita symbol and $\sigma_k$ is the Pauli matrix in spin space. Local smearing is then applied to the density operators,
\begin{equation}
\overline{\sigma}(\mathbf n)
=
\sigma(\mathbf n)
+
s_{\rm L}
\sum_{|\mathbf n'-\mathbf n|=1}
\sigma(\mathbf n'),
\qquad
\sigma=\rho,\rho_s.
\label{eq:local_smearing}
\end{equation}
A static Coulomb force is included perturbatively and is not part of the sign-problem-free auxiliary-field sampling.

All calculations below are performed within a common baseline setup. Unless otherwise specified, we use a spatial lattice spacing $a=1.32~\mathrm{fm}$, a cubic box with $L=11$, and a temporal step $a_t=(1000~\mathrm{MeV})^{-1}$. 

Following Wu and Zhang~\cite{PRB71-155115}, a sufficient condition for sign-problem-free sampling is that the auxiliary-field transformed transfer matrix be invariant under an antiunitary operator $\mathcal T$ satisfying $\mathcal T^2=-1$. For even-even nuclei, we choose the single-particle orbitals in the trial state as Kramers pairs,
\begin{equation}
|\phi_{A/2+k}\rangle=\mathcal T|\phi_k\rangle,
\qquad
k=1,\ldots,A/2,
\label{eq:TR_pairing}
\end{equation}
where the $\phi_k$ are single-particle wave functions consisting of the trial Slater determinant.
For a fixed $s$-field, if the transformed transfer matrix $M(s)$
satisfies
\begin{equation}
[ M(s),\mathcal T]=0,
\end{equation}
then the correlation matrix $Z_{ij}(s)=\langle \phi_i| M(s)|\phi_j\rangle$
takes the specific form
\begin{equation}
Z(s)=
\begin{pmatrix}
U(s) & -V(s)^*\\
V(s) & U(s)^*
\end{pmatrix},
\label{eq:Z_block}
\end{equation}
with $U(s)$ and $V(s)$ complex $A/2\times A/2$ matrices.

Define a block matrix
\begin{equation}
\Sigma=i\sigma_y\otimes I_{A/2}
= \begin{pmatrix}
0 & I_{A/2}\\
-I_{A/2} & 0
\end{pmatrix},
\label{eq:Sigma_def}
\end{equation}
we can verify that
\begin{equation}
Z(s)\Sigma=\Sigma Z(s)^*.
\label{eq:ZSigma}
\end{equation}
For any eigenvalue $\lambda$ and corresponding eigenvector $v$ of $Z(s)$, $Z(s)v=\lambda v$, we have
\begin{equation}
Z(s)\bigl(\Sigma v^*\bigr)=\lambda^* \bigl(\Sigma v^*\bigr).
\end{equation}
Thus complex eigenvalues occur in conjugate pairs. If $\lambda$ is real, $v$ and $\Sigma v^*$ belong to the same eigenspace and are orthogonal because $\Sigma^T=-\Sigma$, so every real eigenvalue is at least doubly degenerate. Therefore,
\begin{equation}
\det Z(s)\ge 0
\label{eq:det_positive}
\end{equation}
for every auxiliary-field configuration. The AFQMC sampling is thus sign-problem-free for even-even nuclei.

For attractive interactions such as the nuclear force ($C_2 < 0$), the two-body interaction in Eq.~(\ref{eq:H_LATOPT1}) can be decoupled via a real auxiliary field. Since both $\overline{\rho}$ and $\overline{\rho}_s$ are even under time reversal, the resulting one-body propagator commutes with $\mathcal{T}$ for each individual auxiliary-field configuration. This is the microscopic reason why the LAT-OPT1 spin-orbit action remains sign-problem-free for even-even nuclei.  The spin-orbit force therefore enters the nonperturbative projection on the same footing as the central force, while the Kramers-pair structure protects determinant positivity.  For odd-$A$ and odd-odd nuclei, this pairing is incomplete, and a residual sign or phase problem may persist; Sec.~\ref{subsec:applicability_beyond_exact_positivity} shows that this residual problem is mild in practice.

In Table~\ref{tab:forces} we summarize the lattice interactions considered in this work.
The \texttt{Isotropic} interaction denotes the two-body interaction with local and nonlocal smearing developed for studies of nuclear clustering in combination with the pinhole algorithm~\cite{PRL119-222505}. It includes the one-pion-exchange potential (OPEP) but omits the Coulomb interaction, whose effects on light- and medium-mass nuclei are relatively small and can largely be absorbed into the effective nuclear interaction. To reduce the computational cost, this interaction is employed with relatively coarse lattice spacings, $a=a_t=(100~\mathrm{MeV})^{-1}$.
The \texttt{EE} interaction denotes the Wigner-SU(4) lattice interaction consisting of two- and three-body forces, designed to reveal the essential elements of the nuclear binding~\cite{Lu2019_PLB797-134863}. The \texttt{LAT-OPT1} interaction is a sign-problem-free interaction that additionally includes a nonperturbative spin-orbit coupling term~\cite{Niu2025_PRL135-222504}. For these latter two interactions we adopt a finer spatial lattice spacing, $a=(150~\mathrm{MeV})^{-1}$, together with the Hamiltonian formulation, corresponding to the limit $a_t^{-1}\rightarrow\infty$ (i.e., the temporal lattice spacing is removed).

The calibration procedure is part of the definition of each regulated lattice theory. In particular, the interaction parameters listed in Table~\ref{tab:forces} are not determined solely by two-nucleon observables. For example, in the  \texttt{EE} interaction the two-body strength and range are constrained by the averaged s-wave scattering length and effective range, while the three-body coupling is fixed to the triton binding energy. The local and nonlocal smearing parameters, $s_{\rm L}$ and $s_{\rm NL}$, are then selected using finite-nucleus information, in particular the liquid-drop systematics of $N=Z$, even-even nuclei in the range $16\le A\le40$. With their respective calibrations, all three interactions are capable of reproducing the binding energies of light- to medium-mass nuclei. It should be emphasized that the interaction parameters depend explicitly on the spatial and temporal lattice spacings. Therefore, meaningful benchmark comparisons require identical interaction parameters together with the same values of $a$ and $a_t$ listed in Table~\ref{tab:forces}. Using the same interaction parameters with different lattice spacings can lead to different effective theories, yielding binding energies that may differ by as much as a factor of two, as demonstrated in the examples presented in the main text.

In discussing the systematic uncertainties of nonperturbative lattice QMC calculations, we consistently employ the most recent \texttt{LAT-OPT1} interaction. We then examine the consequences of using non-negligible temporal cutoffs by considering the \texttt{Isotropic} interaction, which is characterized by finite spatial and temporal cutoffs, also maintaining consistency with the original formulation in the literature. The influence of boundary conditions on nuclear matter calculations is investigated using the \texttt{EE} interaction, again following the setup adopted in the original work. We also discuss the role of nuclear saturation in lattice simulations using newly optimized lattice interactions. In each section, we explicitly specify the interaction employed together with the corresponding lattice cutoffs and other relevant computational settings, such as the kinetic-energy improvement scheme, to ensure the reproducibility of our calculations.

\begin{table*}[t]
\caption{Parameters of the lattice interactions used in this work. $C_2$, $C_3$,
and $C_s$ are the two-body, three-body, and spin-orbit coupling
strengths, respectively, and $s_\mathrm{L}$ and $s_\mathrm{NL}$ are the
local and nonlocal smearing parameters. The OPEP and Coulomb columns
indicate whether these interactions are included in the calibration to
nuclear binding energies. The quantities $a^{-1}$ and $a_t^{-1}$ denote
the spatial and temporal ultraviolet cutoffs, with $a$ and $a_t$ the
corresponding lattice spacings. The limit $a_t^{-1}=\infty$ corresponds
to the Hamiltonian formulation. Benchmark comparisons should employ
identical interaction parameters and lattice cutoffs.}
\label{tab:forces}
\begin{tabular}{
l
S[table-format=-1.2e-2]
S[table-format=-1.2e-2]
S[table-format=1.3]
S[table-format=1.2]
S[table-format=1.2e-2]
c c
S[table-format=3.0]
c
}
\toprule
Force
& {$C_{2}$ (MeV$^{-2}$)}
& {$C_{3}$ (MeV$^{-5}$)}
& {$s_\mathrm{L}$}
& {$s_\mathrm{NL}$}
& {$C_{s}$ (MeV$^{-4}$)}
& OPEP
& Coulomb
& {$a^{-1}$ (MeV)}
& {$a_t^{-1}$ (MeV)} \\
\midrule
\texttt{Isotropic}~\cite{PRL119-222505}
& -1.85e-5
& 0
& {$0.080$}
& {$0.08$}
& 0e0
& Y
& N
& {$100$}
& {$100$}
\\
\texttt{EE}~\cite{Lu2019_PLB797-134863}
& -3.41e-7
& -1.40e-14
& {$0.061$}
& {$0.50$}
& 0
& N
& Y
& {$150$}
& {$\infty$}
\\
\texttt{LAT-OPT1}~\cite{Niu2025_PRL135-222504}
& -4.41e-7
& 1.56e-15
& {$0.081$}
& {$0.45$}
& 8.59e-12
& N
& Y
& {$150$}
& {$\infty$}
\\
\bottomrule
\end{tabular}
\end{table*}

\section{Systematic uncertainties in projection Monte Carlo methods}
\label{sec:Systematic}

The primary objective of this section is to establish that LAT-OPT1 is not merely a useful phenomenological interaction, but a quantitatively controlled nonperturbative benchmark Hamiltonian. 
We therefore focus on the uncertainty budget of the spin-orbit action itself. 
We fix the lattice spacing and the interaction parameters, rather than pursuing the continuum limit. This defines a precise lattice Hamiltonian with a well-defined ground-state energy against which many-body algorithms can be benchmarked. Specifically, we analyze the asymptotic behavior of the imaginary-time projection, finite-volume extrapolations, temporal-step effects, and auxiliary-field-induced operators associated with the square-completed spin-orbit implementation. This comprehensive assessment is carried out for nuclei ranging from $^{4}\mathrm{He}$ to $^{100}\mathrm{Sn}$. 
The Coulomb interaction is omitted in this section so that the entire calculation, including the spin-orbit dynamics, is performed nonperturbatively.

\subsection{Infinite-time and infinite-volume limit}
\label{subsec:baseline_convergence}

In AFQMC simulations, the primary systematic uncertainties arise from extrapolations to the infinite-projection-time and infinite-volume limits. While light nuclei possess compact wave functions that converge rapidly, heavier systems are more susceptible to finite-time and finite-volume effects. To mitigate these deviations, we numerically investigate the asymptotic convergence to optimize key control parameters, such as the spatial box size $L$. This establishes a balance between high-precision theoretical predictions and computational efficiency for subsequent calculations.

We first consider the imaginary-time extrapolation.
At sufficiently large projection time, the excited-state contamination is exponentially suppressed, and the projected energy can be written as a function of imaginary time $\tau$,
\begin{equation}
E(\tau)=E(\infty)+\sum_{n\ge 1} A_n e^{-\tau\Delta_n},
\qquad
\Delta_n\equiv E_n-E_0>0.
\label{eq:E_tau_series}
\end{equation}
In practice, once the asymptotic regime is reached, it is sufficient to retain only the leading contribution,
\begin{equation}
E(\tau)=E(\infty)+C\,e^{-\Delta_1\tau},
\label{eq:exp_extrap_Lt}
\end{equation}
with $\Delta_1$ the immediate energy gap above the ground state.

\begin{figure}[htbp]
    \centering
    \includegraphics[width=0.90\columnwidth]{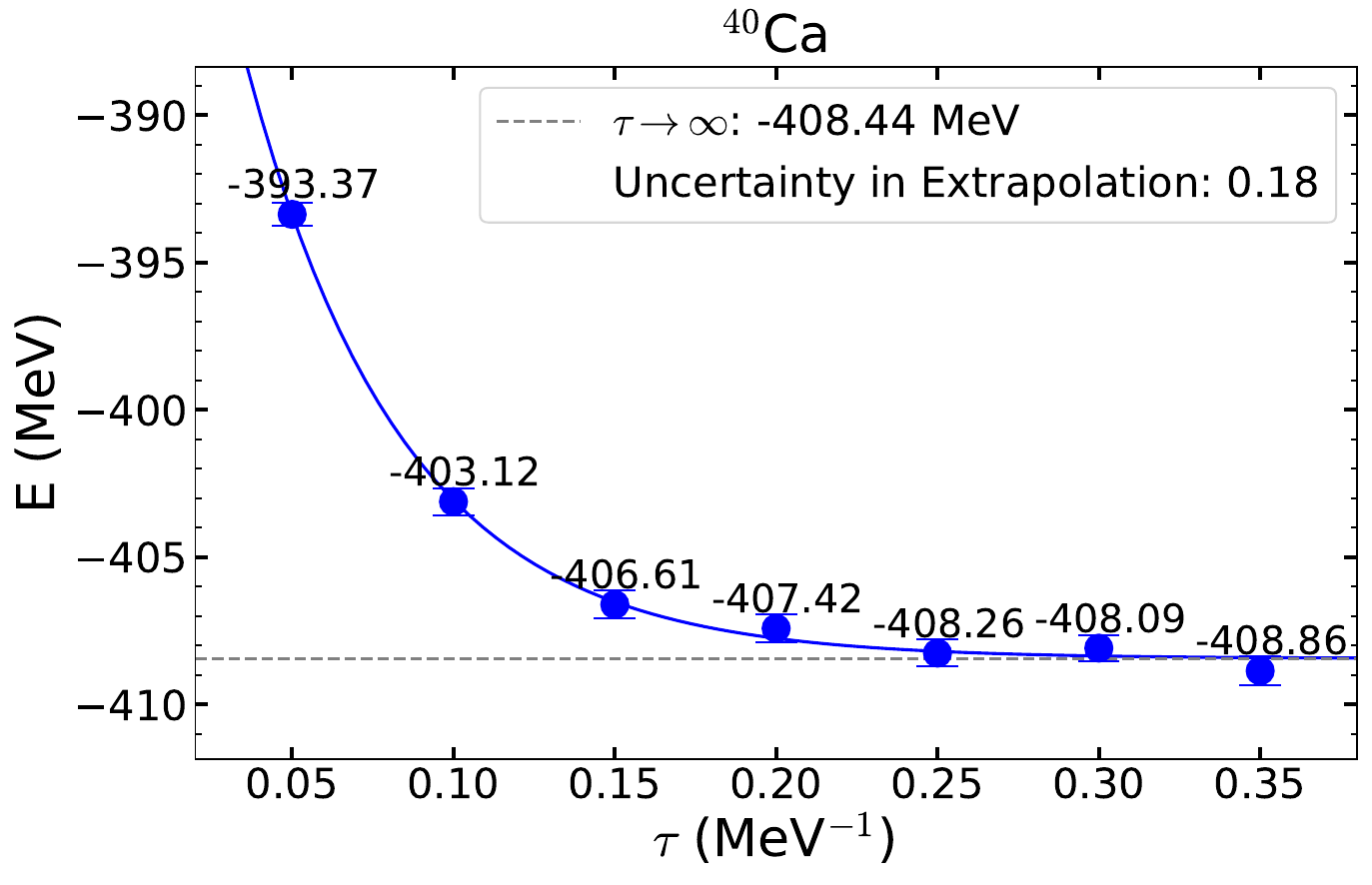}
    \caption{Binding energy of $^{40}\mathrm{Ca}$, excluding Coulomb interactions, as a function of the projection time $\tau$ in a periodic box of size $L=11$. The solid curve denotes the exponential fit according to Eq.~\eqref{eq:exp_extrap_Lt}. }
    \label{fig:Ca40_tau_extrap}
\end{figure}

Fig.~\ref{fig:Ca40_tau_extrap} shows a representative example for $^{40}$Ca. As $\tau$ increases, the projected energies approach a clear plateau, and the extrapolated limit remains stable within a narrow uncertainty band. The fit yields both the limit energy and the combined statistical and $\tau$-extrapolation errors,
\begin{equation}
E(\infty)\bigl(^{40}\mathrm{Ca}\bigr)=-408.44~\mathrm{MeV},
\quad
\delta E_{\rm extr}=0.18~\mathrm{MeV},
\label{eq:Ca40_extrap_value}
\end{equation}
demonstrating that the $\tau\to\infty$ limit can be extracted reliably from the finite-$\tau$ data under the present setup.

\begin{figure}[htbp]
    \centering
    \includegraphics[width=\columnwidth]{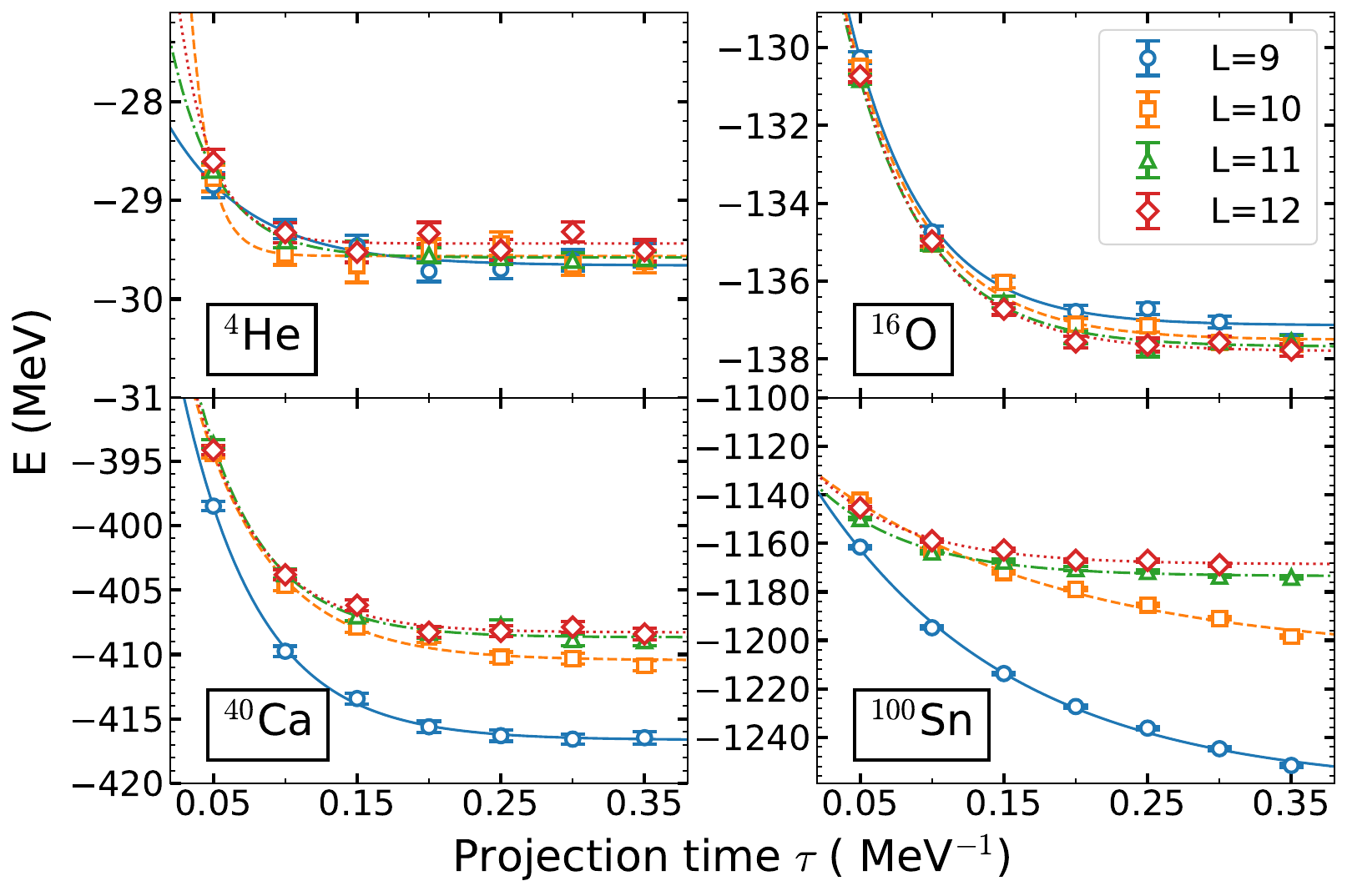}
    \caption{Binding energies, excluding Coulomb interactions, as a function of the projection time $\tau$ in periodic boxes of sizes $L = 9, 10, 11$, and $12$. The lines denote exponential fits according to Eq.~\eqref{eq:exp_extrap_Lt}.
    \label{fig:FVscan_4nuclei}}
\end{figure}

Having quantified the time-extrapolation uncertainties, we next examine the finite-volume dependence. Comparisons across different spatial box sizes are performed only after applying the $\tau\to\infty$ extrapolation for each value of $L$. We define the residual finite-volume effect relative to $L=11$ as
\begin{equation}
\Delta_L \equiv E_{L\rightarrow\infty}(\tau\rightarrow\infty)-E_{L=11}(\tau\rightarrow\infty),
\label{eq:DeltaL}
\end{equation}
and use the energy difference between $L=11$ and a larger box, such as $L=12$, to estimate this residual shift. Fig.~\ref{fig:FVscan_4nuclei} displays the projected energies for $^{4}$He, $^{16}$O, $^{40}$Ca, and $^{100}$Sn at $L=9, 10, 11,$ and $12$. For the heavier $^{40}$Ca and $^{100}$Sn nuclei, the smallest box ($L=9$) already exhibits noticeable finite-volume deviations. In contrast, the extrapolated results for $L=11$ and $L=12$ agree within the respective time-extrapolation uncertainties in all cases. This demonstrates that, for the ground-state energies considered here, the choice of $L=11$ is sufficiently converged with respect to the infinite volume limit. 
In what follows we fix box size to $L=11$.

\subsection{Continuous-time limit}
\label{subsec:atextrapolation}

Next, we examine the dependence on the temporal step size $a_t$. This is a nuanced issue specific to lattice simulations but it is well understood in
the lattice community.
%that has occasionally been a source of confusion in the literature. 
In this work, we operate at a fixed spatial lattice spacing and construct the Hamiltonian using operators defined on the lattice sites. Consequently, the imaginary-time evolution serves merely as a numerical tool to extract the ground state of this specific lattice Hamiltonian. Therefore, the converged physical results must be independent of the imaginary-time discretization, which we will now verify numerically.

The crucial point is that the transfer matrix is not equivalent to the imaginary-time evolution operator over a finite time step $a_t$. Normal ordering is introduced primarily to facilitate the path-integral formulation. Specifically, standard auxiliary-field transformations hold exactly only under normal ordering. Taking the conventional Hubbard-Stratonovich transformation as an example,
\begin{equation}
:e^{-a_t(K + C\rho^2/2)}: = \int_{-\infty}^{+\infty} \frac{ds}{\sqrt{2\pi}} :e^{-s^2/2-a_t K + \sqrt{-a_t C}s\rho}:,   \label{eq:HStransform}
\end{equation}
where $K$ is the kinetic energy operator and $C$ is the coupling constant for the two-body contact interaction. 
For brevity we omit the summation over lattice sites.
Note that Eq.~\eqref{eq:HStransform} is exact to all orders in $a_t$. However, its proof relies fundamentally on the fact that $K$ and $\rho$ commute within the normal-ordering symbols, allowing both operators to be treated as $c$-numbers. In contrast, if we omit normal ordering, Taylor expand both sides of Eq.~\eqref{eq:HStransform}, and integrate out the auxiliary field, the resulting $\mathcal{O}(a_t^2)$ terms are
\begin{align}
\mathrm{l.h.s.} =& \frac{a_t^2}{2}\Big[K^2 + \frac{C}{2}(K\rho^2 + \rho^2K) + \frac{C^2\rho^4}{4}  \Big],   \nonumber \\
\mathrm{r.h.s.} =& \frac{a_t^2}{2}\left[
K^2
+\frac{C}{3}\left(K\rho^2+\rho K\rho+\rho^2K\right)
+\frac{C^2\rho^4}{4}
\right],
\end{align}
respectively. Evidently, these two expressions differ for any $a_t \neq 0$. This discrepancy becomes even more pronounced when introducing more noncommuting interaction terms. Therefore, to ensure that the auxiliary-field transformation remains exact at any finite $a_t$, it is highly advantageous to adopt the transfer matrix rather than the bare, un-ordered imaginary-time evolution operator.

The consequence of employing the transfer-matrix formalism is that the projected ground state is the exact eigenstate of the transfer matrix itself, rather than that of the target Hamiltonian $H$ and the true imaginary-time evolution operator. Nevertheless, we can formally introduce an effective Hamiltonian $H_\mathrm{eff}$ satisfying
\begin{equation}
  \exp(-a_t H_\mathrm{eff}) = :\exp(-a_t H):, \label{eq:Heff}
\end{equation}
whereby the projected ground state is the exact eigenstate of $H_\mathrm{eff}$ instead of $H$. For simple contact interactions, $H_\mathrm{eff}$ can be explicitly derived by taking the logarithm of Eq.~\eqref{eq:Heff}:
\begin{align}
H_\mathrm{eff}  = & -\frac{1}{a_t}\ln\left[ :\exp(-a_tH):\right] \nonumber \\
 = & -\frac{1}{a_t}\ln\left[ 1 - a_tH + \frac{a_t^2}{2}:H^2: + ...\right] \nonumber \\
 = &  H + \frac{a_t}{2}\left(H^2 - :H^2:\right) + \mathcal{O}(a_t^2).    
\end{align}
Here, the induced term proportional to $a_t$ can be recast into local interactions by applying Wick's theorem to bring the terms into normal-ordered form. This procedure not only generates operators at the one-, two-, and many-body levels, but it also induces derivative interactions through the contraction of local contact interactions with the kinetic energy term.

As the projection implicitly evolves the wave function using $H_\mathrm{eff}$ instead of $H$, the resulting wave function $|\Phi_\mathrm{eff}\rangle$ exhibits an order-$\mathcal{O}(a_t)$ deviation from the exact normalized eigenstate $|\Phi\rangle$ of $H$ (with eigenvalue $E$), given by
\begin{equation}
  |\Phi_\mathrm{eff}\rangle   = |\Phi\rangle + a_t |\delta \Phi\rangle + \mathcal{O}(a_t^2).
\end{equation}
The calculated energy is then evaluated via the expectation value of $H$ within $|\Phi_\mathrm{eff}\rangle$,
\begin{equation}
\frac{\langle \Phi_{\mathrm{eff}}|H|\Phi_{\mathrm{eff}}\rangle}
{\langle \Phi_{\mathrm{eff}}|\Phi_{\mathrm{eff}}\rangle}
=E+a_t^2\langle \delta\Phi |(H-E)|\delta\Phi\rangle
+O(a_t^3),
\end{equation}
where the linear $\mathcal{O}(a_t)$ term cancels exactly. Note that the suppression of this linear correction is a direct consequence of stationary perturbation theory and applies exclusively to the Hamiltonian. For general operators inserted at the middle time step, the discretization error remains of order $\mathcal{O}(a_t)$.

In the continuum time limit $a_t \rightarrow 0$, $H_\mathrm{eff}$ coincides with $H$, and the calculated energy becomes exact. This limit can be systematically approached by computing the energies using a series of different $a_t$ values and performing a quadratic extrapolation. Alternatively, one can insert counterterms into the Hamiltonian within the transfer matrix to cancel the induced terms up to a certain order in $a_t$, analogous to the Symanzik improvement program in lattice QCD. In practice, because the leading energy correction is of order $\mathcal{O}(a_t^2)$, a direct calculation employing a reasonably small $a_t$ yields adequate precision, as we will demonstrate numerically.

\begin{table}[htbp]
\centering
\small
\setlength{\tabcolsep}{5pt}
\caption{
Binding energies of selected nuclei, excluding Coulomb interactions, extrapolated to the ground-state limit ($\tau \rightarrow \infty$) for various box sizes ranging from $L=9$ to $12$ under periodic boundary conditions. The $L = \infty$ results represent the infinite-volume values extrapolated according to Eq.~(16). The uncertainty $\delta E$ denotes the combined statistical and $\tau$-extrapolation errors, while $\Delta E \equiv E(L) - E(\infty)$ quantifies the finite-volume shift relative to the infinite-volume limit. All energies are in MeV.
}
\label{tab:fv_time_extrap_values}
\begin{tabular}{lcccc}
\toprule
Nucleus & $L$ & $E$ & $\delta E$ & $\Delta E$ \\
\midrule
$^{4}\mathrm{He}$ & $9$  & $-29.66$ & $0.06$ & $-0.16$ \\
$^{4}\mathrm{He}$ & $10$ & $-29.56$ & $0.04$ & $-0.08$ \\
$^{4}\mathrm{He}$ & $11$ & $-29.58$ & $0.02$ & $-0.10$ \\
$^{4}\mathrm{He}$ & $12$ & $-29.44$ & $0.05$ & $0.04$ \\
$^{4}\mathrm{He}$ & $\infty$ & $-29.48$ & $0.03$ & $0.00$ \\
\midrule
$^{16}\mathrm{O}$ & $9$  & $-137.13$ & $0.16$ & $0.80$ \\
$^{16}\mathrm{O}$ & $10$ & $-137.51$ & $0.18$ & $0.42$ \\
$^{16}\mathrm{O}$ & $11$ & $-137.68$ & $0.09$ & $0.25$ \\
$^{16}\mathrm{O}$ & $12$ & $-137.80$ & $0.08$ & $0.13$ \\
$^{16}\mathrm{O}$ & $\infty$ & $-137.93$ & $0.04$ & $0.00$ \\
\midrule
$^{40}\mathrm{Ca}$ & $9$  & $-416.63$ & $0.17$ & $-8.56$ \\
$^{40}\mathrm{Ca}$ & $10$ & $-410.45$ & $0.34$ & $-2.38$ \\
$^{40}\mathrm{Ca}$ & $11$ & $-408.67$ & $0.24$ & $-0.60$ \\
$^{40}\mathrm{Ca}$ & $12$ & $-408.29$ & $0.25$ & $-0.22$ \\
$^{40}\mathrm{Ca}$ & $\infty$ & $-408.07$ & $0.09$ & $0.00$ \\
\midrule
$^{100}\mathrm{Sn}$ & $9$  & $-1260.75$ & $3.43$ & $-94.86$ \\
$^{100}\mathrm{Sn}$ & $10$ & $-1206.56$ & $7.02$ & $-40.67$ \\
$^{100}\mathrm{Sn}$ & $11$ & $-1173.51$ & $0.75$ & $-7.62$ \\
$^{100}\mathrm{Sn}$ & $12$ & $-1168.61$ & $0.98$ & $-2.72$ \\
$^{100}\mathrm{Sn}$ & $\infty$ & $-1165.89$ & $3.46$ & $0.00$ \\
\bottomrule
\end{tabular}
\end{table}

\begin{figure}[htbp]
    \centering
    \includegraphics[width=\columnwidth]{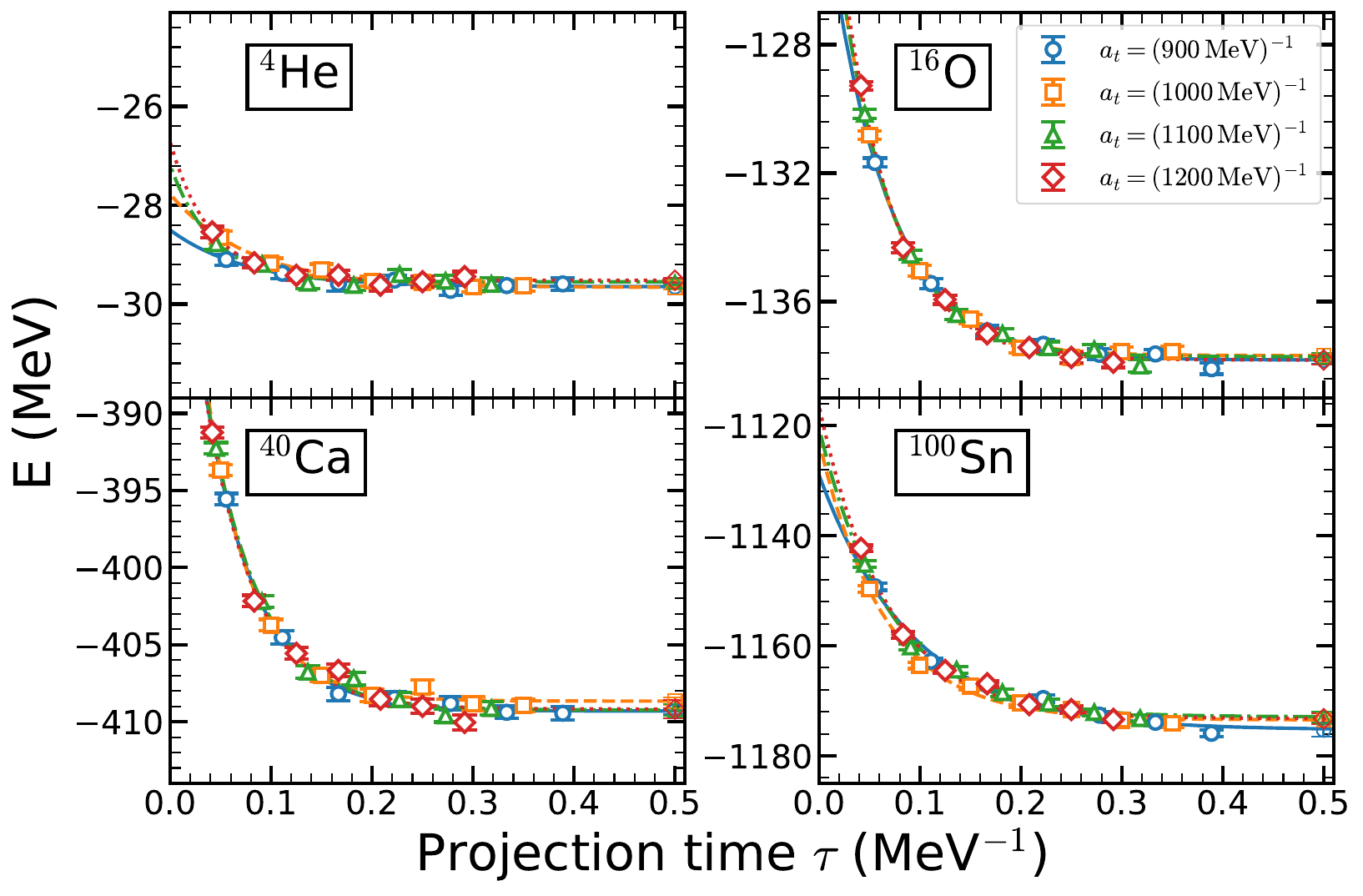}
    \caption{
    Binding energies, excluding Coulomb interactions, as a function of the projection time $\tau$ with different temporal steps $a_t$ in cubic periodic boxes of sizes $L = 11$. 
    The lines represent exponential fits according to Eq.~\eqref{eq:exp_extrap_Lt}.
    }
    \label{fig:at_tcutoff}
\end{figure}

To quantify the uncertainties associated with finite temporal steps, we repeat the projection-and-extrapolation procedure for several representative values of $a_t$ and compare the resulting infinite-projection-time energies, $E(\infty)$. Fig.~\ref{fig:at_tcutoff} shows the results for $^{4}$He, $^{16}$O, $^{40}$Ca, and $^{100}$Sn using $a_t=(900, 1000, 1100, 1200~\mathrm{MeV})^{-1}$. While minor differences between the finite-$\tau$ curves are visible at short projection times, these discrepancies diminish as the projection evolves. The extrapolated values obtained from the large-$\tau$ fits remain consistent within their respective extrapolation uncertainties across the considered range of $a_t$. Since no systematic dependence on $a_t$ can be resolved at the current level of statistical precision, we adopt $a_t=(1000~\mathrm{MeV})^{-1}$ as the standard choice for subsequent calculations.

The reduction in the uncertainty associated with a finite temporal step can be understood within the framework of perturbation theory. Although the effective interaction in the transfer matrix depends on $a_t$, the energy is always evaluated by inserting the exact Hamiltonian at the middle time slice. Consequently, the discrepancy between the $a_t$-dependent effective interaction and the inserted Hamiltonian only introduces a second-order perturbative correction to the binding energies, and is thus highly suppressed.

Taken together, the results in Figs.~\ref{fig:Ca40_tau_extrap}--\ref{fig:at_tcutoff} show that the projected ground-state energies are numerically stable with respect to the control parameters for the representative nuclei studied here.
This establishes the baseline setup for the subsequent analysis, in which the projection length, box size, and temporal step are kept fixed while other aspects of the framework are examined.

\begin{table}[htbp]
\centering
\small
\setlength{\tabcolsep}{5pt}
\caption{
Binding energies of selected nuclei, excluding Coulomb interactions, extrapolated to the ground-state limit ($\tau \rightarrow \infty$) for a box size of $L=11$ under periodic boundary conditions. Results for different temporal lattice spacings from $a_t = (900\text{ MeV})^{-1}$ to $(1200\text{ MeV})^{-1}$ are compared. The entry labeled $\infty$ represents the continuous-time limit ($a_t \rightarrow 0$), obtained via a fit of the form $E(a_t) = E(0) + c a_t^2$. The uncertainty $\delta E$ denotes the combined statistical and $\tau$-extrapolation errors, while $\Delta E \equiv E(a_t) - E(0)$ quantifies the temporal discretization shift relative to the continuous-time limit. All energies are in MeV.
}
\label{tab:at_time_extrap_values}
\begin{tabular}{lcccc}
\toprule
Nucleus & $a_t^{-1}$ (MeV) & $E$ & $\delta E$ & $\Delta E$ \\
\hline
$^{4}\mathrm{He}$ & $900$  & $-29.65$ & $0.06$ & $-0.33$ \\
$^{4}\mathrm{He}$ & $1000$ & $-29.66$ & $0.04$ & $-0.34$ \\
$^{4}\mathrm{He}$ & $1100$ & $-29.55$ & $0.07$ & $-0.23$ \\
$^{4}\mathrm{He}$ & $1200$ & $-29.52$ & $0.04$ & $-0.20$ \\
$^{4}\mathrm{He}$ & $\infty$ & $-29.32$ & $0.08$ & $0.00$ \\
\midrule
$^{16}\mathrm{O}$ & $900$  & $-137.82$ & $0.11$ & $-0.13$ \\
$^{16}\mathrm{O}$ & $1000$ & $-137.68$ & $0.09$ & $0.01$ \\
$^{16}\mathrm{O}$ & $1100$ & $-137.72$ & $0.12$ & $-0.03$ \\
$^{16}\mathrm{O}$ & $1200$ & $-137.82$ & $0.14$ & $-0.13$ \\
$^{16}\mathrm{O}$ & $\infty$ & $-137.69$ & $0.19$ & $0.00$ \\
\midrule
$^{40}\mathrm{Ca}$ & $900$  & $-409.31$ & $0.19$ & $-0.97$ \\
$^{40}\mathrm{Ca}$ & $1000$ & $-408.67$ & $0.24$ & $-0.33$ \\
$^{40}\mathrm{Ca}$ & $1100$ & $-409.24$ & $0.38$ & $-0.90$ \\
$^{40}\mathrm{Ca}$ & $1200$ & $-409.18$ & $0.58$ & $-0.84$ \\
$^{40}\mathrm{Ca}$ & $\infty$ & $-408.34$ & $0.69$ & $0.00$ \\
\midrule
$^{100}\mathrm{Sn}$ & $900$  & $-1175.30$ & $1.23$ & $-4.98$ \\
$^{100}\mathrm{Sn}$ & $1000$ & $-1173.48$ & $0.77$ & $-3.16$ \\
$^{100}\mathrm{Sn}$ & $1100$ & $-1172.92$ & $1.00$ & $-2.60$ \\
$^{100}\mathrm{Sn}$ & $1200$ & $-1173.26$ & $1.00$ & $-2.94$ \\
$^{100}\mathrm{Sn}$ & $\infty$ & $-1170.32$ & $1.38$ & $0.00$ \\
\bottomrule
\end{tabular}
\end{table}

\subsection{Additional auxiliary-field-induced interactions}
\label{subsec:implementation_corrections}

After establishing the numerical baseline for the projected ground-state energies, we next examine the consequences of employing a sign-problem-free realization of the interaction. The positivity requirement imposes stringent constraints on the form of the interaction after the auxiliary-field transformation.
A critical question is whether this transformation generates additional contributions to the effective Hamiltonian acting on the wave functions during the imaginary-time projection, and if so, whether these contributions remain quantitatively under control within the AFQMC calculations. While this issue is similar to the $a_t$-dependence discussed in the previous section, the interactions induced by the auxiliary-field transformation scale differently from the artifacts purely induced by nonzero $a_t$ and warrant a separate discussion.

Two sources of uncertainties of this kind can be identified.
The first is the discrete auxiliary-field realization of the three-body density interaction, which may in principle induce an effective four-body term. The second is specific to the new spin-orbit action: the complete-square realization needed to preserve positivity naturally generates higher-order two-body and three-body structures involving $\rho_s$.  These terms are not a flaw of the construction; they are the controlled cost of incorporating spin-orbit physics nonperturbatively without a sign problem. The main purpose of this subsection is to make this cost explicit and to show that it is numerically small.

For purely attractive two-body interactions, a continuous auxiliary-field method, such as the conventional Hubbard-Stratonovich transformation, is sufficient. However, the inclusion of a three-body density term typically necessitates complex auxiliary fields that introduce severe sign fluctuations. 
We therefore adopt a discrete auxiliary-field construction~\cite{Lu2019_PLB797-134863}, wherein the exponential of the many-body density interaction within a single time slice is recast as a weighted sum over a finite set of real auxiliary fields,
\begin{equation}
\begin{aligned}
: \exp \Bigl(&
-\frac{1}{2}C_2 a_t \rho^2
-\frac{1}{6}C_3 a_t \rho^3   
-\frac{1}{24}C_4 a_t \rho^4
\Bigr): \\
&=
\sum_{k=1}^{N}\omega_k \,
: \exp\!\left(\sqrt{-C_2 a_t}\,\phi_k\, \rho\right): .
\end{aligned}
\label{eq:disc_af_master}
\end{equation}
Here, $N$ denotes the number of discrete auxiliary-field nodes, with $\phi_k$ and $\omega_k$ representing the corresponding field amplitudes and weights, respectively. To ensure a real coupling on the right-hand side, one requires $C_2<0$.
Furthermore, preserving positivity in the Monte Carlo sampling dictates that all weights must be strictly positive ($\omega_k>0$). The term proportional to $C_4$ represents a four-body contact interaction, introduced here primarily as a bookkeeping device to determine whether a minimal four-body force is unavoidable to satisfy these positivity conditions. 

For our present purposes, it is sufficient to expand both sides of Eq.~\eqref{eq:disc_af_master} up to $\mathcal{O}(\rho^4)$ and match the coefficients order by order. This procedure yields a set of constraints for the parameters $\omega_k$ and $\phi_k$. We then adopt a minimal construction with $N=3$ and set $\phi_2=0$, such that the remaining two amplitudes, $\phi_1$ and $\phi_3$ are the roots of the quadratic equation
\begin{equation}
\phi^2 + \frac{C_3}{\sqrt{-a_tC_2^3}} \phi - \frac{C_3^2}{a_tC_2^3} + \left( \frac{C_4}{a_t C_2^2} - 3 \right) = 0,   \label{eq:quadeq}
\end{equation}
while the corresponding weights $\omega_1, \omega_2,$ and $\omega_3$, which sum to unity, can be expressed as
\begin{equation}
\begin{split}
\omega_1 &= \frac{1}{\phi_1 (\phi_1 - \phi_3)}, \qquad
\omega_2 = 1 + \frac{1}{\phi_1 \phi_3},\\
\omega_3 &= \frac{1}{\phi_3 (\phi_3 - \phi_1)} .
\end{split}
\label{eq:weights}
\end{equation}
Using Vieta's formulas, it is straightforward to verify that Eqs.~(\ref{eq:quadeq}--\ref{eq:weights}) satisfy Eq.~(\ref{eq:disc_af_master}) up to $\mathcal{O}(\rho^4)$. For a purely two-body interaction ($C_3 = C_4 = 0$), the solutions simplify to $\phi_1 = -\phi_3 = \sqrt{3}$, $\phi_2 = 0$, $\omega_1 = \omega_3 = 1/6$, and $\omega_2 = 2/3$.

It follows immediately from Eq.~(\ref{eq:weights}) that $\phi_1$ and $\phi_3$ must have opposite signs to ensure $\omega_1 > 0$ and $\omega_3 > 0$. This requirement is satisfied if and only if the left-hand side of Eq.~(\ref{eq:quadeq}) is negative at $\phi = 0$, which leads to the condition
\begin{equation}
C_4 < 3a_t C_2^2+\frac{C_3^2}{C_2}. \label{eq:positivitycondition}
\end{equation}
This inequality establishes an upper bound for the four-body interaction required to maintain strictly positive weights $\omega_k$. For sufficiently small $a_t$, the right-hand side of Eq.~(\ref{eq:positivitycondition}) becomes negative, implying that a nonzero attractive four-body force must be introduced to ensure positivity. However, for a sufficiently weak $C_3$ and a moderately large $a_t$, the right-hand side can be positive, allowing us to choose $C_4 = 0$. This is precisely the scenario when substituting $a_t = (1000~\mathrm{MeV})^{-1}$ and the two- and three-body couplings of LAT-OPT1 into Eq.~(\ref{eq:positivitycondition}). Consequently, for this specific parameter set, we can perform exact simulations without any induced four-body force. In the case of more general interactions, the positivity condition in Eq.~(\ref{eq:positivitycondition}) must be carefully examined. If the three-body force is sufficiently strong, the induced four-body force could substantially distort the wave function and must be rigorously accounted for in practical simulations.

Another source of uncertainty is related to the treatment of the spin-orbit coupling term.
Such a term was introduced by substituting the density operator $\rho$ in Eq.~(\ref{eq:disc_af_master}) by 
\begin{equation}
\rho'(\mathbf n)=\rho(\mathbf n)+\alpha\,\rho_s(\mathbf n),
\qquad
\alpha\equiv \frac{C_s}{2C_2},
\label{eq:rho_shift_alpha}
\end{equation}
where $\rho_s$ is the spin-orbit density defined in Eq.~(\ref{eq:rho_s_def}). One obtains the expansion
\begin{equation}
:\frac{C_2}{2}\rho'^2:
=
:\frac{C_2}{2}\rho^2
+\frac{C_s}{2}\rho\,\rho_s
+\frac{C_2}{2}\alpha^2\rho_s^2:,
\label{eq:complete_square_expand}
\end{equation}
where only the first two terms are desired.
We denote the residual term proportional to $\alpha^2$ in Eq.~(\ref{eq:complete_square_expand}) as $\Delta V_2^{\mathrm{(ind)}}$.
Furthermore, the auxiliary field transformation in Eq.~(\ref{eq:disc_af_master}) also generates additional terms to the three-body forces.
Similarly we define the residual three-body forces as
\begin{equation}
\Delta V_{3}^{(\mathrm{ind})}
=
:\frac{C_3}{6}\Big[\big(\rho+\alpha\rho_s\big)^3-\rho^3\Big]: .
\label{eq:V3_ind_def}
\end{equation}

These terms are suppressed for $C_s \ll C_2$ and expected to remain negligible.
The coefficient of the induced two-body term is proportional to $C_s^2$, and is therefore suppressed relative to the linear spin-orbit contribution. In addition, $\rho_s$ itself contains gradient operators, so that the $\rho_s$-dependent pieces of $\Delta V_{3}^{(\mathrm{ind})}$ correspond to higher-order short-range structures. They are therefore expected to contribute much less to the low-energy bulk observables considered here than the leading density and spin-orbit terms.

\begin{table}[htbp]
\centering
\small
\caption{
Breakdown of the various energy contributions originating from the individual components of the Hamiltonian Eq.~(\ref{eq:H_LATOPT1}) and the additional terms generated during the auxiliary-field transformation. $E_{2}$ denotes the leading two-body contact contribution arising from the two-body Wigner SU(4) term. $E_{\mathrm{sl}}$ is the total spin-orbit contribution. $\Delta E_{\rho_s^2}$ represents the contribution from the induced two-body term proportional to $\rho_s^2$. $E_{3}$ is the leading Wigner SU(4) three-body contribution proportional to $\rho^3$, and $\Delta E_{3}^{(\mathrm{ind})}$ stands for the induced three-body contribution generated by the shift $\rho \rightarrow \rho + \alpha \rho_s$. All energies are in MeV.
}
\label{tab:so_decomp_extended}
\begin{tabular}{lccccc}
\toprule
Nucleus &
$E_{2}$ &
$E_{\mathrm{sl}}$ &
$\Delta E_2^\mathrm{(ind)}$ &
$E_{3}$ &
$\Delta E_{3}^{(\mathrm{ind})}$ \\
\midrule
$^{4}\mathrm{He}$   & $-75.1$   & $-0.3$   & $-0.1$ & $0.5$  & $0.0$ \\
$^{12}\mathrm{C}$   & $-261.6$  & $-13.3$  & $-0.7$ & $2.0$  & $0.5$ \\
$^{14}\mathrm{C}$   & $-308.7$  & $-12.7$  & $-0.2$ & $2.4$  & $0.3$ \\
$^{16}\mathrm{O}$   & $-376.6$  & $-5.6$   & $0.1$  & $3.1$  & $0.1$ \\
$^{40}\mathrm{Ca}$  & $-1062.0$ & $-13.6$  & $1.8$  & $20.4$ & $0.3$ \\
$^{48}\mathrm{Ca}$  & $-1260.2$ & $-42.3$  & $1.7$  & $12.3$ & $1.0$ \\
$^{56}\mathrm{Ni}$  & $-1516.0$ & $-74.6$  & $1.1$  & $15.4$ & $1.8$ \\
$^{80}\mathrm{Zr}$  & $-2334.6$ & $-23.3$  & $5.0$  & $25.1$ & $0.8$ \\
$^{90}\mathrm{Zr}$  & $-2632.0$ & $-64.8$  & $5.4$  & $27.8$ & $1.5$ \\
$^{100}\mathrm{Sn}$ & $-3003.0$ & $-105.0$ & $5.4$  & $31.5$ & $2.2$ \\
\bottomrule
\end{tabular}
\end{table}

To quantify the magnitudes of these effects, we evaluate the ground-state expectation values of the various interaction terms within wave functions projected with full LAT-OPT1. 
The results are summarized in Table~\ref{tab:so_decomp_extended}. Here, $E_2$ and $E_3$ denote the Wigner-SU(4) two- and three-body interaction energies, respectively, and $E_{\mathrm{sl}}$ represents the contribution of the linear spin-orbit term, all originating from Eq.~\eqref{eq:H_LATOPT1}. The quantities $\Delta E_{2,3}^{(\mathrm{ind})}$ represent the expectation values of the residual corrections $\Delta V_{2,3}^{(\mathrm{ind})}$. We find that the induced terms remain small relative to the dominant interaction components across all nuclei considered. As expected for a short-range, density-dependent contribution, the induced two-body energy $\Delta E_2^{\mathrm{(ind)}}$ increases gradually with the mass number, yet remains on the order of only a few MeV even in heavy systems. This contribution is also modest compared to the total spin-orbit energy.
For instance, in $^{100}\mathrm{Sn}$,
\begin{equation}
\frac{|\Delta E_2^\mathrm{(ind)}|}{|E_{\mathrm{sl}}|}
\approx
\frac{5.4}{105.0}\sim 5\%
\label{eq:ratio_rhos2_esl}
\end{equation}
confirming its role as a higher-order correction. The induced three-body contribution $\Delta E_{3}^{(\mathrm{ind})}$ exhibits a similar behavior and never exceeds a few MeV. Given the relative magnitudes and structural complexity of these operators, for all LAT-OPT1 calculations performed in this work, we explicitly include the induced two-body term $\Delta V_2^\mathrm{(ind)}$ in the Hamiltonian Eq.~\eqref{eq:H_LATOPT1} while neglecting its three-body counterpart $\Delta V_3^\mathrm{(ind)}$.

In summary, the auxiliary-field transformation adopted here does not perfectly reproduce the interactions of the target Hamiltonian in Eq.~(\ref{eq:H_LATOPT1}).
Consequently, the effective Hamiltonian driving the wave-function propagation differs slightly from the target one. We numerically examined the magnitudes of these induced terms, confirming the absence of an induced four-body force, and demonstrating that the induced two- and three-body forces are heavily suppressed relative to the dominant central and spin-orbit interactions. We also note that their ultimate impact on the binding energies is further mitigated at the first-order perturbative level, analogous to our previous analysis of nonzero-$a_t$ effects. Therefore, the expectation values presented in Table~\ref{tab:so_decomp_extended} do not represent the true corrections from the induced terms, provided the exact target Hamiltonian is consistently inserted at the middle time slice during the evaluation of the binding energy.

\subsection{Non-perturbative effects of spin-orbit coupling}
\label{subsec:spin_orbit_in_propagation}

The most direct test of the new action is whether the spin-orbit force must be present during the Euclidean-time projection, rather than inserted only as an expectation value after the wave function has been generated. Excluding the spin-orbit term from the projection and evaluating it solely at the middle time step reduces its effect to a first-order perturbative correction on an SU(4)-projected wave function. Embedding it directly into the projection Hamiltonian instead allows the spin-orbit interaction to reorganize the many-body wave function and generate the correct shell structure. By the variational principle, propagation with an incomplete Hamiltonian yields an approximate wave function whose energy expectation value is an upper bound on the exact ground-state energy of the full Hamiltonian. 

To verify this point, we compare three projection schemes: (1) full LAT-OPT1 propagation; (2) propagation restricted to the SU(4) component of LAT-OPT1; and (3) propagation using the Wigner-SU(4) interaction from Ref.~\cite{Lu2019_PLB797-134863}. 
The energies for all schemes are measured using the full LAT-OPT1 Hamiltonian at the middle time step. Consequently, any energy variations stem solely from differences in the projected wave functions.
As presented in Table~\ref{tab:latopt1_totalE_rowmodel_symbol}, we observe that these variations far exceed the negligible effects induced interactions discussed earlier, yielding substantial nonperturbative shifts in the total binding energies.

\begin{table}[htbp]
\centering
\small
\setlength{\tabcolsep}{4pt}
\caption{
Comparison of binding energies, excluding Coulomb interactions, obtained from three different projected wave functions. All energies are evaluated using the same Hamiltonian, Eq.~(\ref{eq:H_LATOPT1}), inserted at the middle time step. The three projection schemes are: (1) projection with the full LAT-OPT1 Hamiltonian including the spin-orbit term; (2) projection with only the SU(4) symmetric part of LAT-OPT1; and (3) projection using the Wigner--SU(4) parameter set from Ref.~\cite{Lu2019_PLB797-134863}. All energies are in MeV.
}
\label{tab:latopt1_totalE_rowmodel_symbol}
\begin{tabular}{lcccccc}
\toprule
 & $^{4}\mathrm{He}$ & $^{16}\mathrm{O}$ & $^{24}\mathrm{Mg}$ & $^{28}\mathrm{Si}$ & $^{32}\mathrm{S}$ & $^{40}\mathrm{Ca}$ \\
\midrule
(1) & $-29.4$ & $-138.3$ & $-220.5$ & $-266.4$ & $-314.3$ & $-408.4$ \\
(2) & $-29.3$ & $-135.0$ & $-213.4$ & $-257.7$ & $-303.3$ & $-401.1$ \\
(3) & $-29.0$ & $-133.6$ & $-209.1$ & $-253.2$ & $-298.7$ & $-395.8$ \\
\bottomrule
\end{tabular}
\end{table}

The resulting binding energies consistently satisfy
\begin{equation}
E^{(1)} < E^{(2)} < E^{(3)},
\label{eq:energy_ordering_projection}
\end{equation}
demonstrating that the full LAT-OPT1 propagation yields the most energetically favorable wave function. The comparison between schemes (1) and (2) is particularly instructive. As scheme (1) treats the spin-orbit interaction non-perturbatively throughout the imaginary-time evolution, the difference
\begin{equation}
\Delta E_{\mathrm{LS}} \equiv E^{(2)} - E^{(1)}
\label{eq:DeltaELS_def}
\end{equation}
directly quantifies its non-perturbative effects. 
For the representative nuclei in Table~\ref{tab:latopt1_totalE_rowmodel_symbol}, $\Delta E_{\mathrm{LS}}$ remains small for $^{4}\mathrm{He}$, but reaches $3.30$~MeV for $^{16}\mathrm{O}$, $7.06$~MeV for $^{24}\mathrm{Mg}$, $8.68$~MeV for $^{28}\mathrm{Si}$, $10.97$~MeV for $^{32}\mathrm{S}$, and $7.37$~MeV for $^{40}\mathrm{Ca}$. These shifts are of the same order as shell-correction energies.  They show that the spin-orbit term is not a small perturbative decoration of an SU(4) wave function.  Rather, it is a dynamical ingredient of the action and must be included in the projection in order to benchmark shell structure and open-shell dynamics.

\subsection{Residual sign problem from unpaired nucleons}
\label{subsec:applicability_beyond_exact_positivity}

The exact positivity condition established in Sec.~\ref{sec:Method} applies strictly to even-even nuclei, where the single-particle trial states organize into time-reversal pairs. For odd-$A$ and odd-odd nuclei, unpaired nucleons break this sign-problem-free structure. The crucial question is thus whether the ensuing phase fluctuations remain sufficiently small to permit stable, high-precision ground-state energy calculations.
To quantify this residual phase problem, we calculate the average phase
\begin{equation}
\left\langle e^{i\theta}\right\rangle_{|w|} \equiv \frac{\sum_c w(c)}{\sum_c |w(c)|},
\label{eq:avg_phase_def}
\end{equation}
where $w(c)$ is the amplitude of the auxiliary-field configuration $c$ containing the fermionic determinant. Deviations from the sign-problem-free limit ($\langle e^{i\theta}\rangle_{|w|}=1$) measure the severity of sign fluctuations.

\begin{figure}[htbp]
\centering
\includegraphics[width=0.95\columnwidth]{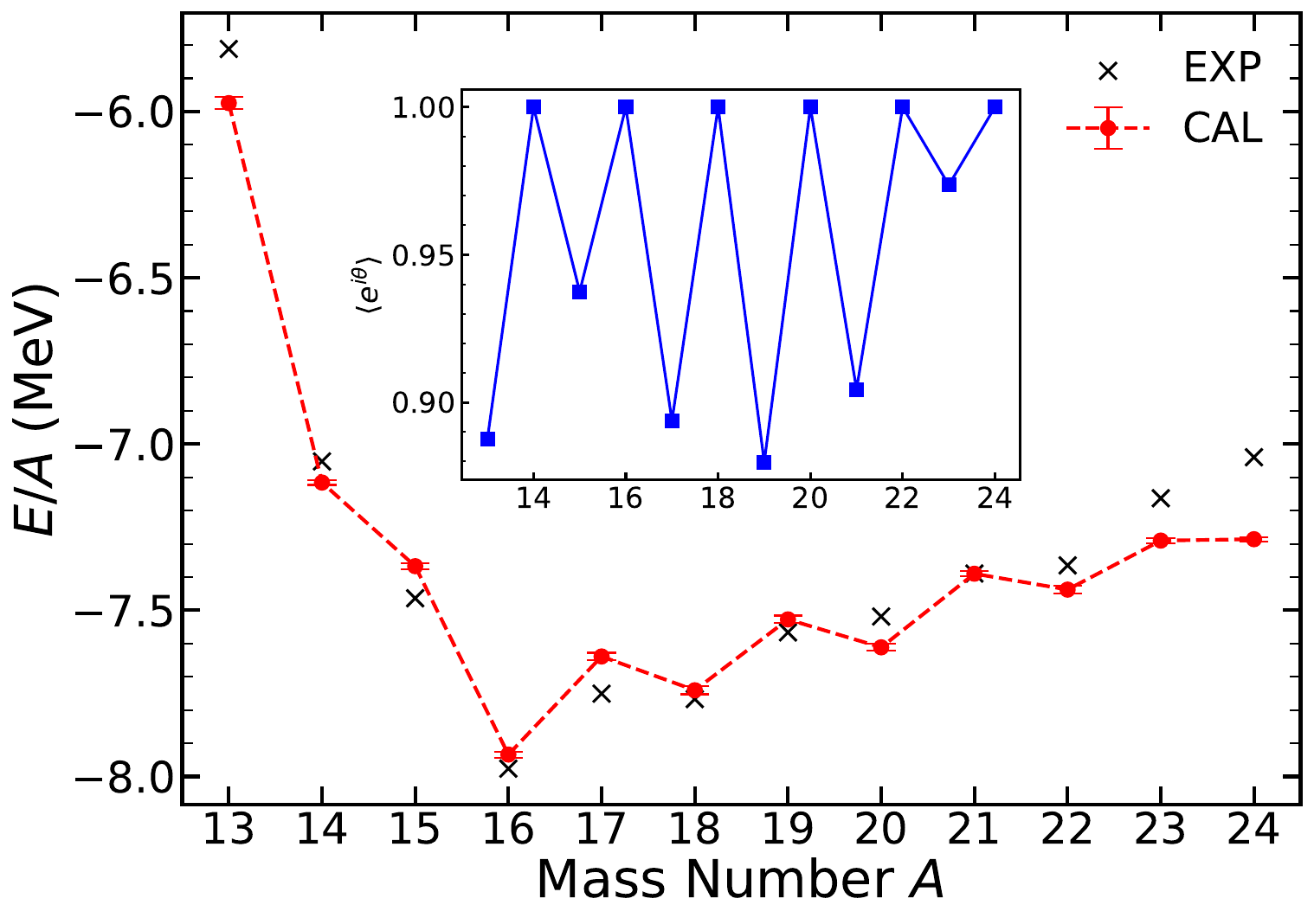}
\caption{Binding energy per nucleon along the oxygen isotopic chain. Experimental values (black crosses) are compared with the calculated results (red dots). The inset shows the average phase. }
\label{fig:Oxygen}
\end{figure}

Fig.~\ref{fig:Oxygen} illustrates the results for the oxygen isotopic chain. The odd-$A$ isotopes exhibit a mild sign problem due to the presence of a single unpaired neutron. As shown in the inset, the average phase $\langle e^{i\theta}\rangle_{|w|}$ drops slightly below unity for these odd-$A$ nuclei. Nevertheless, the statistical uncertainties of their calculated energies remain comparable to those of the neighboring even-even nuclei, demonstrating that the practical loss of precision is minimal and fully under control.

Table~\ref{tab:new-energies} extends this analysis to several heavy odd-$A$ and odd-odd nuclei. Because all calculations employ the same number of configurations, the Monte Carlo statistical errors directly reflect the severity of the sign problem, with sign-problem-free even-even nuclei serving as references ($\langle e^{i\theta}\rangle_{|w|}=1$). For odd-$A$ nuclei, the average phase remains close to unity ($0.94$--$0.98$), yielding energy uncertainties comparable to neighboring even-even systems and confirming that residual sign fluctuations remain mild.

\begin{table}[htbp]
\centering
\small
\setlength{\tabcolsep}{6pt}
\caption{
Calculated and experimental ground-state energies of representative nuclei compared with the experiment, and corresponding average phase $\langle e^{i\theta}\rangle_{|w|}$. 
The Coulomb force is included perturbatively for the convenience of comparison with the experiment.
The quoted uncertainties refer to the combined statistical and $\tau$-extrapolation errors.
}
\label{tab:new-energies}
\begin{tabular}{lccc}
\toprule
\textbf{Nucleus} &
\textbf{$E$ (MeV)} &
\textbf{Exp (MeV)} &
$\langle e^{i\theta}\rangle_{|w|}$ \\
\midrule

\multicolumn{4}{l}{Even-even nuclei} \\
$^{16}$O    & $-126.9(2)$  & $-127.6$ & 1 \\
$^{40}$Ca   & $-343.0(2)$  & $-342.1$ & 1 \\
$^{80}$Zr   & $-672.1(8)$  & $-669.2$ & 1 \\
$^{100}$Sn  & $-823.2(9)$  & $-825.2$ & 1 \\
\midrule

\multicolumn{4}{l}{Odd-$A$ nuclei} \\
$^{33}$S    & $-278.8(1)$ & $-280.4$ & 0.94 \\
$^{41}$Ca   & $-350.8(3)$ & $-350.4$ & 0.94 \\
$^{57}$Ni   & $-489.4(3)$ & $-494.2$ & 0.98 \\
$^{31}$P    & $-260.4(2)$ & $-262.9$ & 0.97 \\
$^{39}$K    & $-334.0(1)$ & $-333.7$ & 0.94 \\
$^{55}$Co   & $-471.1(3)$ & $-476.8$ & 0.95 \\
\midrule

\multicolumn{4}{l}{Odd-odd nuclei} \\
$^{10}$B    & $-62.5(3)$  & $-64.8$  & 0.66 \\
$^{18}$F    & $-132.8(3)$ & $-137.4$ & 0.70 \\
$^{38}$K    & $-318.3(1)$ & $-320.7$ & 0.79 \\
$^{58}$Cu   & $-492.4(4)$ & $-497.1$ & 0.90 \\
$^{78}$Y    & $-649.4(3)$ & $-651.2$ & 0.91 \\
\bottomrule
\end{tabular}
\end{table}

For odd-odd nuclei, we observe a seemingly counter-intuitive trend.
The sign problem is more pronounced in lighter systems but is clearly mitigated with increasing mass number. In particular, the average phase approaches unity for heavier systems such as $^{58}$Cu and $^{78}$Y. For $A \gtrsim 40$, the statistical uncertainties become comparable to those of even-even nuclei. This behavior suggests that, in heavier nuclei, the fraction of unpaired nucleons that break time-reversal pairing diminishes relative to the total particle number. Their impact on the overall phase is thus diluted, thereby weakening the residual phase problem. Consequently, while exact positivity is restricted to even-even nuclei, the practical applicability of this framework extends well beyond this sector, rendering both odd-$A$ and heavy odd-odd systems accessible to high-precision calculations.

\section{NLEFT Benchmarks}
\label{sec:NLEFT}

The preceding section established that the spin-orbit action has a controlled error budget.  We now turn to benchmark comparisons with earlier NLEFT actions.  This section is deliberately diagnostic: it identifies which comparisons are well defined and which are not.  The central point is simple but essential.  A lattice interaction is not specified by its coupling constants alone.  It is defined together with its regulator scheme, including the spatial lattice spacing, temporal lattice spacing or Hamiltonian limit, kinetic-energy discretization, smearing prescription, and operator content.  Changing these ingredients without refitting the couplings generally defines a different Hamiltonian or transfer matrix.  Separately, even for a fixed interaction, finite-volume benchmarks must compare the same volume, particle number, and boundary conditions; otherwise one is comparing different finite-volume approximations to the thermodynamic or infinite-volume limit.  Several of the apparent discrepancies raised in Ref.~\cite{Rothman2026_NuLattice} originate from mixing these distinct comparisons.

The authors of Ref.~\cite{Rothman2026_NuLattice} criticize both the 2016 clustering action of Ref.~\cite{Elhatisari2016_PRL117-132501}, which used $a=(100~\mathrm{MeV})^{-1}$ and $a_t=(150~\mathrm{MeV})^{-1}$, and the 2017 clustering action of Ref.~\cite{PRL119-222505}, which used $a=(100~\mathrm{MeV})^{-1}$ and $a_t=(100~\mathrm{MeV})^{-1}$.  The new finite-nucleus and nuclear-matter data presented below focus on the 2017 action.  This choice is practical rather than conceptual: the 2016 action has a more severe Monte Carlo sign problem, making high-precision saturation studies more demanding, while the regulator-consistency issue raised by Ref.~\cite{Rothman2026_NuLattice} is the same for both actions.  Repeating the same matched-comparison analysis for the 2016 action would lead to the same conclusion: the interactions are well-defined finite-cutoff lattice actions, and the apparent contradictions arise from benchmarking them against a different Hamiltonian-limit problem.

\subsection{Consequences of a non-negligible temporal lattice spacing}
\label{sec:largetempstep}

The historical development of NLEFT has 
undergone a paradigm shift from the transfer-matrix formalism to the Hamiltonian 
formalism. Calculations following Ref.~\cite{Lu2019_PLB797-134863} generally 
adopt a small temporal lattice spacing of $a_t \approx (1000~\mathrm{MeV})^{-1}$ 
and extract energies by inserting the Hamiltonian at the midpoint of the 
Euclidean-time projection. In contrast, earlier NLEFT clustering studies employed larger temporal spacings and used the temporal spacing as part of the ultraviolet regulator.  In particular, the 2016 clustering action of Ref.~\cite{Elhatisari2016_PRL117-132501} used $a=(100~\mathrm{MeV})^{-1}$ and $a_t=(150~\mathrm{MeV})^{-1}$, while the 2017 clustering action of Ref.~\cite{PRL119-222505} used $a=(100~\mathrm{MeV})^{-1}$ and $a_t=(100~\mathrm{MeV})^{-1}$.  This approach involves calculating energies from the logarithm of the eigenvalues of the normal-ordered transfer matrix. Note that 
this calculational paradigm change was also noted in the NLEFT book in 2019~\cite{Lahde:2019npb}.

There are certain computational advantages to the transfer-matrix approach.  For ultracold atom calculations involving two-component fermions in the unitary limit, the two-body interaction is a strong zero-range interaction corresponding to a single-site interaction on the lattice.  As a result, taking the limit $a_t \rightarrow 0$ requires taking $a_t$ much smaller than $ma^2$, where $m$ is the particle mass.  This is both computationally demanding and completely unnecessary.  As shown in Refs.~\cite{Lee:2005it,Lee:2005fk,Lee:2008xsa,Bour:2011xt}, this computational bottleneck problem can be avoided by treating the temporal lattice spacing $a_t$ as part of the ultraviolet regulator, working in concert with the spatial lattice spacing $a$.

For calculations of chiral effective field theory, there were also advantages to keeping $a_t$ nonzero.  As described in Ref.~\cite{Lee:2005nm} and Ref.~\cite{Epelbaum2009_EPJA41-125}, the overall magnitude of the required three-body interactions could be reduced by tuning the value of $a_t$.  As explained in Sec.~\ref{subsec:atextrapolation}, the transfer-matrix approach yields the ground-state energy of the induced effective Hamiltonian $H_{\mathrm{eff}}$ defined in Eq.~\eqref{eq:Heff} rather than that 
of the bare Hamiltonian $H$. These two Hamiltonians differ by induced interactions, leading to substantially different binding energies for non-negligible $a_t$. 

By 2018 or so, the NLEFT collaboration had achieved an understanding of how to control the size of three-body interactions using locally-smeared and nonlocally-smeared interactions.  New progress in efficient Monte Carlo importance sampling also made it possible to perform simulations with $a_t = (1000~\mathrm{MeV})^{-1}$ or even smaller.  These two developments as well as the focus on performing chiral effective field theory calculations at higher orders prompted the switch to the Hamiltonian lattice approach.  This was achieved by taking the temporal lattice as small as possible and computing expectation values of the Hamiltonian.  

In the following, we perform a direct numerical comparison of the transfer-matrix and Hamiltonian formalisms.  The purpose is not to assign a preference to one formulation in isolation, but to demonstrate that the two cannot be mixed after the couplings have been calibrated.  A transfer-matrix action fitted at nonzero $a_t$ and a Hamiltonian action obtained by taking $a_t\to0$ are different regulated theories unless the interaction parameters are refit.  To make this point with the least ambiguity, we use the lattice action of Ref.~\cite{PRL119-222505}, for which the original temporal regulator is known.  The qualitative conclusions remain robust for all nonperturbative sign-problem-free interactions developed within NLEFT.

We define the Hamiltonian energy $E_H$ as the expectation value of the 
Hamiltonian operator with respect to either a projected or a trial wave 
function $|\Phi\rangle$,
\begin{equation}
E_H = \langle \Phi | H | \Phi \rangle . \label{eq:HamiltonianEnergy}
\end{equation}
By contrast, the transfer-matrix energy is defined as
\begin{equation}
E_M(a_t) = -\frac{1}{a_t}\ln\langle \Phi | :\exp(-a_t H): |\Phi \rangle ,
\label{eq:transferEnergy}
\end{equation}
where $a_t$ is a finite temporal lattice spacing. We emphasize that the 
value of $a_t$ appearing in Eq.~\eqref{eq:transferEnergy} need not be 
identical to the temporal step size used in the imaginary-time projection 
to obtain $|\Phi\rangle$. For the present purpose, we only require the 
wave functions in Eqs.~\eqref{eq:HamiltonianEnergy} and \eqref{eq:transferEnergy} 
 to be identical. For any $|\Phi\rangle$, the Hamiltonian energy is 
recovered in the continuous-time limit,
\begin{equation}
E_H = \lim_{a_t \rightarrow 0} E_M(a_t).
\end{equation}
However, for non-negligible temporal lattice spacings, 
the two energies can differ substantially.

It should be emphasized that the legacy clustering actions were originally formulated and calibrated within the finite-temporal-spacing transfer-matrix framework, with $a_t$ serving as an ultraviolet regulator in the language of EFT.  The 2016 action and 2017 action used different finite temporal regulators, $a_t=(150~\mathrm{MeV})^{-1}$ and $a_t=(100~\mathrm{MeV})^{-1}$, respectively.  These are not incidental numerical step sizes that can be removed after the fit, they are part of the definition of the regulated action.  Such an action cannot, in principle, be directly applied to the Hamiltonian formalism without refitting the interaction parameters.  For the purpose of the current comparison, we deliberately employ the 2017 interaction across both formalisms to demonstrate how an inconsistent application of the theoretical framework can lead to unphysical artifacts.  The same regulator argument applies to the 2016 action, although its more severe sign problem makes a comparably detailed saturation study less convenient.  This underscores a crucial caveat for cross-model benchmarking in lattice calculations.

\begin{figure}[htbp]
\centering
\includegraphics[width=0.9\columnwidth]{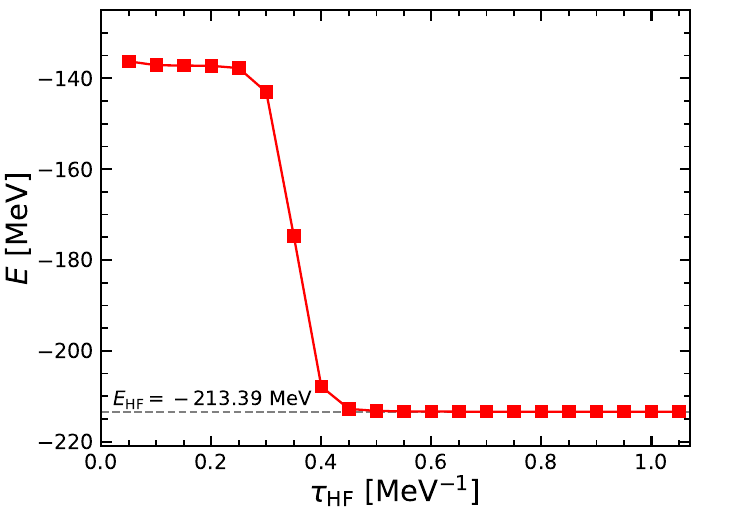}
\caption{Convergence of the HF Hamiltonian energy of $^{16}$O as a function of the iteration time $\tau_\mathrm{{HF}} = L_{\text{iter}} \delta t_{\text{HF}}$ using the shifted power method with a numerical step size of $\delta t_{\text{HF}}^{-1}=1000~{\rm MeV}$. The red squares show the evolution of the energy during the HF iteration, and the dashed line denotes the fully converged HF energy limit.}
\label{fig:HF_iter}
\end{figure}

For a comprehensive benchmark, we compare expectation values evaluated using two sets of wave functions: variational Hartree-Fock (HF) states, denoted by $|\Phi_{\mathrm{HF}}\rangle$, and exactly projected AFQMC states, denoted by $|\Phi_{\mathrm{exact}}\rangle$. To avoid any confusion with the normal-ordered transfer matrix discussed previously, we emphasize that the HF solutions are obtained by directly solving the self-consistent nonlinear HF equations,
\begin{equation}
h(\psi_m) \psi_n = E_n \psi_n,
\end{equation}
where $h(\psi_m)$ is the single-particle Hamiltonian that depends on all single-particle wave functions $\psi_m$ ($m=1, \dots, A$). These equations are derived from the exact bare lattice Hamiltonian with no normal ordering applied. The system is solved using a shifted power method, a standard technique in the nuclear mean-field literature~\cite{Ring1980}. We iteratively apply the evolution operator $U = 1 - \delta t_{\text{HF}} h$ to a set of single-particle wave functions, performing Gram-Schmidt orthogonalization at each step to maintain antisymmetrization. Here, the numerical step size $\delta t_{\text{HF}}$ is strictly a convenience for the iteration algorithm; it is entirely distinct from the physical temporal lattice spacing $a_t$. We use this exact same HF solver approach regardless of whether the final ground-state energy is being evaluated using the Hamiltonian or transfer-matrix formalisms. Fig.~\ref{fig:HF_iter} illustrates the smooth convergence of this HF iteration for the Hamiltonian energy of $^{16}$O as a function of the iteration time $\tau_{\mathrm{HF}} = L_{\text{iter}} \delta t_{\text{HF}}$ for $\delta t_{\text{HF}}^{-1}=1000~{\rm MeV}$.

Once the HF states are obtained, they serve as the initial trial wave functions for the full AFQMC calculations. For these exact projections, we employ an identical spatial computational setup with a lattice spacing of $a=1.97$~fm and a box size of $L=6$. The imaginary-time projection is performed using the interaction from Ref.~\cite{PRL119-222505} with a temporal step $a_t=(1000~\mathrm{MeV})^{-1}$ and a total projection time $\tau=L_t a_t=0.35~\mathrm{MeV}^{-1}$. 

Furthermore, to fully isolate the numerical consequences of utilizing different computational formalisms, we compare the energies obtained using different discretization schemes for the kinetic energy operator. Early NLEFT calculations typically employed finite-difference (FD) approximations on the lattice to evaluate the spatial derivatives in the kinetic energy term. More recently, several NLEFT studies have utilized the fast Fourier transform (FFT) method to further suppress spatial discretization artifacts. To maintain consistency with Ref.~\cite{PRL119-222505}, we adopt the $n=3$ improved FD scheme for our baseline comparison, which employs a seven-point stencil to evaluate the second derivative along each spatial direction. Analogous to the comparison between transfer-matrix and Hamiltonian energies, we maintain the exact same interaction parameters throughout this section, regardless of the chosen kinetic energy scheme.

Table~\ref{tab:Finite_Nu_HF} presents the calculated energies for four representative nuclei ($^{4}\text{He}$, $^{8}\text{Be}$, $^{12}\text{C}$, and $^{16}\text{O}$) across all methodological variations. For each nucleus, we perform eight distinct calculations encompassing three pairs of comparisons: (i) the Hamiltonian energy $E_H$ versus the transfer-matrix energy $E_M$, (ii) the HF state $|\Phi_{\mathrm{HF}}\rangle$ versus the exact AFQMC projected wave function $|\Phi_{\mathrm{exact}}\rangle$, and (iii) the finite-difference (FD) scheme versus the fast Fourier transform (FFT) method. To evaluate $E_M$, the temporal lattice spacing in Eq.~\eqref{eq:transferEnergy} is kept finite, as in the original transfer-matrix formulation. The important point for this benchmark is not the specific numerical value attached to this cutoff, but that the exact same nonzero-$a_t$ energy definition is used for both the Hartree-Fock and the fully projected calculations to ensure a rigorous comparison.

\begin{table}[htbp]
\caption{Comparison of Hamiltonian energies ($E_H$, Eq.~\eqref{eq:HamiltonianEnergy}) and transfer-matrix energies ($E_M$, Eq.~\eqref{eq:transferEnergy}) obtained with HF ($E^{\mathrm{HF}}$) and AFQMC projected ($E^{\mathrm{proj}}$) states. The finite-difference (FD) and fast Fourier transform (FFT) schemes for discretizing the kinetic energy operator are compared. All energies are in MeV. Interactions are adopted from Ref.~\cite{PRL119-222505}.
The asterisk ($\ast$) indicates results reproduced according to Ref.~\cite{PRL119-222505}.   \label{tab:Finite_Nu_HF}}
\smallskip
\setlength{\tabcolsep}{0pt}
\footnotesize
\begin{tabular*}{\columnwidth}{@{\extracolsep{\fill}}llrrrr}
\toprule
\multicolumn{1}{c}{Method}
& \multicolumn{1}{c}{Nucleus}
& \multicolumn{1}{c}{$E_H^{\mathrm{HF}}$}
& \multicolumn{1}{c}{$E_H^{\mathrm{proj}}$}
& \multicolumn{1}{c}{$E_M^{\mathrm{HF}}$}
& \multicolumn{1}{c}{$E_M^{\mathrm{proj}}$*} \\
\midrule
FD
& $^{4}$He  & $-19.82$  & $-33.71(22)$  & $-23.69(10)$ & $-25.40(22)$ \\
& $^{8}$Be  & $-54.36$  & $-73.57(23)$  & $-35.31(9)$  & $-52.26(31)$ \\
& $^{12}$C  & $-137.30$ & $-146.00(27)$ & $-69.04(8)$  & $-83.82(34)$ \\
& $^{16}$O  & $-213.39$ & $-234.57(31)$ & $-108.50(4)$ & $-128.61(35)$ \\
\midrule
FFT
& $^{4}$He  & $-11.73$  & $-28.62(16)$  & $-21.49(7)$  & $-22.54(19)$ \\
& $^{8}$Be  & $-39.42$  & $-60.08(16)$  & $-27.43(8)$  & $-46.61(26)$ \\
& $^{12}$C  & $-98.69$  & $-112.71(20)$ & $-46.14(8)$  & $-73.53(34)$ \\
& $^{16}$O  & $-166.38$ & $-189.89(30)$ & $-85.90(8)$  & $-109.67(34)$ \\
\bottomrule
\end{tabular*}
\end{table}

We begin by analyzing the results obtained via the FD scheme. Here, the 
transfer-matrix energy evaluated with the projected wave function, 
$E_M^{\mathrm{proj}}$, is chosen as our reference benchmark. These energies 
are calculated under identical conditions to those in 
Ref.~\cite{PRL119-222505} and successfully reproduce their 
reported values, which are known to be in reasonable agreement with 
experimental binding energies. Note that Coulomb contributions are not 
explicitly included in these calculations, as they have already been absorbed 
into the interaction calibration procedure. The corresponding HF 
energies, $E_M^{\mathrm{HF}}$, are slightly higher, providing strict upper 
bounds for the exact projection energies. The energy difference between 
$E_M^{\mathrm{HF}}$ and $E_M^{\mathrm{proj}}$ quantifies the 
beyond-mean-field correlation energy, accounting for approximately $20\%$ 
of the total binding energy for medium-mass nuclei such as ${}^{12}\text{C}$ 
and ${}^{16}\text{O}$.

However, when evaluating the energies within the Hamiltonian formalism, a 
severe overbinding is observed. 
Specifically, the 
Hamiltonian energy $E_H^{\mathrm{proj}}$ for ${}^{16}\text{O}$ is roughly 
twice as large as its transfer-matrix counterpart $E_M^{\mathrm{proj}}$. 
This spurious attraction is driven by removing the finite-$a_t$ regulator
while keeping couplings that were calibrated in the nonzero-$a_t$
transfer-matrix formulation. Because a non-negligible temporal lattice spacing is employed for calculating $E_M^{\mathrm{proj}}$ and $E_M^{\mathrm{HF}}$ here, these induced interactions become highly significant, reaching a magnitude comparable to the bare interaction. The 
HF energies within the Hamiltonian formalism, $E_H^{\mathrm{HF}}$, 
exhibit a similar overbinding when compared to the transfer-matrix values 
$E_M^{\mathrm{HF}}$. Nevertheless, a comparison between the two Hamiltonian 
energies $E_H^{\mathrm{HF}}$ and $E_H^{\mathrm{proj}}$ confirms that the variational principle remains strictly preserved. 
Finally, while the FFT scheme generally yields somewhat smaller binding 
energies than the FD scheme, the overall qualitative trends, including the 
prominent discrepancies between the transfer-matrix and Hamiltonian formalisms, 
as well as between the HF and projected states, remain completely robust.

\begin{table}[htbp]
\centering
\caption{Transfer-matrix energy $E_M^{\mathrm{HF}}$ obtained with the
HF state for $^{16}$O as a function of the temporal cutoff ($t_{\mathrm{cutoff}} = 1/a_t$). }
\begin{tabular}{cc cc}
\toprule
$t_{\mathrm{cutoff}}$ [MeV] & $E_M^{\mathrm{HF}}$ [MeV]
& $t_{\mathrm{cutoff}}$ [MeV] & $E_M^{\mathrm{HF}}$ [MeV] \\
\midrule
100     & $-108.50(4)$ & 2000    & $-204.05(4)$ \\
150     & $-131.43(4)$ & 2400    & $-205.55(4)$ \\
200     & $-145.95(4)$ & 3000    & $-207.06(4)$ \\
250     & $-156.09(4)$ & 4000    & $-208.61(5)$ \\
300     & $-163.51(4)$ & 5000    & $-209.53(5)$ \\
400     & $-173.69(4)$ & 7000    & $-210.63(5)$ \\
500     & $-180.38(4)$ & 10000    & $-211.44(5)$ \\
600     & $-185.16(4)$ & 20000    & $-212.43(5)$ \\
800     & $-191.50(4)$ & 30000    & $-212.74(5)$ \\
1000    & $-195.50(4)$ & 60000    & $-213.09(5)$ \\
1200    & $-198.25(4)$ & 100000    & $-213.21(5)$ \\
1400    & $-200.28(4)$ & 200000    & $-213.30(5)$ \\
1600    & $-201.82(4)$ & 500000    & $-213.34(6)$ \\
1800    & $-203.06(4)$ & 1000000    & $-213.39(6)$ \\
\bottomrule
\end{tabular}

\end{table}

\begin{figure}[htbp]
\centering
\includegraphics[width=0.9\columnwidth]{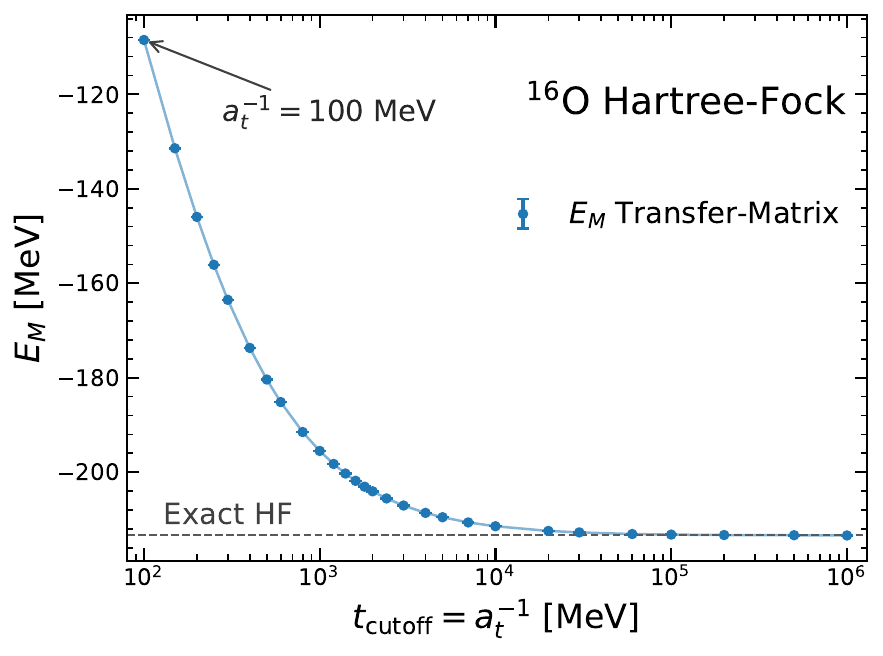}
\caption{HF energy in transfer-matrix formalism $E_M$ for $^{16}$O as a function of the temporal cutoff ($t_{\mathrm{cutoff}} = 1/a_t$).}
\label{fig:HF_Tc_check}
\end{figure}

To illustrate how a finite temporal lattice spacing changes the transfer-matrix 
energy from the Hamiltonian expectation value, Fig.~\ref{fig:HF_Tc_check} displays 
the transfer-matrix energy $E_M^{\mathrm{HF}}$ of ${}^{16}\text{O}$ as a function 
of the temporal cutoff $t_{\mathrm{cutoff}} = 1/a_t$. For comparison, the 
Hamiltonian energy $E_H^{\mathrm{HF}} = -213.39$~MeV is indicated by a horizontal 
dashed line. As shown in Table~\ref{tab:Finite_Nu_HF}, $E_M^{\mathrm{HF}}$ is 
substantially less bound than $E_H^{\mathrm{HF}}$ at the finite temporal spacing of the transfer-matrix action. 
As $t_{\mathrm{cutoff}}$ increases, $E_M^{\mathrm{HF}}$ becomes progressively 
more bound and asymptotically converges to $E_H^{\mathrm{HF}}$ in the 
continuous-time limit ($a_t \rightarrow 0$). This behavior directly reflects the 
nonzero-$a_t$ artifacts inherent to the transfer-matrix formulation.
At finite temporal
spacing, the transfer matrix induces an effective $\mathcal{O}(a_t)$ repulsive 
contribution that causes prominent discrepancies at non-negligible $a_t$.
Consequently, direct cross-formalism comparisons are not valid unless the interaction parameters are refit in the new formulation or the same regulated lattice formulation is used throughout.

Lastly, we discuss the physical implications of the temporal cutoff from a 
modern EFT perspective. Such a cutoff is introduced to regularize unphysical 
high-energy degrees of freedom, yielding an effective theory tailored for 
low-energy physics. Consequently, the effective interaction must be 
cutoff-dependent to absorb scale variations and preserve low-energy observables. 
For each choice of $a_t$, the interaction parameters must be recalibrated to the chosen physical input observables.
Otherwise, the framework becomes 
inconsistent, leading to spurious predictions. In particular, at a low temporal 
cutoff (non-negligible $a_t$), short-range physics above the scale $a_t^{-1}$ is entirely 
absorbed into the effective couplings, giving rise to an effective Hamiltonian 
distinctly different from its high-cutoff counterparts. Within the EFT framework, 
interaction strengths are intrinsically tied to the renormalization scale, 
making an interaction specified without its associated cutoff physically 
ill-defined. Note that the standard Hamiltonian formalism corresponds to the 
infinite-cutoff limit ($a_t \rightarrow 0$).
Any meaningful cross-model benchmark with NLEFT calculations therefore requires strict alignment of the regulated lattice formalism and interaction parameters tuned within that same formalism. Otherwise the comparison is between different theories, not between different many-body approximations to the same theory.

\subsection{Correlation energy of finite nuclei}

The correlation energy, defined as the difference between the Hartree--Fock (HF) energy and the true ground-state energy, provides an energy-based measure of the lowering due to correlations beyond the mean-field approximation. It is a useful diagnostic of beyond-mean-field effects, but it is not a complete characterization of the correlated many-body wave function.  With all systematic errors well controlled, NLEFT calculations provide an ideal avenue for investigating these complex quantum effects. In this work, we employ the \texttt{EE} interaction as a representative example~\cite{Lu2019_PLB797-134863}. For simplicity, the Coulomb interaction is neglected in this discussion. We solve the HF equations using the imaginary-time projection method described in the previous section, while the AFQMC projected energies are obtained using the same settings as in Ref.~\cite{Lu2019_PLB797-134863}. The corresponding results are compiled in Table~\ref{tab:HF_energy_comparison_for_EE}.

Table~\ref{tab:HF_energy_comparison_for_EE} compares the HF energies with the exact AFQMC energies for light $\alpha$-conjugate nuclei up to $^{16}$O, alongside the calculated correlation energies, $E_{\mathrm{corr}}$. As expected from the variational principle, the HF approach consistently underbinds these nuclei relative to the AFQMC ground-state energies. Furthermore, the correlation energies constitute a significant fraction of the total binding energies, typically accounting for more than 30\% for nuclei considered here. For light systems such as $^4$He and $^8$Be, the nuclei remain nearly unbound at the HF level. This is consistent with the strong clustering tendencies expected in these light nuclei, which can only be properly captured by incorporating many-body correlations beyond the mean-field approximation. In contrast, for medium-mass nuclei like $^{16}$O, although the correlation energy remains substantial, the HF energy becomes the dominant component of the total energy. Consequently, it is generally feasible to construct phenomenological mean-field models for medium-mass and heavy nuclei by readjusting the effective interactions to implicitly absorb these correlation energies into pure HF calculations~\cite{Bender2003_RMP75-121, Meng2006_PPNP57-470, Robledo2019_JPG46-013001}.

It is worth noting that a portion of the correlation energy can be captured within the mean-field framework by employing beyond-mean-field techniques~\cite{Ring1980, Guo2024_AtomDataNuclDataTab158-101661}. For instance, restoring the broken symmetries, such as projecting onto zero total momentum to remove the center-of-mass energy, or projecting onto zero total angular momentum to account for rotational energy in deformed nuclei, contributes approximately 10~MeV to the correlation energy in typical medium-mass systems. Additionally, pairing correlations further lower the total energy by a few MeV. Shape fluctuations can also be incorporated via the generator coordinate method (GCM), which further deepens the binding by several MeV. However, for typical doubly magic nuclei like $^{16}$O, these rotational, pairing, and shape correlation energies are negligible, leaving around 30~MeV of the correlation energy unaccounted for within standard beyond-mean-field frameworks. It has long been recognized that such beyond-mean-field calculations do not exhaust the total correlation energy and are traditionally restricted to applications using phenomenologically readjusted interactions. To extract the exact correlation energies directly from first-principles interactions, performing exact nuclear \textit{ab initio} calculations, such as AFQMC accompanied with rigorous uncertainty quantification, remains the only viable approach.

\begin{table}[htbp]
\centering
\caption{Comparison of Hartree-Fock energies ($E^{\mathrm{HF}}_H$) and AFQMC projected energies ($E^{\mathrm{proj}}_H$) within Hamiltonian framework, excluding Coulomb, for \texttt{EE} interaction in a $L = 8$ box. $E_\mathrm{corr} =E_H^{\mathrm{proj}} - E_H^{\mathrm{HF}} $ denotes the correlation energies. All energies are in MeV.}
\label{tab:HF_energy_comparison_for_EE}
\begin{tabular*}{\columnwidth}{@{\extracolsep{\fill}}llrrr}
\toprule
\multicolumn{1}{c}{Method}
& \multicolumn{1}{c}{Nucleus}
& \multicolumn{1}{c}{$E_H^{\mathrm{HF}}$}
& \multicolumn{1}{c}{$E_H^{\mathrm{proj}}$}
& \multicolumn{1}{c}{$E_\mathrm{corr}$} \\
\midrule
FFT & $^{4}$He  & $ -5.80$ & $-29.43(14)$  & $-23.63$ \\
    & $^{8}$Be  & $-14.08$ & $-56.15(17)$  & $-42.07$ \\
    & $^{12}$C  & $-39.57$ & $-87.47(29)$  & $-47.90$ \\
    & $^{16}$O  & $-86.82$ & $-131.24(32)$ & $-44.42$ \\
\bottomrule
\end{tabular*}
\end{table}

\subsection{Thermodynamic limit}
\label{sec:thermolimit}

Unlike finite nuclei which extrapolate to the infinite-volume vacuum at a fixed particle number, nuclear and neutron matter require the thermodynamic limit (infinite volume and particle number at constant density). 
In the latter, finite-volume effects combined with specific boundary conditions induce fictitious fermion shell structures, causing unphysical kinks in observables at emergent lattice magic numbers. 
Although simulating 66 single-species fermions can closely approach the thermodynamic limit~\cite{Forbes2011_PRL106-235303, Carlson2011_PRA84-061602R}, the fixed lattice spacing in NLEFT prohibits continuous volume adjustments, leaving large gaps during density scans. 
To alternatively explore discrete densities by varying nucleon numbers within a fixed volume, one must explicitly eliminate these fictitious shell effects.

Finite-volume shell effects arise from momentum discretization under specific boundary conditions.
For instance, periodic boundary conditions (PBC) in a cubic box restrict particles to $\bm{p} = ({2\pi}/{L})\bm{n}$, inducing artificial shell closures at magic numbers 2, 14, 38, ... for single-species fermions. 
Twisted boundary conditions (TBC)~\cite{Byers1961_PRL7-46} resolve this by shifting the wave function phases at boundaries, enabling particles to sample arbitrary momenta between the PBC grid points.
Averaging over all twist angles then provides a direct route to the infinite-volume limit~\cite{Loh1988_SyntheticMetals27-A499, Valenti1991_PRB44-13203}. 
First applied to exactly solvable systems~\cite{Loh1988_SyntheticMetals27-A499, Valenti1991_PRB44-13203, Gros1992_ZPB86-359, Gammel1993_SyntheticMetals55-4437, Gros1996_PRB53-6865} and quantum Monte Carlo methods~\cite{Lin2001PRE64-016702}, TBC has successfully accessed infinite-volume results in lattice QCD~\cite{Bedaque2004_PLB593-82, Divitiis2004_PLB595-408, Bedaque2005PLB616-208, Sachrajda2004_PLB609-73}, though its use in NLEFT remains sparse~\cite{Korber2016PRC93-054002, PhysRevLett.125.192502}. 
Here, we discuss how to apply TBC within lattice Monte Carlo simulations to eliminate the unphysical shell effects in computing the equation of state for symmetric nuclear matter.

We impose twisted boundary conditions on the single-particle wave functions according to
\begin{align}
    \psi (x + L, y, z, \sigma, \tau) &= e^{i 2\sigma \theta_x} \psi (x, y, z, \sigma, \tau), \nonumber \\
    \psi (x, y + L, z, \sigma, \tau) &= e^{i 2\sigma \theta_y} \psi (x, y, z, \sigma, \tau), \nonumber \\
    \psi (x, y, z + L, \sigma, \tau) &= e^{i 2\sigma \theta_z} \psi (x, y, z, \sigma, \tau),
\end{align}
where $\theta_i \in [-\pi, \pi]$ ($i=x,y,z$) are the independent twist angles in the three spatial directions. Crucially, opposite twist phases are assigned to opposite spin projections ($\sigma = \pm 1/2$) to circumvent the sign problem. In this work, we employ the average twisted boundary conditions (ATBC) by integrating over the entire twist parameter space. Within the Monte Carlo framework, this integration is straightforwardly implemented by assigning each computational thread a random twist triplet uniformly sampled from $[-\pi, \pi]^3$.

\begin{figure}[htbp]
\begin{centering}
\includegraphics[width=1.0\columnwidth]{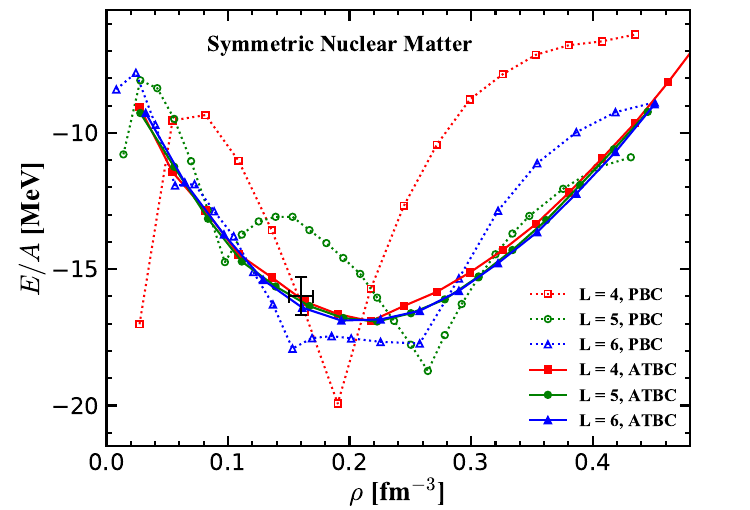}
\par\end{centering}
\caption{\label{fig:ATBC}
The binding energy per nucleon calculated with periodic boundary conditions (open symbols) and average twisted boundary conditions (full symbols, as explained below). 
The squares, circles and triangles show the results calculated with box sizes $L=$ 4, 5 and 6, respectively.
The cross symbol marks the empirical saturation density and binding energy.}
\end{figure}

The equation of state for symmetric nuclear matter under different boundary conditions is compared in Fig.~\ref{fig:ATBC}. Open and filled symbols represent PBC and ATBC calculations, respectively, where different box sizes at a fixed density imply varying nucleon numbers. The PBC results exhibit severe shell effects characterized by energy oscillations that peak or dip at lattice magic numbers ($A = 4, 28, 76, \dots$). These fictitious features damp out with increasing volume but persist clearly at $L = 6$, rendering the apparent energy minimum at $\rho \approx 0.15\text{ fm}^{-3}$ (associated with the $A = 76$ shell) misleading. Conversely, ATBC eradicates all unphysical kinks and collapses the multi-volume data onto a single curve, with even the smallest box ($L = 4$) providing excellent convergence.

In lattice nuclear matter calculations, the kinetic energy eigenstates are plane waves characterized by discrete lattice momenta. Consequently, the HF wave function $|\Psi_{\text{HF}}\rangle$ is constructed as a Slater determinant by filling the lowest-momentum states up to the Fermi surface. This state serves as the initial trial wave function $|\Psi_0\rangle$ at Euclidean time $\tau = 0$, with the baseline expectation value given by $E_{\text{HF}} = \langle \Psi_{\text{HF}} | H | \Psi_{\text{HF}} \rangle$. When $|\Psi_{\text{HF}}\rangle$ is evolved in AFQMC simulations to project out the correlated ground state, the energy at $\tau = 0$ corresponds to $E_{\text{HF}}$, whereas the asymptotic limit $\tau \rightarrow \infty$ yields the exact ground-state energy $E_{\infty}$. The correlation energy is then extracted via the difference $E_{\text{corr}} = E_{\infty} - E_0$.

\begin{figure}[t]
\centering

\includegraphics[width=1.0\columnwidth]{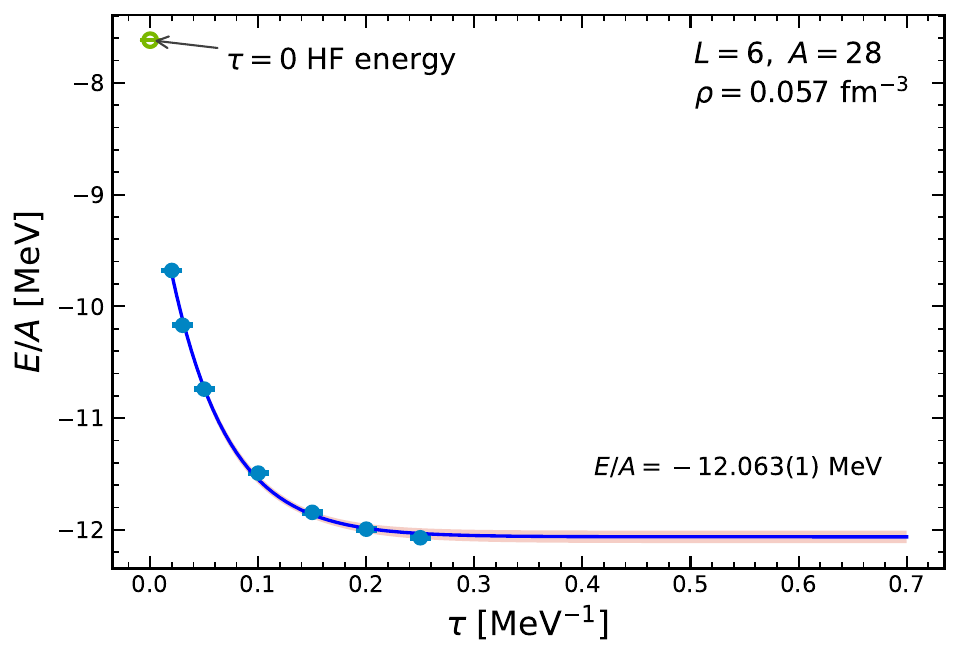}

\includegraphics[width=1.0\columnwidth]{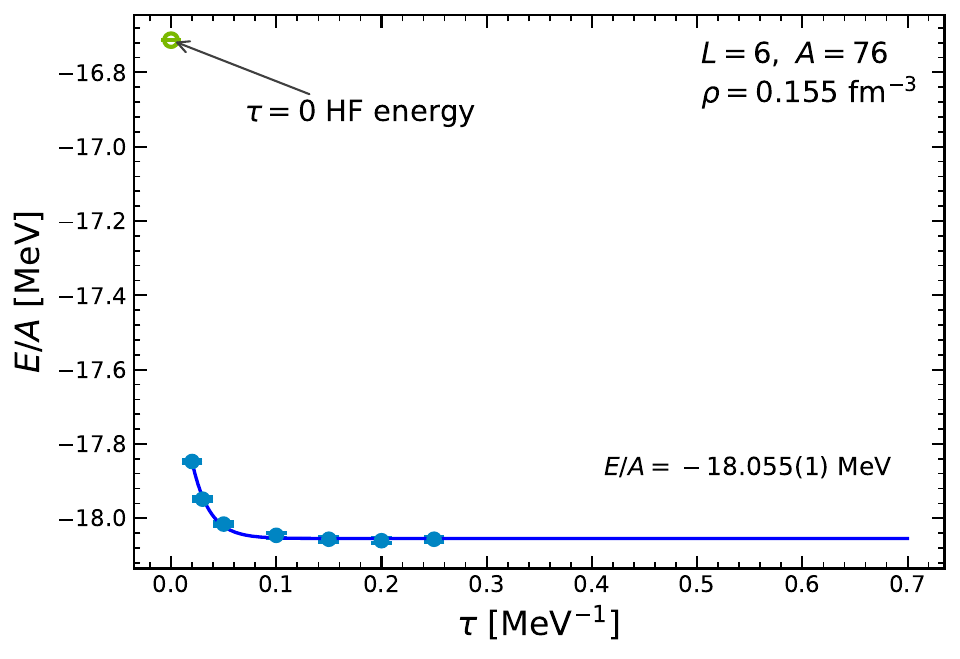}

\caption{
   Imaginary-time extrapolation of nuclear matter calculations at densities $\rho = 0.06\text{ fm}^{-3}$ and $0.16\text{ fm}^{-3}$ in an $L=6$ box under periodic boundary conditions. The \texttt{EE} Hamiltonian is adopted from Ref.~\cite{Lu2019_PLB797-134863}. The projections start from plane-wave states at $\tau=0$, which correspond to the respective Hartree-Fock energies.
}
\label{fig:correlation_energy}
\end{figure}

Fig.~\ref{fig:correlation_energy} illustrates the imaginary-time extrapolations for a box size $L = 6$ with nucleon numbers $A = 28$ and $76$, utilizing a lattice spacing of $a = 1.32~\text{fm}$ and the essential element interaction from Ref.~\cite{Lu2019_PLB797-134863}. 
In this section, we calculate the energies using the Hamiltonian formalism and take a small temporal step $a_t = (1000~\mathrm{MeV})^{-1}$ to reduce the discretization error for large nucleon numbers.
The latter configuration ($A=76$) gives a density close to the empirical saturation point. 
Under the PBC, the AFQMC projected energies converge rapidly with $\tau$. 
The extracted correlation energy exceeds $4.4~\text{MeV}$ at the sub-saturation density but decreases to roughly $1.3~\text{MeV}$ near the saturation point. 
For both systems, the HF energy provides an upper bound to the fully projected energy, demonstrating strict adherence to the variational principle.

\begin{figure}[htbp]
    \centering
    \includegraphics[width=1.0\columnwidth]{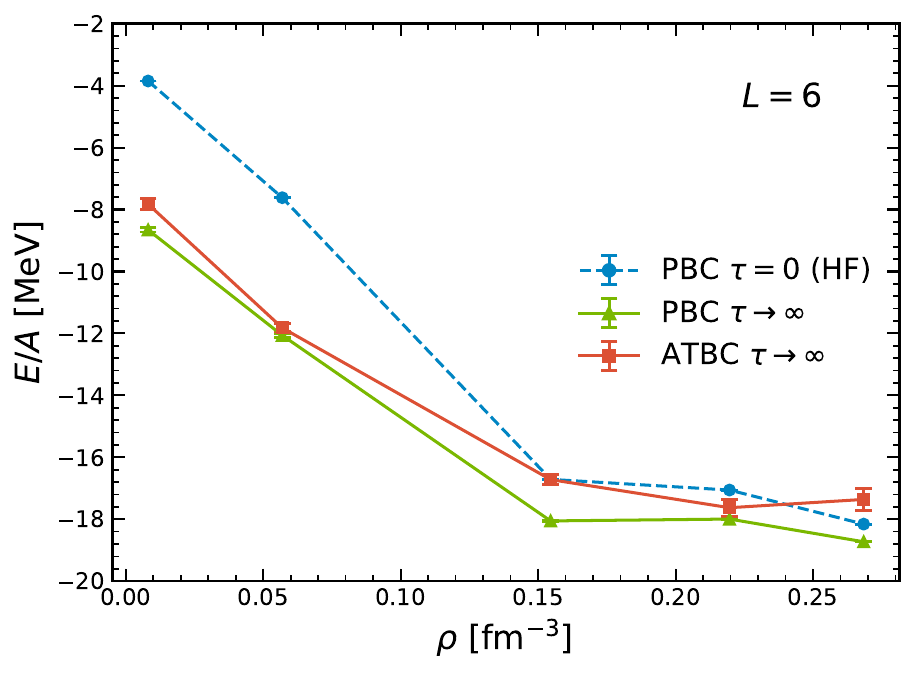}
    \caption{Energy per nucleon in a $L=6$ periodic box calculated using HF approximation and AFQMC projection under periodic boundary conditions. ATBC denotes the AFQMC results with average twisted boundary condition.}
    \label{fig:nuclear_matter_L6}
\end{figure}

At artificial magic numbers in a periodic box, HF calculations typically overestimate the binding energy relative to the thermodynamic limit due to additional shell correction energies at the shell closure.
Conversely, they underestimate the binding energy compared to exact many-body methods due to the variational principle. For benchmarking purposes, we perform calculations for SNM in an $L=6$ box with a lattice spacing of $a=1.32$~fm, utilizing the interaction from Ref.~\cite{Lu2022_PRL128-242501} at artificial PBC magic numbers $A=4,28,76,108,132,228,324,$ and $372$. 
Fig.~\ref{fig:nuclear_matter_L6} compares the energy per nucleon obtained at the HF level with the exact AFQMC ground-state energies under the PBC.
For the projected energies, we also include results using ATBC to estimate the thermodynamic limit. 
Under the PBC, the exact AFQMC energies are consistently lower than the corresponding HF energies, confirming that the variational bound is satisfied.
Notably, there is no rigorous theoretical basis for directly comparing the finite-box HF energy at magic numbers with the exact energy at the thermodynamic limit.
While the shell closure artificially lowers the finite-box HF energy below the thermodynamic limit HF energy, the variational principle ensures that the thermodynamic limit HF energy remains above the exact thermodynamic limit ground state. 
Because these two approximations introduce competing systematic errors, the finite-box HF energy at artificial magic numbers provides no formal constraint on the exact energy at the thermodynamic limit, as partially illustrated by comparing the HF and ATBC results in Fig.~\ref{fig:nuclear_matter_L6}.

\subsection{Many-body correlations in neutron matter}
\label{sec:many-body-correlations}
Ref.~\cite{Rothman2026_NuLattice} observes that, in neutron matter, the \texttt{EE} interaction gives small correlation energies when compared with other interactions. We agree that this interaction is soft in the sense that it is not designed to reproduce high-momentum two-body phase shifts. Instead, it was constructed to capture finite nuclei and nuclear matter with a minimal set of parameters. However, a small correlation energy, by itself, is not evidence that the many-body wave function is trivial or lacking important low-energy physics. Correlation energy is a single energy diagnostic and can be small even when the many-body wave function exhibits nontrivial irreducible two-body and four-body correlations.

\begin{figure}[htbp]
    \centering
    \includegraphics[width=1.0\columnwidth]{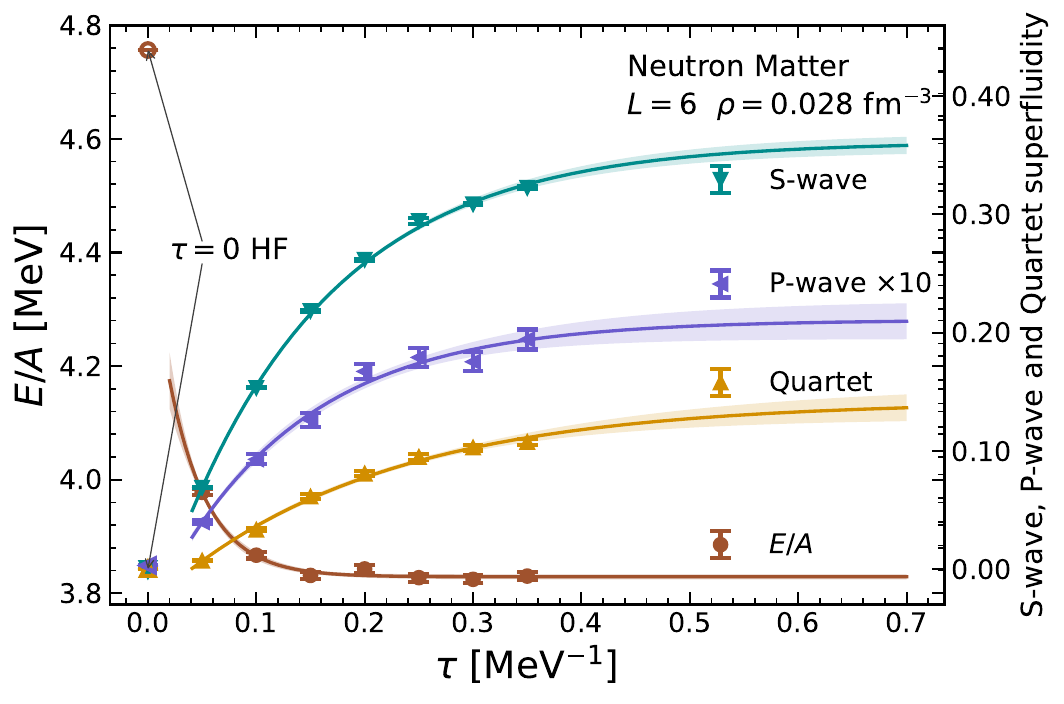}
    \caption{Euclidean-time projection of the energy per neutron and the condensate signals for $s$-wave pairs, $p$-wave pairs, and quartets for 14 neutrons in an $L=6$ periodic box using the essential element (\texttt{EE}) Hamiltonian.  For visibility, the $p$-wave signal is multiplied by a factor of 10.}
    \label{fig:EE_L6_S_QT}
\end{figure}

In Fig.~\ref{fig:EE_L6_S_QT}, we present pure neutron matter calculations with the essential element (\texttt{EE}) Hamiltonian.
Besides the system's energy, we also measure the condensation of s-wave pairs, p-wave double pairs, and quartets associated with multimodal superfluidity \cite{Ma2026}.  For these calculations we compute two-body and four-body irreducible densities or cumulants. The irreducible one-, two-, and three-body density operators are defined as 
\begin{equation}
\begin{aligned}
    \rho^{\mathrm{I}}_{1} &= \rho_{1}\\
    \rho_{12}^{\mathrm{II}} &= \rho_{12} - \rho_1 \rho_2 \\
    \rho_{123}^{\mathrm{III}} 
    &= \rho_{123} - (\rho_1 \rho^{\mathrm{II}}_{23}+\rho_2 \rho^{\mathrm{II}}_{13} + \rho_3 \rho^{\mathrm{II}}_{12}) 
       - \rho_1 \rho_2 \rho_3  \\
    &= \rho_{123} - (\rho_1 \rho_{23} + \rho_2 \rho_{13} + \rho_3 \rho_{12}) + 2\rho_1 \rho_2 \rho_3
\end{aligned}
\end{equation}
in which
\begin{equation}
    \begin{aligned}
        \rho_{1} &= \langle \Psi| a_{j_1}^{\dagger}(\vec{k}_1) a_{j_1}(\vec{k}_1)|\Psi\rangle \\
        \rho_{12} &=  \langle\Psi|a_{j_1}^{\dagger}(\vec{k}_1) a_{j_2}^{\dagger}(\vec{k}_2) a_{j_2}(\vec{k}_2) a_{j_1}(\vec{k}_1)| \Psi \rangle \\
        \rho_{123} &=  \langle\Psi|a_{j_1}^{\dagger}(\vec{k}_1) a_{j_2}^{\dagger}(\vec{k}_2) a_{j_3}^{\dagger}(\vec{k}_3) a_{j_3}(\vec{k}_3) a_{j_2}(\vec{k}_2) a_{j_1}(\vec{k}_1)| \Psi \rangle.
    \end{aligned}
\end{equation}
and the fourth-order cumulant of four-body density operators can be written as,
\begin{equation}
\label{eq:rhoIV_irre}
\begin{aligned}
\rho_{1234}^{\mathrm{IV}}
={}& \rho_{1234}  \\
&-\Bigl(
 \rho_{12}^{\mathrm{II}}\rho_{34}^{\mathrm{II}}
+\rho_{13}^{\mathrm{II}}\rho_{24}^{\mathrm{II}}
+\rho_{14}^{\mathrm{II}}\rho_{23}^{\mathrm{II}}
 \Bigr) \\
&-\Bigl(
 \rho_1\rho_{234}^{\mathrm{III}}
+\rho_2\rho_{134}^{\mathrm{III}}
+\rho_3\rho_{124}^{\mathrm{III}}
+\rho_4\rho_{123}^{\mathrm{III}}
 \Bigr) \\
&-\Bigl(
 \rho_1\rho_2\rho_{34}^{\mathrm{II}}
+\rho_1\rho_3\rho_{24}^{\mathrm{II}}
+\rho_1\rho_4\rho_{23}^{\mathrm{II}}
 \Bigr) \\
&-\Bigl(
 \rho_2\rho_3\rho_{14}^{\mathrm{II}}
+\rho_2\rho_4\rho_{13}^{\mathrm{II}}
+\rho_3\rho_4\rho_{12}^{\mathrm{II}}
 \Bigr) \\
&-\rho_1\rho_2\rho_3\rho_4 .
\end{aligned}
\end{equation}
with the ``raw'' four-body density 
\begin{equation}
\label{eq:rhoIV_raw}
\begin{aligned}
\rho_{1234}
={}&
\Bigl\langle \Psi \Bigm|
a_{j_1}^{\dagger}(\vec{k}_1)
a_{j_2}^{\dagger}(\vec{k}_2)
a_{j_3}^{\dagger}(\vec{k}_3)
a_{j_4}^{\dagger}(\vec{k}_4)
\\
&\qquad {}\times
a_{j_4}(\vec{k}_4)
a_{j_3}(\vec{k}_3)
a_{j_2}(\vec{k}_2)
a_{j_1}(\vec{k}_1)
\Bigm| \Psi \Bigr\rangle .
\end{aligned}
\end{equation}
The dominant contributions to s-wave pairing come from 
$\rho^{\mathrm{II}}_{\uparrow\downarrow}$ and 
$\rho^{\mathrm{II}}_{\downarrow\uparrow}$, 
whereas the dominant contributions to p-wave pairing come from 
$\rho^{\mathrm{II}}_{\uparrow\uparrow}$ and 
$\rho^{\mathrm{II}}_{\downarrow\downarrow}$.
In the dilute limit, the four-body cumulant provides an accurate measure of the number of quartets~\cite{Ma2026}.
Further details on lattice calculations of multimodal superfluid correlations can be found in Ref.~\cite{Ma2026}.

The initial wave function for the Euclidean-time projection is chosen to be the Hartree--Fock Slater determinant in momentum space.
Therefore, the observables at the Hartree--Fock level are obtained at $\tau=0$ MeV$^{-1}$.
At this point, the Hartree--Fock energy is $4.76$ MeV per nucleon, while both the two-body (s-wave, p-wave) and four-body (quartet) superfluid signals vanish.
As the Euclidean time $\tau$ increases, many-body correlations are gradually built up.
At $\tau=0.15$ MeV$^{-1}$, the energy starts to converge to $3.82$ MeV per nucleon.
However, the two-body pair and four-body quartet correlations continue to grow with increasing projection time.
Extrapolating to large $\tau$, we measure $0.36$ s-wave pairs, $0.02$ p-wave pairs and $0.14$ quartets in the $L=6$, $A=14$ system.
These results show that the correlation energy alone can substantially understate the quantum correlations and entanglement present in the many-body wave function.

\subsection{Nuclear saturation in the lattice transfer-matrix formalism}

To illustrate the consequences of inconsistent benchmarking in the saturation problem, we examine symmetric nuclear matter using the \texttt{Isotropic} interaction listed in Table~\ref{tab:forces}. This interaction was calibrated within the transfer-matrix formalism at finite temporal lattice spacing.  Its couplings are therefore not Hamiltonian-limit couplings; they include the finite temporal regulator as part of the definition of the effective theory.

\begin{figure}[htbp]
\begin{centering}
\includegraphics[width=1.0\columnwidth]{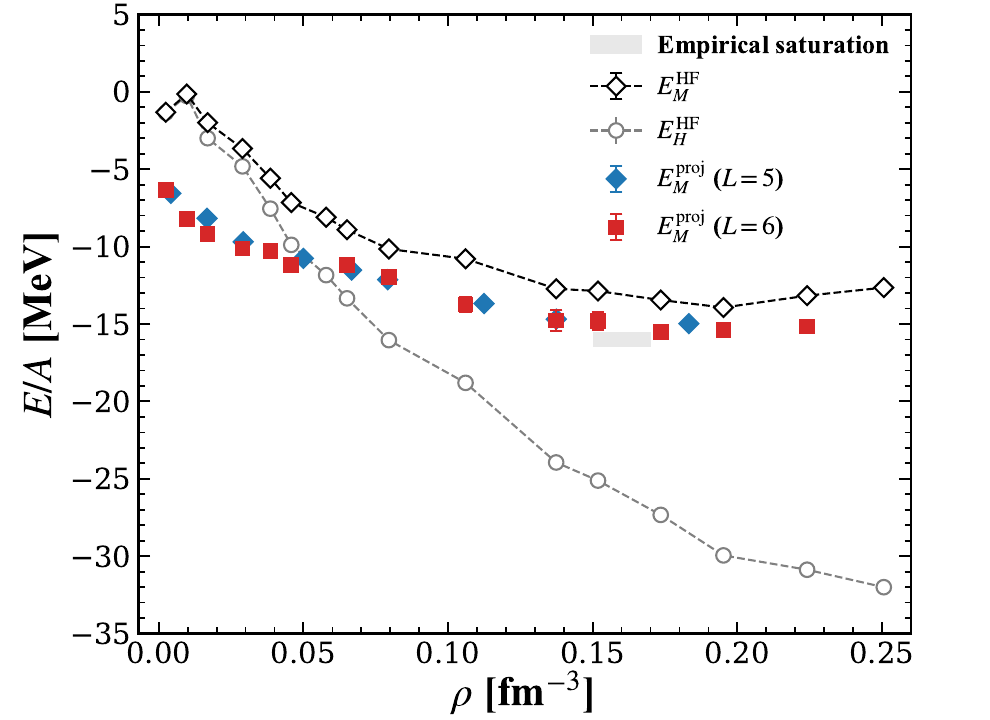}
\par\end{centering}
\caption{Energy per nucleon $E/A$ of symmetric nuclear matter as a function of density $\rho$ using the \texttt{Isotropic} interaction. The black diamonds ($E_M^{\mathrm{HF}}$) represent Hartree-Fock energies evaluated consistently within the transfer-matrix formalism. The blue diamonds and red squares ($E_M^{\mathrm{proj}}$) denote the exact ground-state energies obtained via auxiliary-field Monte Carlo simulations. The gray circles ($E_H^{\mathrm{HF}}$) represent Hartree-Fock energies calculated in the continuous-time Hamiltonian formalism using the identical interaction parameters, reproducing the methodology of Ref.~\cite{Rothman2026_NuLattice}. The massive spurious overbinding of $E_H^{\mathrm{HF}}$ highlights the systematic error of applying a temporally regulated interaction to a continuous-time formalism without renormalizing the effective coupling constants.\label{fig:EA_vs_rho}}
\end{figure}

In Fig.~\ref{fig:EA_vs_rho}, we present the energy per nucleon $E/A$ as a function of density $\rho$. The black diamonds denote the Hartree-Fock energies, $E_M^{\mathrm{HF}}$, evaluated consistently within the transfer-matrix formalism. The exact ground-state energies, $E_M^{\mathrm{proj}}$, obtained via auxiliary-field Monte Carlo simulations in the transfer-matrix formalism for periodic box sizes $L=5$ and $L=6$, are shown as blue diamonds and red squares, respectively. These exact AFQMC calculations demonstrate that the \texttt{Isotropic} interaction produces a physically reasonable saturation curve that passes close to the empirical saturation region. As expected from the variational principle, the transfer-matrix Hartree-Fock result ($E_M^{\mathrm{HF}}$) provides an upper bound to these ground state energies.

In contrast, the gray circles denote the Hartree-Fock energies, $E_H^{\mathrm{HF}}$, evaluated within the continuous-time Hamiltonian formalism ($a_t \to 0$) using the exact same un-renormalized interaction parameters. This curve explicitly reproduces the methodology and results presented in Ref.~\cite{Rothman2026_NuLattice}. By removing the temporal regulator without simultaneously recalibrating the interaction parameters, the Hamiltonian calculation generates a strong spurious overbinding. The artificial depth of the $E_H^{\mathrm{HF}}$ curve is not a physical prediction of the interaction.  It is the direct result of removing the temporal regulator while keeping couplings that were fitted with that regulator present.  This comparison demonstrates that transferring an interaction across different computational formalisms without recalibration invalidates the benchmark and leads to incorrect conclusions regarding nuclear saturation.

\subsection{Nuclear saturation in the lattice Hamiltonian formalism}

In this section, we investigate nuclear saturation on an $L=6$ lattice by introducing three leading-order chiral Hamiltonians.
Following the construction of Ref.~\cite{Elhatisari2016_PRL117-132501} the leading-order chiral Hamiltonian is written as:
\begin{equation}
    H = H_{\mathrm{free}} + V_{c_0}^{s_{\rm L}, s_{\rm NL}} + V_{\mathrm{ope}}.
\end{equation}
where $H_{\mathrm{free}}$ denotes the kinetic-energy operator, $V_{c_0}^{s_{\rm L},s_{\rm NL}}$ is the smeared contact interaction, and $V_{\mathrm{OPE}}$ is the one-pion-exchange interaction.
We employ the same spatial lattice spacing, $a=(100~\mathrm{MeV})^{-1}$, as in Ref.~\cite{Elhatisari2016_PRL117-132501}, while adopting a smaller temporal lattice spacing, $a_t=(1000~\mathrm{MeV})^{-1}$, corresponding to a higher temporal cutoff. For the kinetic-energy operator, we use the $n=3$ improved finite-difference discretization.
% The local smearing parameter is fixed at $s_{\rm L}=0.05$, while three values of the nonlocal smearing parameter $s_{\text{NL}} = \{0.1,0.2,0.3\}$ are considered.
The nonlocal smearing parameter $s_{\text{NL}} = \{0.1,0.2,0.3\}$
and local smearing $s_{\text{L}} = \{0.08,0.05,0.02\}$ are considered.
For each choice of $s_{\rm NL}$ and $s_{\rm L}$ , the contact coupling $c_0$ is adjusted to reproduce the binding energy of $^{4}$He on the $L=6$ lattice using Euclidean-time projection at $\tau=0.35~\mathrm{MeV}^{-1}$. 
The resulting Hamiltonians are denoted by $H^\text{A}$, $H^\text{B}$ and $H^\text{C}$. 
Their corresponding parameters and $^{4}$He binding energies are summarized in Table~\ref{tab:L6_He4_ABC}, where the first-order Coulomb contribution is included.
Since the projected energy $E(\tau \to \infty)$ is lower than $E(\tau=0.35)$, the target energy used in the tuning procedure is chosen to be slightly less bound than the physical $^{4}$He ground-state energy.

\begin{table}[htbp]
\centering
\caption{Binding energy of $^4$He, with $H^\text{A}$, $H^\text{B}$ and $H^\text{C}$ in $L=6$ box with different $s_{\text{NL}}$ and two-body coupling $c_0$, Coulomb energy is also listed for reference. \label{tab:L6_He4_ABC}}
\begin{tabular}{c c c c c c }
\toprule
   & $s_{\text{NL}}$ & $s_{\text{L}}$ & $c_0$ (MeV$^{-2}$) & $E^{L=6}_{\tau=0.35}$ (MeV)&  $E^{L=6}_{\tau=0.35}+\langle V_{\rm cou}\rangle$   \\
\midrule
$H^{\text{A}}$ & 0.1 & 0.08 & $-$1.46E-5 & $-$28.81 (8)  & $-$28.00 (8)    \\
$H^\text{B}$ & 0.2 & 0.05 & $-$0.68E-5 & $-$28.88 (8)  & $-$28.20 (8)    \\
$H^\text{C}$ & 0.3 & 0.02 & $-$3.72E-6 & $-$28.86 (8)  & $-$28.22 (8)   \\
\bottomrule
\end{tabular}
\end{table}

For completeness, we also calculate the $H^{\rm B}$ binding energy of $^4$He, $^8$Be, $^{12}$C and $^{16}$O in $a=1.32$ fm $L=6$ box at Euclidean time $\tau=0.35$ MeV$^{-1}$ with temporal lattice spacing $a_t = 1/(1000 \mathrm{MeV})$, which is listed in Table~\ref{tab:E_finite_ABC}.
\begin{table}[t]
\centering
\caption{$H^{\rm B}$ binding energy of $^4$He, $^8$Be, $^{12}$C and $^{16}$O. \label{tab:E_finite_ABC}}
\begin{tabular}{c c c}
\toprule
Nucleus & $E^{\mathrm{HF}}_{\tau=0}$ [$L = 6$] &  $E_{\tau=0.35}$ [$L = 6$]  \\
\midrule
$^{4}$He & -11.07  & -28.88 (08)  \\
$^{8}$Be & -27.79 &  -55.70 (19)  \\
$^{12}$C &  -65.74 & -90.83 (12)  \\ 
$^{16}$O & -93.58  & -130.43 (16)  \\
\bottomrule
\end{tabular}
\end{table}

\begin{figure}[htbp]
    \centering
    \includegraphics[width=1.0\columnwidth]{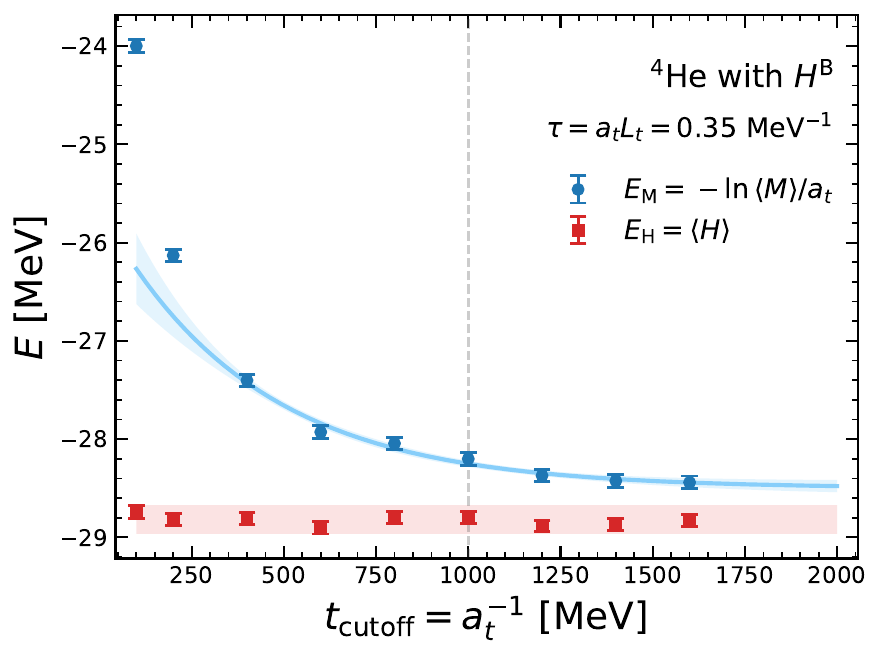}
    \caption{Lattice calculation of $^4$He energy in an $a=1.32$ fm $L=6$ box at Euclidean time $\tau=0.35$ MeV$^{-1}$ with Hamiltonian $H^\text{B}$ and temporal lattice spacing $a_t = 1/(100 \mathrm{MeV}) \to 1/(1600 \mathrm{MeV})$. }
    \label{fig:He4_tcutoff}
\end{figure}

It should be mentioned that the energies discussed above are all the expectation values of the Hamiltonian, $E_\mathrm{H}$.
To further illustrate the relationship between the transfer-matrix energy $E_{\mathrm{M}}$ and the Hamiltonian expectation value $E_{\mathrm{H}}$, we perform calculations for $^{4}$He at fixed Euclidean time $\tau=0.35~\mathrm{MeV}^{-1}$ while varying the temporal lattice spacing from $a_t=(100~\mathrm{MeV})^{-1}$ to $(1600~\mathrm{MeV})^{-1}$. The results are shown in Fig.~\ref{fig:He4_tcutoff}. As expected, the difference between $E_{\mathrm{M}}$ and $E_{\mathrm{H}}$ decreases with decreasing $a_t$, and the two definitions become consistent in the continuum-time limit $a_t\rightarrow0$. These results also indicate that, when $a_t$ is sufficiently small, the
Hamiltonian expectation value $E_{\mathrm H}$ evaluated with the finite-$a_t$
transfer-matrix wave function provides an accurate estimate of the $a_t\to0$ Hamiltonian-limit energy.

\begin{figure}[htbp]
\centering

\includegraphics[width=1.0\columnwidth]{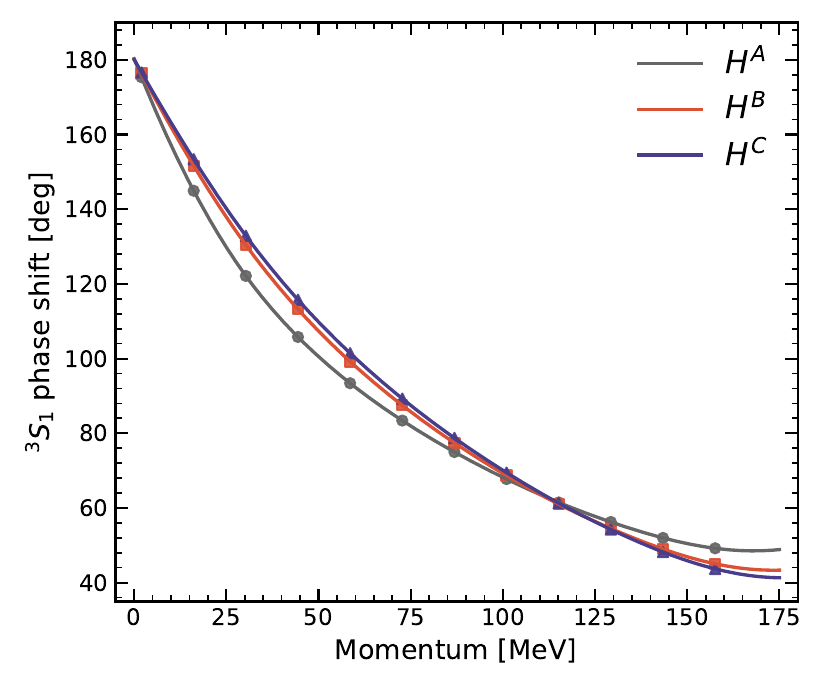}

\includegraphics[width=1.0\columnwidth]
{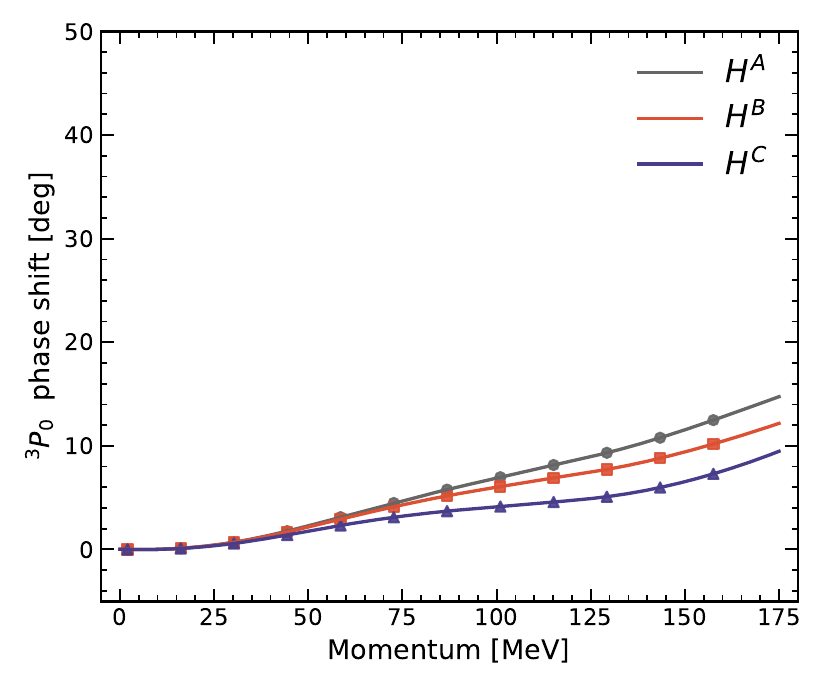}
\caption{
Comparison of the three leading-order chiral Hamiltonians: $H^{\rm A}$, $H^{\rm B}$ and $H^{\rm C}$.
Top: $^3S_1$ phase shifts.
Bottom:  $^3P_0$ phase shifts. 
}
\label{fig:compare_HABC}
\end{figure}

In Fig.~\ref{fig:compare_HABC}, we plot the $^3S_1$ and $^3P_0$ phase shifts with three leading-order chiral Hamiltonians: $H^\text{A}$, $H^\text{B}$, and $H^\text{C}$. 
It should be noted that all three LO interactions are close to the SU(4) symmetry limit. This implies that the $^1P_1$ and $^3P_1$ channels do not exhibit strong repulsive behavior, resulting in an overall soft interaction. The non-local smearing parameter $s_{\text{NL}}$ acts as a regulator and modifies the curvature of the S-wave interaction. At the same time, the local smearing parameter $s_{\text{L}}$ has a significant impact on the P-wave attraction.
The important point for the present discussion is that the low-energy two-body phase shifts change only moderately across $H^{\rm A}$, $H^{\rm B}$, and $H^{\rm C}$, while the nuclear-matter saturation properties change dramatically. This is not a paradox. It is precisely the kind of sensitivity emphasized in Ref.~\cite{Elhatisari2016_PRL117-132501}: the many-body transition from a dilute Bose gas of weakly interacting $\alpha$ particles to a self-bound nuclear liquid is controlled not only by low-energy two-nucleon observables, but also by the range, locality, and momentum dependence of the short-range interaction. In particular, the strength and range of the locally smeared part of the interaction strongly influence how $\alpha$ clusters interact and merge as the density is increased.

\begin{figure}[htbp]
    \centering
    \includegraphics[width=1.0\columnwidth]{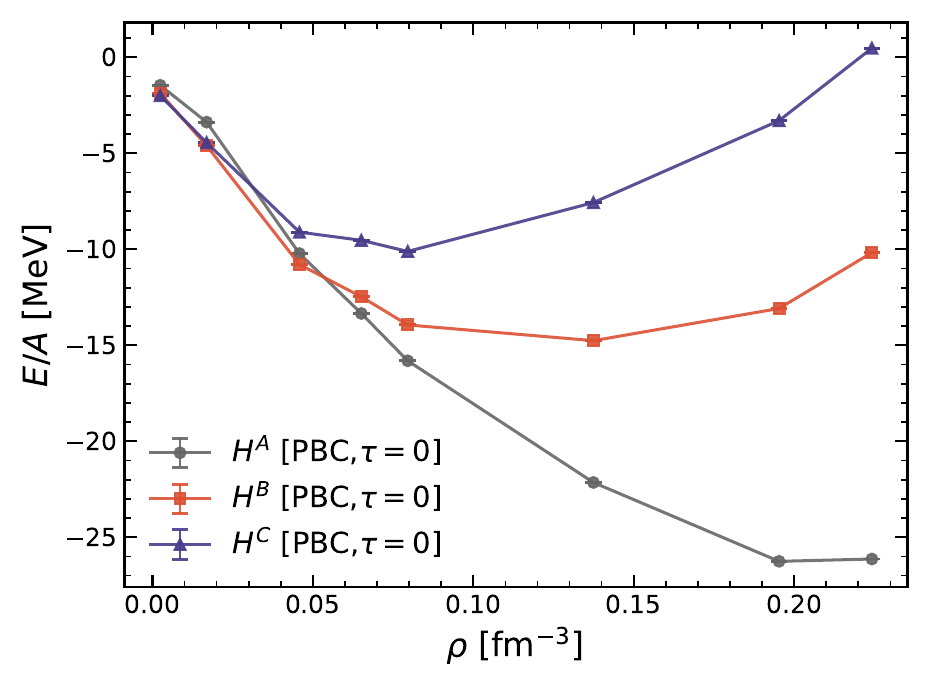}
    \caption{Nuclear matter energy per particle $E/A$  at $\tau=0$ in a $L=6$ lattice. }
    \label{fig:compare_HABC_EA}
\end{figure}

We now turn to the discussion of nuclear saturation on the lattice.
In Fig.~\ref{fig:compare_HABC_EA}, we plot the $\tau=0$ Hartree--Fock (HF) energies per particle for nuclear matter in an $L=6$ box with periodic boundary conditions using $H^\text{A}$, $H^\text{B}$, and $H^\text{C}$.
Although the phase shifts in Fig.~\ref{fig:compare_HABC} are broadly similar, the corresponding equations of state are very different.
Moving from $H^\text{A}$ to $H^\text{B}$ to $H^\text{C}$, the nonlocal smearing parameter is increased while the local smearing parameter is decreased.  While all three Hamiltonians are constrained to give approximately the same $^4$He energy, $H^\text{A}$ remains too attractive in nuclear matter and produces an overbound liquid with saturation at too high a density, whereas $H^\text{C}$ is much less attractive and turns upward already at relatively low density, placing it close to the quantum phase transition boundary to an $\alpha$-cluster Bose gas.  The intermediate Hamiltonian $H^\text{B}$ strikes a good balance and its minimum is close to the empirical saturation point.  This strong many-body sensitivity confirms the observation made in Ref.~\cite{Elhatisari2016_PRL117-132501} that the transition between an $\alpha$-cluster Bose gas and a nuclear liquid is sensitive to the strength and range of the local part of the interaction and is not fully determined by the low-energy nucleon-nucleon phase shifts.

\subsection{Interaction energy at high density}
\label{sec:high_density_scaling}
In Ref.~\cite{Rothman2026_NuLattice}, it was suggested that the saturation of the interaction energy per particle at high densities observed in recent NLEFT calculations is a lattice artifact, driven by lattice-packing effects and occupancy constraints. We show here that this inference is not supported by the evidence. The flattening of the interaction energy per particle at high density is not, by itself, a diagnostic signature of lattice packing. The same qualitative behavior occurs already in continuous space for a Galilean-invariant interaction with a relative-momentum regulator. In order to identify a lattice artifact, one must show that the effect tracks with the lattice cutoff or site-occupancy scale rather than some other physical scale of the interaction.

We consider the interaction energy in the laboratory rest frame for symmetric nuclear matter at zero temperature. Symmetric nuclear matter consists of four nucleon species (proton spin-up, proton spin-down, neutron spin-up, neutron spin-down), yielding a spin-isospin degeneracy of $\nu = 4$. The total number density is $\rho = \nu k_F^3 / 6\pi^2 = 2k_F^3 / 3\pi^2$, where $k_F$ is the Fermi momentum.

\subsubsection{Nonlocal Gaussian Interaction in Continuous Space}
\label{subsec:Gaussian}
We consider a nonlocal Gaussian interaction in continuous space that produces an S-wave interaction. In order to preserve Galilean invariance, the interaction depends only on the relative momenta. In a general two-body scattering process, the interaction matrix element $\langle \mathbf{p}' | V | \mathbf{p} \rangle$ depends on both the incoming relative momentum $\mathbf{p} = (\mathbf{k}_1 - \mathbf{k}_2)/2$ and the outgoing relative momentum $\mathbf{p}' = (\mathbf{k}'_1 - \mathbf{k}'_2)/2$, and is independent of the total laboratory momentum $\mathbf{P} = \mathbf{k}_1 + \mathbf{k}_2$.

However, in infinite homogeneous nuclear matter, the Hartree-Fock energy is the expectation value of the Hamiltonian with respect to a Slater determinant of orthogonal plane waves. Momentum conservation for the direct (Hartree) and exchange (Fock) terms restricts the non-zero contributions to forward scattering ($\mathbf{p}' = \mathbf{p}$) and exchange scattering ($\mathbf{p}' = -\mathbf{p}$), respectively. For a parity-even S-wave interaction, the matrix elements for both terms are identical and depend solely on the magnitude of the initial relative momentum,
\begin{equation}
\langle -\mathbf{p} | V | \mathbf{p} \rangle = \langle \mathbf{p} | V | \mathbf{p} \rangle \equiv V(p).
\end{equation}
Summing over the $\nu = 4$ spin-isospin states, the direct term yields a degeneracy factor of $\nu^2 = 16$, while the exchange term yields a factor of $\nu = 4$ for a spin-isospin independent interaction. The net spin-isospin weighting is therefore $16 - 4 = 12$.

The relevant diagonal elements of the interaction can be parameterized using a Gaussian regulator to suppress high-momentum relative modes:
\begin{equation}
V(p) = V_0 \exp\left(-\frac{p^2}{\Lambda^2}\right),
\label{eq:pot_gauss}
\end{equation}
where $\Lambda$ is the momentum regulator scale and $V_0$ parameterizes the interaction strength.

Accounting for the global $1/2$ factor in the two-body energy sum and the net spin-isospin factor of $12$, the interaction energy per particle is given by
\begin{equation}
\frac{E_{\text{int}}}{A} = \frac{6}{\rho} \int \frac{d^3k_1 \, d^3k_2}{(2\pi)^6} \, V(p) \, \Theta(k_F - k_1) \Theta(k_F - k_2).
\label{eq:e_int_start}
\end{equation}
We transform the integration variables to the total laboratory momentum $\mathbf{P}$ and the relative momentum $\mathbf{p}$. The Jacobian determinant for this linear transformation is unity,
\begin{equation}
d^3k_1 \, d^3k_2 = d^3P \, d^3p.
\end{equation}
The Fermi sea boundaries are enforced by
\begin{equation}
\begin{aligned}
\Theta_F(\mathbf{P}, \mathbf{p}) ={}& \Theta\left(k_F - \left|\frac{\mathbf{P}}{2} + \mathbf{p}\right|\right) \\
&\times \Theta\left(k_F - \left|\frac{\mathbf{P}}{2} - \mathbf{p}\right|\right).
\end{aligned}
\end{equation}
For a fixed relative momentum $\mathbf{p}$, the allowed region in $\mathbf{P}/2$ is the geometric overlap of two Fermi spheres of radius $k_F$ separated by a distance $2p$. Letting $\mathbf{u} = \mathbf{P}/2$, the differential volume scales as $d^3P = 8 \, d^3u$. Thus, the integral over $\mathbf{P}$ yields exactly 8 times the standard spherical overlap volume:
\begin{equation}
\mathcal{V}_{\text{over}}(p) = 8 \left( \frac{4\pi k_F^3}{3} \right) \left[ 1 - \frac{3p}{2k_F} + \frac{1}{2}\left(\frac{p}{k_F}\right)^3 \right] \Theta(k_F - p).
\label{eq:v_overlap}
\end{equation}
Substituting Eq.~\eqref{eq:v_overlap} into Eq.~\eqref{eq:e_int_start} and integrating over the angular components of $\mathbf{p}$, which contributes a factor of $4\pi$, we use the density relation $\rho = 2k_F^3 / 3\pi^2$. All $k_F^3$ terms cancel. The product of the analytical constants reduces to the prefactor $6 V_0 / \pi^2$, yielding
\begin{equation}
\frac{E_{\text{int}}}{A} = \frac{6 V_0}{\pi^2} \int_0^{k_F} dp \, p^2 e^{-p^2/\Lambda^2} \left[ 1 - \frac{3p}{2k_F} + \frac{1}{2}\left(\frac{p}{k_F}\right)^3 \right].
\label{eq:e_int_simplified}
\end{equation}

We evaluate this expression in two density regimes.

\textbf{The Low-Density Limit ($k_F \ll \Lambda$):}
The relative momentum $p$ is bounded by $k_F \ll \Lambda$, making the regulator approximately inactive, $e^{-p^2/\Lambda^2} \approx 1$. The remaining polynomial integral is
\begin{equation}
\begin{aligned}
I_{\rm low} &= \int_0^{k_F} dp \, p^2 \left[ 1 - \frac{3p}{2k_F} + \frac{1}{2}\left(\frac{p}{k_F}\right)^3 \right] \\
&= k_F^3 \left( \frac{1}{3} -\frac{3}{8} +\frac{1}{12} \right) = \frac{k_F^3}{24}.
\end{aligned}
\label{eq:low_density_integral}
\end{equation}
Because $\rho = 2k_F^3 / 3\pi^2$, we recover the expected contact-interaction scaling:
\begin{equation}
\frac{E_{\text{int}}}{A} \approx \frac{6 V_0}{\pi^2} \left( \frac{k_F^3}{24} \right) = \frac{V_0 k_F^3}{4\pi^2} = \frac{3V_0}{8}\rho.
\label{eq:low_density_limit}
\end{equation}
At low densities, the interaction energy per particle grows linearly with density.

\textbf{The High-Density Limit ($k_F \gg \Lambda$):}
As the density increases, $k_F$ exceeds $\Lambda$. The Gaussian regulator restricts the important part of the integrand to $p \lesssim \Lambda \ll k_F$. In this domain, $p/k_F$ is small, and the upper limit of the integral can be extended to infinity. More explicitly, Eq.~\eqref{eq:e_int_simplified} gives
\begin{equation}
\frac{E_{\text{int}}}{A} = \frac{6 V_0}{\pi^2} \left[ \frac{\sqrt{\pi}}{4}\Lambda^3 -\frac{3}{4}\frac{\Lambda^4}{k_F} +\frac{1}{2}\frac{\Lambda^6}{k_F^3} +\cdots \right],
\label{eq:high_density_expansion}
\end{equation}
and therefore
\begin{equation}
\lim_{k_F \to \infty} \frac{E_{\text{int}}}{A} = \frac{6 V_0}{\pi^2} \int_0^\infty dp \, p^2 e^{-p^2/\Lambda^2} = \frac{3 V_0}{2 \pi^{3/2}} \Lambda^3.
\label{eq:high_density_limit}
\end{equation}
This derivation demonstrates that the flattening of the interaction energy per particle at high densities is a consequence of the finite relative-momentum regulator in this continuous-space model. At $k_F \gg \Lambda$, only a fraction of pairs of order $(\Lambda/k_F)^3$ have relative momenta inside the regulator support. This phase-space suppression exactly compensates for the growth of the density after dividing by the particle number, leaving an interaction energy per particle of order $V_0 \Lambda^3$. No spatial lattice, Brillouin zone, or one-particle-per-site occupancy constraint is involved in this argument. The flattening of the interaction energy is a property of the regulated finite-resolution interaction, not evidence of a lattice artifact.

We emphasize that this argument concerns the interaction-energy component. The saturation point of the full equation of state also depends on the kinetic energy, which grows with $k_F^2$, as well as on the density dependence of all other interaction terms. Nevertheless, the central point remains: the high-density flattening of the interaction energy is not a unique signature of lattice packing.

\subsubsection{Lattice Smearing and the Effective Momentum Cutoff}
\label{subsec:lattice_smearing}
To connect this continuous-space result to the lattice formalism, we examine the nonlocal smearing parameter $s_{\rm NL}$ used in Ref.~\cite{Lu2019_PLB797-134863} and other NLEFT publications. On the lattice, the nonlocally smeared annihilation operator is
\begin{equation}
\overline{a}(\mathbf{n}) = a(\mathbf{n}) + s_{\rm NL} \sum_{|\mathbf{n}'-\mathbf{n}|=1} a(\mathbf{n}').
\label{eq:smearing_real_space}
\end{equation}
Applying a Fourier transform, the operator in momentum space becomes $\overline{a}(\mathbf{k}) = f(\mathbf{k}) a(\mathbf{k})$, where the form factor $f(\mathbf{k})$ is
\begin{equation}
f(\mathbf{k}) = 1 + 2s_{\rm NL} \sum_{i=x,y,z} \cos(k_i a).
\label{eq:form_factor}
\end{equation}
For low momenta, $ka \ll 1$, expanding the cosine terms yields
\begin{equation}
f(\mathbf{k}) \approx (1 + 6s_{\rm NL}) \left[ 1 - \frac{s_{\rm NL} a^2}{1 + 6s_{\rm NL}} \mathbf{k}^2 \right].
\label{eq:form_factor_expanded}
\end{equation}
Using $1 - x \approx \exp(-x)$ for small $x$, this maps onto a continuous Gaussian smearing function,
\begin{equation}
f(\mathbf{k}) \approx (1 + 6s_{\rm NL}) \exp\left( - \frac{s_{\rm NL} a^2}{1 + 6s_{\rm NL}} \mathbf{k}^2 \right).
\label{eq:form_factor_gaussian}
\end{equation}
In the two-body center-of-mass frame, the two-body interaction involves four smeared operators. At low momenta, the effective interaction matrix element for a pair with relative momentum $\mathbf{p}$ is modulated by $[f(\mathbf{p})]^4$:
\begin{equation}
[f(\mathbf{p})]^4 \propto \exp\left( - \frac{4 s_{\rm NL} a^2}{1 + 6s_{\rm NL}} \mathbf{p}^2 \right).
\label{eq:form_factor_fourth}
\end{equation}
Equating this momentum curvature to the continuous Gaussian regulator $\exp(-p^2/\Lambda^2)$ introduced in Eq.~\eqref{eq:pot_gauss}, we identify the effective momentum scale
\begin{equation}
\Lambda = \frac{1}{2a} \sqrt{\frac{1 + 6s_{\rm NL}}{s_{\rm NL}}}.
\label{eq:lambda_relation}
\end{equation}
This estimate should be interpreted as the momentum regulator scale associated with the smearing. The lattice form factor in Eq.~\eqref{eq:form_factor} is periodic and non-Gaussian near the edge of the Brillouin zone, but its low momentum curvature determines the scale at which the interaction begins to weaken with increasing relative momentum. For example, the calculations in Ref.~\cite{Lu2019_PLB797-134863} utilize a nonlocal smearing parameter of $s_{\rm NL} = 0.5$. Substituting this into Eq.~\eqref{eq:lambda_relation} yields
\begin{equation}
\Lambda = \frac{1}{2a} \sqrt{\frac{1 + 6(0.5)}{0.5}} \approx \frac{1.41}{a}.
\end{equation}
This scale is significantly below the lattice momentum cutoff $\pi/a$. Thus the weakening of the interaction with increasing relative momentum begins at a scale set by the smearing of the interaction. With nearest-neighbor smearing, $s_{\rm NL}$ changes the effective regulator scale by a factor relative to $1/a$. More extended smearing kernels, involving sites beyond nearest neighbors in Eq.~\eqref{eq:smearing_real_space}, can separate the physical regulator scale from the lattice cutoff more significantly.

The conclusion is that nonlocal smearing introduces a finite momentum scale in the interaction. The suppression of the high-density interaction energy should therefore be compared against this physical regulator scale. It cannot be identified as a lattice-packing artifact merely from the observation that the interaction energy per particle flattens at large density.

\subsubsection{Secondary Flattening Mechanisms and Absolute-Momentum Regulators}
\label{subsec:wang_regulators}
While Galilean-invariant nonlocal interactions produce interaction-energy flattening through the suppression of large relative momentum $\mathbf{p}$, alternative regulator schemes can provide additional sources of interaction-energy suppression. We consider the recent work of Wang et al.~\cite{Wang2026_arXiv2604.20681}, which investigates saturation properties at fine lattice spacings.

In Ref.~\cite{Wang2026_arXiv2604.20681}, a spatial lattice spacing of $0.987\ \text{fm}$ is used. On this grid, one nucleon per site corresponds to a physical density of $1.04\ \text{fm}^{-3}$, roughly seven times higher than the empirical nuclear saturation density. Saturation is achieved utilizing an attractive two-body interaction with no three-body forces included. Since the site-occupancy scale lies far above the empirical saturation density, lattice packing cannot by itself explain the appearance of saturation on this fine grid. The mechanism must instead be associated with the momentum structure of the regulated interaction.

The model in Ref.~\cite{Wang2026_arXiv2604.20681} employs an absolute-momentum regulator, which acts on the individual nucleon laboratory momenta $\mathbf{k}_1$ and $\mathbf{k}_2$ rather than solely on the relative momentum $\mathbf{p}$. Such a regulator depends on the total center-of-mass momentum $\mathbf{P}$ and therefore is not exactly Galilean invariant at all momenta. In the Supplemental Material of that work, it is shown that Galilean symmetry can be restored up to $\mathcal{O}(Q^4)$ utilizing a Symanzik improvement scheme.

Although Galilean invariance at low momenta is preserved through this improvement, the interaction is weakened when the total momentum of the two-nucleon system is large. At high densities, the expansion of the Fermi sphere ensures that both the relative momentum $\mathbf{p}$ and the total momentum $\mathbf{P}$ become large for many nucleon pairs. The absolute-momentum regulator suppresses these large-laboratory-momentum states, providing an additional mechanism for the flattening of the interaction energy per particle in symmetric nuclear matter. This highlights a methodological difference between lattice formalisms utilizing absolute-momentum regulators and formulations where Galilean invariance is preserved at all momentum scales.

Whether through relative-momentum suppression in a Galilean-invariant finite-resolution interaction or through additional total-momentum suppression from an absolute-momentum regulator, the flattening of the interaction energy at high densities is driven by the momentum structure of the regulated interaction. It is therefore not valid to infer, from the flattening alone, that the effect is caused by lattice-packing limits or occupancy constraints. A genuine lattice artifact would have to be demonstrated by showing scaling with the lattice cutoff or site-occupancy density under lattice refinement, rather than with the physical regulator scale of the interaction.

\section{Benchmarking Lessons Learned}
\label{sec:benchmarking_lessons}

The authors of Ref.~\cite{Rothman2026_NuLattice} recently performed Hartree-Fock calculations using several legacy NLEFT interactions, including the 2016 and 2017 clustering actions, and argued that these results reveal problems with published AFQMC calculations, variational bounds, correlation energies, and saturation mechanisms.  After examining the analysis in detail, we find that the numerical Hartree--Fock calculations provide useful diagnostics, but that several of the resulting physical interpretations are erroneous and misleading.

The central issue is that several comparisons in Ref.~\cite{Rothman2026_NuLattice} do not keep the regulated lattice problem fixed.  A regulated lattice interaction is not specified by its coupling constants or spatial Hamiltonian density alone; it is defined together with the regulator scheme used to calibrate it, including the spatial lattice spacing, temporal regulator or Hamiltonian limit, kinetic-energy discretization, smearing prescription, and operator content.  Changing these ingredients without a corresponding refit generally defines a different Hamiltonian or transfer matrix.  A separate issue arises in finite-volume benchmarks: even for a fixed interaction, one must compare calculations at the same volume, particle number, and boundary conditions, or else account explicitly for the extrapolation to the infinite-volume or thermodynamic limit.  Several of the apparent discrepancies discussed in Ref.~\cite{Rothman2026_NuLattice} result from mixing these distinct regulated theories or finite-volume limits.

\subsection{What Hamiltonian is being benchmarked?}

An important issue for the legacy NLEFT calculations of
Refs.~\cite{Elhatisari2016_PRL117-132501,PRL119-222505}
is the distinction between a lattice Hamiltonian and a lattice transfer
matrix.  The published AFQMC calculations were performed with a
normal-ordered transfer matrix at nonzero temporal lattice spacing $a_t$.
The corresponding energy is extracted from the transfer matrix,
schematically as
\begin{equation}
E_M=-\frac{1}{a_t}\log \lambda_M ,
\end{equation}
where $\lambda_M$ is the relevant eigenvalue of the transfer matrix.  This
is not, at nonzero $a_t$, identical to the expectation value
\begin{equation}
E_H=\langle H\rangle
\end{equation}
of the $a_t\to0$ lattice Hamiltonian constructed from the same bare
couplings.

Equivalently, the nonzero-$a_t$ transfer matrix defines an induced
effective Hamiltonian $H_{\rm eff}(a_t)$ through
\begin{equation}
M=\exp[-a_t H_{\rm eff}(a_t)] .
\end{equation}
For a normal-ordered interacting transfer matrix,
\begin{equation}
M=:\!\exp(-a_t H)\!: ,
\end{equation}
and therefore
\begin{equation}
H_{\rm eff}(a_t)=H+\mathcal{O}(a_t),
\end{equation}
with induced one-, two-, three-, and higher-body operators generated by
normal ordering, noncommutativity, and the auxiliary-field representation.
For the temporal lattice spacings used in the legacy transfer-matrix actions,
these induced terms are not negligible; they are part of the calibrated
regulated action.  Thus the nonzero-$a_t$ transfer matrix is not merely a
numerical approximation to the $a_t=0$ Hamiltonian.  It is its own
well-defined regulated lattice action.

This distinction is already implicit in Ref.~\cite{Rothman2026_NuLattice},
where it is noted that the AFQMC calculations of
Refs.~\cite{Elhatisari2016_PRL117-132501,PRL119-222505}
were performed using a transfer-matrix formalism with nonzero temporal
lattice spacing and are not equivalent to the $a_t\to0$ Hamiltonian limit.
That observation is correct, but it changes the interpretation of the
benchmark.  A calculation performed with the $a_t\to0$ Hamiltonian is not
a benchmark of the nonzero-$a_t$ transfer-matrix theory unless the temporal
regulator is removed consistently or the couplings are refit.

The same regulator-consistency issue applies to both the 2016 action of
Ref.~\cite{Elhatisari2016_PRL117-132501}, with
$a_t=(150~\mathrm{MeV})^{-1}$, and the 2017 action of
Ref.~\cite{PRL119-222505}, with
$a_t=(100~\mathrm{MeV})^{-1}$.  We present new numerical saturation data
for the 2017 action because its sign problem is milder and therefore gives
a cleaner diagnostic test.  The 2016 action has stronger sign oscillations,
but the formal issue is identical: changing the temporal regulator without
refitting changes the regulated theory being solved.

\subsection{Finite-nucleus Hartree--Fock bounds}

The finite-nucleus tables in Ref.~\cite{Rothman2026_NuLattice} show that
Hartree--Fock energies computed from the $a_t\to0$ Hamiltonian can lie
below published AFQMC energies obtained from nonzero-$a_t$ transfer
matrices.  Once the distinction in the previous subsection is made, this
does not represent a variational paradox.  A Hartree--Fock variational
upper bound applies only within a single regulated quantum system: the same
Hamiltonian or transfer matrix, the same regulator, the same volume, the
same boundary conditions, and the same energy definition.  It cannot be
used to compare $E_H$ for the $a_t\to0$ Hamiltonian against $E_M$ for a nonzero-$a_t$ transfer matrix.

Our explicit comparisons in Table~\ref{tab:Finite_Nu_HF} and
Fig.~\ref{fig:HF_Tc_check} demonstrate this point directly.  At fixed wave
function, $E_M(a_t)$ approaches $E_H$ only as $a_t\to0$.  At the nonzero
temporal spacings used in the legacy calculations, the difference can be
large because the transfer matrix contains induced $a_t$-dependent
operators.  However, when the comparison is made within one formulation,
the expected variational ordering is restored: transfer-matrix
Hartree--Fock energies bound transfer-matrix projected energies, and
Hamiltonian Hartree--Fock energies bound Hamiltonian projected energies.

Therefore the finite-nucleus tables of Ref.~\cite{Rothman2026_NuLattice}
do not show that the AFQMC calculations failed and do not show that the
nonzero-$a_t$ interactions are pathological.  They show that bare couplings
from a nonzero-$a_t$ transfer-matrix action cannot be inserted
into the $a_t\to0$ Hamiltonian limit and then used as a variational
benchmark without regulator matching or refitting.

\subsection{Correlation energy is not a complete measure of correlations}
\label{subsec:correlations}

Ref.~\cite{Rothman2026_NuLattice} observes that, in neutron matter, the \texttt{EE} interaction gives small correlation energies compared with other interactions. This observation is not a deficiency of the interaction or of the many-body method.  The \texttt{EE} Hamiltonian was designed as an ``essential elements'' interaction with exact Wigner SU(4) symmetry, a small number of parameters, and a mild sign problem.  It is intentionally soft and does not aim to reproduce all high-momentum two-body phase shifts.

More importantly, the correlation energy is not a complete diagnostic of the many-body wave function.  Our neutron-matter calculation in Fig.~\ref{fig:EE_L6_S_QT} starts from a Hartree-Fock state in which the irreducible s-wave pair, p-wave pair, and quartet signals vanish.  During Euclidean-time projection the energy changes modestly, but the irreducible two-body and four-body densities grow continuously.  At large $\tau$, the system exhibits multimodal superfluidity with s-wave pair, p-wave pair, and quartet condensation.  Thus, a small Hartree-Fock-to-AFQMC energy difference does not imply a simple wave function devoid of nontrivial correlations and entanglement.

\subsection{Thermodynamic-limit comparisons require matched limits}

A separate benchmarking issue concerns nuclear matter.  Ref.~\cite{Rothman2026_NuLattice} compares finite periodic-box Hartree-Fock energies at artificial closed shells with AFQMC results that are often closer to a twist-averaged or thermodynamic-limit estimate.  This is not a controlled variational comparison.  Periodic boundary conditions generate artificial shell closures at special particle numbers.  At those closed shells, finite-box Hartree-Fock energies can be lowered by shell effects relative to the thermodynamic-limit Hartree-Fock energy.  This finite-size lowering competes with the many-body correlation energy.  Therefore a finite periodic-box Hartree-Fock value at an artificial magic number is not a rigorous upper bound on a twist-averaged or thermodynamic-limit AFQMC result.

Average twisted boundary conditions remove these shell structures, as shown in Fig.~\ref{fig:ATBC}.  When comparisons are made at fixed $A$, $L$, boundary condition, action, and energy definition, the projected AFQMC energies lie below the corresponding Hartree-Fock values, as required by the variational principle.  Apparent violations arise only when finite-size shell corrections and many-body correlation energies are mixed across different limits.

\subsection{Momentum regulation versus lattice packing}

The authors of Ref.~\cite{Rothman2026_NuLattice} argue that the high-density flattening of the interaction energy for the \texttt{EE} interaction is caused by dense lattice packing and Pauli blocking of hopping terms.  Such effects can certainly appear at very high occupation.  However, this does not establish that the empirical saturation region or the interaction-energy flattening relevant to NLEFT is a lattice-packing artifact.

There are two reasons.  First, the density scale at which lattice packing and Pauli-blocking effects could plausibly occur is significantly higher than the empirical saturation density.  For the \texttt{EE} lattice spacing $a=1.32~\mathrm{fm}$ ($a^{-1}=150~\mathrm{MeV}$), the average site occupation at the empirical saturation density $\rho_0=0.16~\mathrm{fm}^{-3}$ is $\bar n = \rho_0 a^3\approx0.36$, a factor of three lower than an occupancy of $\bar n = 1$ and a factor of eleven smaller than the maximum filling $\bar n = 4$.  For the finer lattice spacing $a=0.987~\mathrm{fm}$ used in Ref.~\cite{Wang2026_arXiv2604.20681}, the average site occupation at the empirical saturation density is $\bar n = \rho_0 a^3\approx0.15$, a factor of six lower than an occupancy of $\bar n = 1$ and a factor of twenty-six smaller than the maximum filling $\bar n = 4$.

Second, Sec.~\ref{sec:high_density_scaling} shows analytically that interaction-energy flattening occurs already in continuous space for a Galilean-invariant nonlocal Gaussian interaction.  At large Fermi momentum, only a fraction of pairs of order $(\Lambda/k_F)^3$ have relative momentum below the regulator scale; this phase-space suppression cancels the density growth in $E_{\rm int}/A$ and produces a constant high-density limit proportional to $V_0\Lambda^3$.  No lattice sites, Brillouin zone, or one-particle-per-site occupancy limit enter this derivation.

The lattice smearing used in NLEFT introduces precisely such a finite momentum scale.  For nearest-neighbor nonlocal smearing, the low-momentum curvature corresponds to
\begin{equation}
\Lambda = \frac{1}{2a}\sqrt{\frac{1+6s_{\rm NL}}{s_{\rm NL}}},
\end{equation}
which is below the Brillouin-zone cutoff for typical parameters.  This sensitivity to the regulator rather than to the lattice spacing is borne out directly by the Hamiltonians $H^{\rm A}$, $H^{\rm B}$, and $H^{\rm C}$ of Sec.~\ref{sec:NLEFT}.  At fixed lattice spacing $a=(100~\mathrm{MeV})^{-1}$, and hence fixed site-occupancy scale, varying $s_\mathrm{NL}$ and $s_\mathrm{L}$ produces the markedly different saturation curves shown in Fig.~\ref{fig:compare_HABC_EA}.  Smaller $s_\mathrm{NL}$ together with stronger local attraction shifts the saturation point to higher density with stronger binding.  This confirms numerically that, in the density regime relevant to typical NLEFT saturation calculations, the saturation properties track with the smearing-induced regulator scale rather than the lattice-packing scale.

\subsection{Continuum limit and incorrect contact-interaction renormalization}
\label{subsec:continuum_limit_response}

Ref.~\cite{Rothman2026_NuLattice} presents a formal continuum-limit argument against the lattice interaction of Ref.~\cite{Lu2019_PLB797-134863}.  The argument considers taking $a\to0$ at fixed physical volume and concludes that the two-body interaction ``leads to a collapse of the system'' and that the lattice potential ``cannot simply be used in computations that employ continuum Hamiltonians.''  The central step in this argument is the assertion that, after a low-density matching condition is imposed,
\begin{equation}
-\frac{\langle V_0\rangle}{A}\propto a^{-3}
\qquad\hbox{while}\qquad
\frac{\langle T\rangle}{A}\propto a^{-2}.
\end{equation}
The resulting collapse is then attributed to the continuum limit of the lattice interaction.

This scaling is not the correct renormalization of a three-dimensional short-range interaction.  The $a^{-3}$ behavior follows from holding fixed a lowest-density Hartree--Fock, or Born-level, matrix element.  Equivalently, it treats the volume integral of the short-range interaction as the quantity to be kept fixed.  But in three dimensions a contact interaction is not renormalized by fixing its Born matrix element.  The physical renormalization condition must be imposed on the two-body scattering amplitude, or equivalently on a bound-state or virtual-state pole.

The standard example is the 3D attractive Fermi Hubbard interaction tuned to infinite scattering length.  In the Hamiltonian lattice formulation, the on-site coupling at unitarity scales as
\begin{equation}
U_c \equiv \frac{C_c}{a^3}
\simeq -\frac{3.957}{m a^2},
\label{eq:unitary_hubbard_scaling}
\end{equation}
for the simplest lattice kinetic-energy action~\cite{Bour:2011xt,Burovski:2006ty}.  Thus the corresponding continuum coefficient scales as
\begin{equation}
C_c \simeq -\frac{3.957}{m}\,a,
\label{eq:continuum_contact_scaling}
\end{equation}
rather than remaining fixed.  The same result follows from an attractive finite well of range $R$: at unitarity the well depth scales as $1/(mR^2)$, so its volume integral scales as $R/m$ and vanishes as $R\to0$.
This $a\to0$ scaling is not special to the unitary limit. For a contact interaction renormalized to any fixed finite scattering length $a_s$, the linearly divergent lattice self-energy dominates the coupling as $a\to0$, giving the same leading behavior $C_c\propto a$.

Therefore the $a^{-3}$ scaling used in Ref.~\cite{Rothman2026_NuLattice} is not the scaling of a properly renormalized three-dimensional contact interaction.  It is the scaling obtained from an inappropriate Born-level matching prescription.  The collapse that follows from this prescription is consequently an artifact of the continuum extrapolation being performed, not a pathology of the finite-cutoff NLEFT interaction of Ref.~\cite{Lu2019_PLB797-134863}.  A correct change of cutoff requires renormalizing to physical scattering observables, and in the many-body nuclear problem generally requires the corresponding three- and higher-body counterterms as well.

\section{Summary and Outlook}
\label{sec:Summary}

Any numerical calculation aiming at an exact solution must confront systematic uncertainties.  These uncertainties are especially challenging when they originate from approximations introduced to make the many-body problem computationally tractable.  Experience from lattice QCD and other fields shows that insufficient control of such effects can produce apparent discrepancies between different theoretical frameworks, or between theory and experiment, that are later resolved once the relevant systematics are quantified.

In this work, we have addressed this challenge in nuclear lattice QMC by carrying out a detailed uncertainty analysis of the recently developed sign-problem-free spin-orbit action LAT-OPT1, while keeping the broader objective of full uncertainty quantification for sign-problem-free NLEFT benchmarks.  The central result is that spin-orbit physics can be incorporated nonperturbatively into the Euclidean-time projection while preserving determinant positivity for even-even nuclei.  We quantified the main sources of computational uncertainty, including finite-projection-time and finite-volume extrapolations, finite-temporal-step effects, auxiliary-field-induced operators, and residual phase fluctuations in odd systems.  The induced operators associated with the square-completed spin-orbit construction are small, the residual sign problem remains mild beyond the even-even sector, and the total many-body computational uncertainty is well below the percent level for doubly magic nuclei up to $^{100}$Sn.

The spin-orbit analysis also demonstrates the physical importance of the new action.  When included nonperturbatively, the spin-orbit term changes projected binding energies by several to ten MeV in medium-mass nuclei, a scale comparable to shell-correction energies.  Thus LAT-OPT1 is not simply an SU(4) action supplemented by a perturbative spin-orbit insertion.  It is a sign-problem-free action in which shell structure is generated dynamically during the projection.

We also analyzed recent criticisms of legacy NLEFT calculations.  Ref.~\cite{Rothman2026_NuLattice} compares finite-temporal-spacing transfer-matrix results with Hamiltonian-limit calculations and interprets the resulting differences as failures of the original AFQMC calculations.  This interpretation is incorrect: the comparison is between inequivalent regulated theories, not between two many-body solutions of the same lattice action.  Severe overbinding appears when interactions calibrated in the transfer-matrix formulation are evaluated in the Hamiltonian limit without refitting.  When Hartree--Fock and AFQMC calculations use the same regulated lattice formulation and the same tuned interaction parameters, the expected variational ordering is restored.  Likewise, finite periodic-box Hartree--Fock energies cannot be used as bounds on twist-averaged thermodynamic-limit results, and small correlation energies do not imply trivial many-body wave functions.  Finally, the high-density flattening of the interaction energy follows from momentum-space regulation, through either Galilean-invariant relative-momentum regulators or absolute-momentum regulators, and is therefore not by itself evidence of lattice packing.

Taken together, these results show that the examined systematic errors are quantitatively controlled across a broad mass range from $^{4}$He to $^{100}$Sn.  They also clarify which benchmark comparisons are meaningful: the regulated lattice formulation, tuned interaction parameters, finite-volume setup, and thermodynamic-limit procedure must be kept consistent.  With these conditions satisfied, the apparent discrepancies discussed above are resolved, and the sign-problem-free NLEFT calculations provide rigorous benchmarks for nuclear many-body methods.

Looking forward, the controlled spin-orbit action provides a practical path toward sign-problem-free studies of shell evolution, spectroscopy, and electroweak transitions in medium-mass and heavy nuclei.  It also provides a stable foundation for adding chiral corrections order by order and for separating genuine nuclear-force deficiencies from many-body solver errors.  Finally, the regulator-consistent thermodynamic-limit analysis developed here provides a framework for high-precision studies of dense matter and for connecting NLEFT calculations to neutron-star and multi-messenger constraints.

\section*{Acknowledgments}
We are grateful for discussions with members of the NLEFT Collaboration.
This work was supported by the Science Challenge Project (No. TZ2025012), NSAF No. U2330401 and National Natural Science Foundation of China with Grant Nos. 12275259, 12547105, the European
Research Council (ERC) under the European Union's Horizon 2020 research
and innovation programme (grant agreement No. 101018170),
and by the CAS President's International Fellowship Initiative (PIFI) (Grant No.~2025PD0022).


\begin{thebibliography}{999}

\bibitem{Rothman2026_NuLattice}
M. Rothman, G. Hagen, M. Heinz, and T. Papenbrock, Saturation of Nuclear Binding from Lattice Hamiltonians, \href{https://arxiv.org/abs/2606.12166v1}
{arXiv:2606.12166v1 [nucl-th]}.

\bibitem{Epelbaum2009_RMP81-1773}
E. Epelbaum, H.-W. Hammer, and Ulf-G. Mei{\ss}ner, Modern theory of nuclear forces, \href{https://doi.org/10.1103/RevModPhys.81.1773}{Rev. Mod. Phys. \textbf{81}, 1773 (2009)}.

\bibitem{Hammer2020_RMP92-025004}
H.-W. Hammer, S. K{\"o}nig, and U. van Kolck, Nuclear effective field theory: Status and perspectives, \href{https://doi.org/10.1103/RevModPhys.92.025004}{Rev. Mod. Phys. \textbf{92}, 025004 (2020)}.

\bibitem{Epelbaum2012PPNP67-343}
E. Epelbaum, Nuclear forces from chiral effective field theory, \href{https://doi.org/10.1016/j.ppnp.2011.12.041}{Prog. Part. Nucl. Phys. \textbf{67}, 343 (2012)}.

\bibitem{Machleidt2014_EPJConf66-01011}
R. Machleidt, Chiral effective field theory for nuclear forces: Achievements and challenges, \href{https://doi.org/10.1051/epjconf/20146601011}{EPJ Web Conf. \textbf{66}, 01011 (2014)}.

\bibitem{Machleidt2024_PPNP137-104117}
R. Machleidt and F. Sammarruca, Recent advances in chiral EFT based nuclear forces and their applications, \href{https://doi.org/10.1016/j.ppnp.2024.104117}{Prog. Part. Nucl. Phys. \textbf{137}, 104117 (2024)}.

\bibitem{Machleidt2011PhysRept503-1}
R. Machleidt and D. R. Entem, Chiral effective field theory and nuclear forces, \href{https://doi.org/10.1016/j.physrep.2011.02.001}{Phys. Rep. \textbf{503}, 1 (2011)}.

\bibitem{Epelbaum2020_FrontPhys8-98}
E. Epelbaum, H. Krebs, and P. Reinert, High-precision nuclear forces from chiral EFT: State-of-the-art, challenges, and outlook, \href{https://doi.org/10.3389/fphy.2020.00098}{Front. Phys. \textbf{8}, 98 (2020)}.

\bibitem{Carlson2015_RMP87-1067}
J. Carlson, S. Gandolfi, F. Pederiva, S. C. Pieper, R. Schiavilla, K. E. Schmidt, and R. B. Wiringa, Quantum Monte Carlo methods for nuclear physics, \href{https://doi.org/10.1103/RevModPhys.87.1067}{Rev. Mod. Phys. \textbf{87}, 1067 (2015)}.

\bibitem{Barrett2013_PPNP69-131}
B. R. Barrett, P. Navr{\'a}til, and J. P. Vary, The ab initio no-core shell model, \href{https://doi.org/10.1016/j.ppnp.2012.10.003}{Prog. Part. Nucl. Phys. \textbf{69}, 131 (2013)}.

\bibitem{Hagen2014_ReptProgPhys77-096302}
G. Hagen, T. Papenbrock, M. Hjorth-Jensen, and D. J. Dean, Coupled-cluster computations of atomic nuclei, \href{https://doi.org/10.1088/0034-4885/77/9/096302}{Rep. Prog. Phys. \textbf{77}, 096302 (2014)}.

\bibitem{Hergert2017_PhysScripta92-023002}
H. Hergert, In-medium similarity renormalization group for closed and open-shell nuclei, \href{https://doi.org/10.1088/1402-4896/92/2/023002}{Phys. Scr. \textbf{92}, 023002 (2017)}.

\bibitem{Bogner2020_FrontinPhys8-379}
H. Hergert, A guided tour of ab initio nuclear many-body theory, \href{https://doi.org/10.3389/fphy.2020.00379}{Front. Phys. \textbf{8}, 379 (2020)}.

\bibitem{Barbieri2004_PPNP52-377}
W. H. Dickhoff and C. Barbieri, Self-consistent Green's function method for nuclei and nuclear matter, \href{https://doi.org/10.1016/j.ppnp.2004.02.038}{Prog. Part. Nucl. Phys. \textbf{52}, 377 (2004)}.

\bibitem{Lee2009_PPNP63-117}
D. Lee, Lattice simulations for few- and many-body systems, \href{https://doi.org/10.1016/j.ppnp.2008.12.001}{Prog. Part. Nucl. Phys. \textbf{63}, 117 (2009)}.

\bibitem{Ekstrom2023_FrontinPhys11-1129094}
A. Ekstr{\"o}m et al., What is ab initio in nuclear theory?, \href{https://doi.org/10.3389/fphy.2023.1129094}{Front. Phys. \textbf{11}, 1129094 (2023)}.

\bibitem{Maris2023_FrontinPhys11-1098262}
P. Maris, H. Le, A. Nogga, R. Roth, and J. P. Vary, Uncertainties in ab initio nuclear structure calculations with chiral interactions, \href{https://doi.org/10.3389/fphy.2023.1098262}{Front. Phys. \textbf{11}, 1098262 (2023)}.

\bibitem{Machleidt2023_FewBodySyst64-77}
R. Machleidt, What is ab initio?, \href{https://doi.org/10.1007/s00601-023-01857-2}{Few-Body Syst. \textbf{64}, 77 (2023)}.

\bibitem{Morris2018_PRL120-152503}
T. D. Morris, J. Simonis, S. R. Stroberg, C. Stumpf, G. Hagen, J. D. Holt, G. R. Jansen, T. Papenbrock, R. Roth, and A. Schwenk, Structure of the lightest tin isotopes, \href{https://doi.org/10.1103/PhysRevLett.120.152503}{Phys. Rev. Lett. \textbf{120}, 152503 (2018)}.

\bibitem{Mougeot2021_NatPhys17-1099}
M. Mougeot et al., Mass measurements of $^{99-101}$In challenge ab initio nuclear theory of the nuclide $^{100}$Sn, \href{https://doi.org/10.1038/s41567-021-01326-9}{Nat. Phys. \textbf{17}, 1099 (2021)}.

\bibitem{Arthuis2020_PRL125-182501}
P. Arthuis, C. Barbieri, M. Vorabbi, and P. Finelli, Ab initio computation of charge densities for Sn and Xe isotopes, \href{https://doi.org/10.1103/PhysRevLett.125.182501}{Phys. Rev. Lett. \textbf{125}, 182501 (2020)}.

\bibitem{Tichai2024_PLB851-138571}
A. Tichai, P. Demol, and T. Duguet, Towards heavy-mass ab initio nuclear structure: Open-shell Ca, Ni and Sn isotopes from Bogoliubov coupled-cluster theory, \href{https://doi.org/10.1016/j.physletb.2024.138571}{Phys. Lett. B \textbf{851}, 138571 (2024)}.

\bibitem{Niu2025_PRL135-222504}
Z.-W. Niu and B.-N. Lu, Sign-problem-free nuclear quantum Monte Carlo simulation, \href{https://doi.org/10.1103/pn99-6dxt}{Phys. Rev. Lett. \textbf{135}, 222504 (2025)}.

\bibitem{Hildenbrand2026_PRL136-062501}
F. Hildenbrand, S. Elhatisari, Ulf-G. Mei{\ss}ner, H. Meyer, Z. Ren, A. Herten, and M. Bode, Lattice calculation of the Sn isotopes near the proton dripline, \href{https://doi.org/10.1103/n7nt-s64t}{Phys. Rev. Lett. \textbf{136}, 062501 (2026)}.

\bibitem{Hu2022_NatPhys18-1196}
B. Hu et al., Ab initio predictions link the neutron skin of $^{208}$Pb to nuclear forces, \href{https://doi.org/10.1038/s41567-022-01715-8}{Nat. Phys. \textbf{18}, 1196 (2022)}.

\bibitem{Hagen2012_PRL109-032502}
G. Hagen, M. Hjorth-Jensen, G. R. Jansen, R. Machleidt, and T. Papenbrock, Evolution of shell structure in neutron-rich calcium isotopes, \href{https://doi.org/10.1103/PhysRevLett.109.032502}{Phys. Rev. Lett. \textbf{109}, 032502 (2012)}.

\bibitem{Otsuka2010_PRL105-032501}
T. Otsuka, T. Suzuki, J. D. Holt, A. Schwenk, and Y. Akaishi, Three-body forces and the limit of oxygen isotopes, \href{https://doi.org/10.1103/PhysRevLett.105.032501}{Phys. Rev. Lett. \textbf{105}, 032501 (2010)}.

\bibitem{Hebeler2010_PRL105-161102}
K. Hebeler, J. M. Lattimer, C. J. Pethick, and A. Schwenk, Constraints on neutron star radii based on chiral effective field theory interactions, \href{https://doi.org/10.1103/PhysRevLett.105.161102}{Phys. Rev. Lett. \textbf{105}, 161102 (2010)}.

\bibitem{Hebeler2011_PRC83-031301}
K. Hebeler, S. K. Bogner, R. J. Furnstahl, A. Nogga, and A. Schwenk, Improved nuclear matter calculations from chiral low-momentum interactions, \href{https://doi.org/10.1103/PhysRevC.83.031301}{Phys. Rev. C \textbf{83}, 031301(R) (2011)}.

\bibitem{Binder2014_PLB736-119}
S. Binder, J. Langhammer, A. Calci, and R. Roth, Ab initio path to heavy nuclei, \href{https://doi.org/10.1016/j.physletb.2014.07.010}{Phys. Lett. B \textbf{736}, 119 (2014)}.

\bibitem{Gazit2009_PRL103-102502}
D. Gazit, S. Quaglioni, and P. Navr{\'a}til, Three-nucleon low-energy constants from the consistency of interactions and currents in chiral effective field theory, \href{https://doi.org/10.1103/PhysRevLett.103.102502}{Phys. Rev. Lett. \textbf{103}, 102502 (2009)}.

\bibitem{Ekstrom2015_PRC91-051301R}
A. Ekstr{\"o}m et al., Accurate nuclear radii and binding energies from a chiral interaction, \href{https://doi.org/10.1103/PhysRevC.91.051301}{Phys. Rev. C \textbf{91}, 051301(R) (2015)}.

\bibitem{Ekstrom2018_PRC97-024332}
A. Ekstr{\"o}m, G. Hagen, T. D. Morris, T. Papenbrock, and P. D. Schwartz, Delta isobars and nuclear saturation, \href{https://doi.org/10.1103/PhysRevC.97.024332}{Phys. Rev. C \textbf{97}, 024332 (2018)}.

\bibitem{Roth2012_PRL109-052501}
R. Roth, S. Binder, K. Vobig, A. Calci, J. Langhammer, and P. Navr{\'a}til, Medium-mass nuclei with normal-ordered chiral NN+3N interactions, \href{https://doi.org/10.1103/PhysRevLett.109.052501}{Phys. Rev. Lett. \textbf{109}, 052501 (2012)}.

\bibitem{Simonis2017_PRC96-014303}
J. Simonis, S. R. Stroberg, K. Hebeler, J. D. Holt, and A. Schwenk, Saturation with chiral interactions and consequences for finite nuclei, \href{https://doi.org/10.1103/PhysRevC.96.014303}{Phys. Rev. C \textbf{96}, 014303 (2017)}.

\bibitem{Bender2003_RMP75-121}
M. Bender, P.-H. Heenen, and P.-G. Reinhard, Self-consistent mean-field models for nuclear structure, \href{https://doi.org/10.1103/RevModPhys.75.121}{Rev. Mod. Phys. \textbf{75}, 121 (2003)}.

\bibitem{Meng2006_PPNP57-470}
J. Meng, H. Toki, S.-G. Zhou, S.-Q. Zhang, W.-H. Long, and L.-S. Geng, Relativistic continuum Hartree-Bogoliubov theory for ground-state properties of exotic nuclei, \href{https://doi.org/10.1016/j.ppnp.2005.06.001}{Prog. Part. Nucl. Phys. \textbf{57}, 470 (2006)}.

\bibitem{Robledo2019_JPG46-013001}
L. M. Robledo, T. R. Rodr{\'i}guez, and R. R. Rodr{\'i}guez-Guzm{\'a}n, Mean field and beyond description of nuclear structure with the Gogny force: A review, \href{https://doi.org/10.1088/1361-6471/aadebd}{J. Phys. G \textbf{46}, 013001 (2019)}.

\bibitem{Drischler2021_PPNP121-103888}
C. Drischler et al., Towards grounding nuclear physics in QCD, \href{https://doi.org/10.1016/j.ppnp.2021.103888}{Prog. Part. Nucl. Phys. \textbf{121}, 103888 (2021)}.

\bibitem{Hergert2016_PhysRept621-165}
H. Hergert, S. K. Bogner, T. D. Morris, A. Schwenk, and K. Hebeler, The in-medium similarity renormalization group: A novel ab initio method for nuclei, \href{https://doi.org/10.1016/j.physrep.2015.12.007}{Phys. Rep. \textbf{621}, 165 (2016)}.

\bibitem{Borasoy2006_NPA768-179}
B. Borasoy, H. Krebs, D. Lee, and Ulf-G. Mei{\ss}ner, The triton and three-nucleon force in nuclear lattice simulations, \href{https://doi.org/10.1016/j.nuclphysa.2006.01.009}{Nucl. Phys. A \textbf{768}, 179 (2006)}.

\bibitem{Epelbaum2009_EPJA41-125}
E. Epelbaum, H. Krebs, D. Lee, and Ulf-G. Mei{\ss}ner, Nuclear lattice simulations with chiral effective field theory, \href{https://doi.org/10.1140/epja/i2009-10755-0}{Eur. Phys. J. A \textbf{41}, 125 (2009)}.

\bibitem{Elhatisari2024_PLB859-139086}
S. Elhatisari, F. Hildenbrand, and Ulf-G. Mei{\ss}ner, The triton lifetime from nuclear lattice effective field theory, \href{https://doi.org/10.1016/j.physletb.2024.139086}{Phys. Lett. B \textbf{859}, 139086 (2024)}.

\bibitem{Wang2025_PRC112-025502}
T. Wang, X. Feng, and B.-N. Lu, Investigating nuclear $\beta$ decay using a lattice quantum Monte Carlo approach, \href{https://doi.org/10.1103/nv3t-fq87}{Phys. Rev. C \textbf{112}, 025502 (2025)}.

\bibitem{Zhang2025_PRD111-036002}
Z. Zhang, X.-Y. Hu, G. He, J. Liu, J.-A. Shi, B.-N. Lu, and Q. Wang, Binding of the three-hadron $DD^{*}K$ system from the lattice effective field theory, \href{https://doi.org/10.1103/PhysRevD.111.036002}{Phys. Rev. D \textbf{111}, 036002 (2025)}.

\bibitem{Troyer2005_PRL94-170201}
M. Troyer and U.-J. Wiese, Computational complexity and fundamental limitations to fermionic quantum Monte Carlo simulations, \href{https://doi.org/10.1103/PhysRevLett.94.170201}{Phys. Rev. Lett. \textbf{94}, 170201 (2005)}.

\bibitem{ProgTheorPhys110-615}
S. Muroya, A. Nakamura, C. Nonaka, and T. Takaishi, Lattice QCD at finite density: An introductory review, \href{https://doi.org/10.1143/PTP.110.615}{Prog. Theor. Phys. \textbf{110}, 615 (2003)}.

\bibitem{PRB71-155115}
C. Wu and S.-C. Zhang, Sufficient condition for absence of the sign problem in the fermionic quantum Monte Carlo algorithm, \href{https://doi.org/10.1103/PhysRevB.71.155115}{Phys. Rev. B \textbf{71}, 155115 (2005)}.

\bibitem{PRB91-241117}
Z.-X. Li, Y.-F. Jiang, and H. Yao, Solving the fermion sign problem in quantum Monte Carlo simulations by Majorana representation, \href{https://doi.org/10.1103/PhysRevB.91.241117}{Phys. Rev. B \textbf{91}, 241117(R) (2015)}.

\bibitem{AnnuRevCondMattPhys10-337}
Z.-X. Li and H. Yao, Sign-problem-free fermionic quantum Monte Carlo: Developments and applications, \href{https://doi.org/10.1146/annurev-conmatphys-033117-054307}{Annu. Rev. Condens. Matter Phys. \textbf{10}, 337 (2019)}.

\bibitem{Lahde2015_EPJA51-92}
T. A. L{\"a}hde, D. Lee, Ulf-G. Mei{\ss}ner, G. Rupak, and X. Zhang, Symmetry-sign extrapolation for auxiliary-field Monte Carlo simulations, \href{https://doi.org/10.1140/epja/i2015-15092-1}{Eur. Phys. J. A \textbf{51}, 92 (2015)}.

\bibitem{Lu2022_PRL128-242501}
B.-N. Lu, N. Li, S. Elhatisari, Y.-Z. Ma, D. Lee, and Ulf-G. Mei{\ss}ner, Perturbative quantum Monte Carlo method for nuclear physics, \href{https://doi.org/10.1103/PhysRevLett.128.242501}{Phys. Rev. Lett. \textbf{128}, 242501 (2022)}.

\bibitem{Elhatisari2024_Nature630-59}
S. Elhatisari et al., Wavefunction matching for solving quantum many-body problems, \href{https://doi.org/10.1038/s41586-024-07422-z}{Nature \textbf{630}, 59 (2024)}.

\bibitem{Ma2024_PRL132-232502}
Y.-Z. Ma, Z. Lin, B.-N. Lu, S. Elhatisari, D. Lee, N. Li, Ulf-G. Mei{\ss}ner, A. W. Steiner, and Q. Wang, Structure factors for hot neutron matter from ab initio lattice simulations with high-fidelity chiral interactions, \href{https://doi.org/10.1103/PhysRevLett.132.232502}{Phys. Rev. Lett. \textbf{132}, 232502 (2024)}.

\bibitem{FrontPhys8-174}
D. Lee, Recent progress in nuclear lattice simulations, \href{https://doi.org/10.3389/fphy.2020.00174}{Front. Phys. \textbf{8}, 174 (2020)}.

\bibitem{AnnRevNuclPartSci75-109}
D. Lee, Lattice effective field theory simulations of nuclei, \href{https://doi.org/10.1146/annurev-nucl-101918-023343}{Annu. Rev. Nucl. Part. Sci. \textbf{75}, 109 (2025)}.

\bibitem{EPJA31-105}
B. Borasoy, E. Epelbaum, H. Krebs, D. Lee, and Ulf-G. Mei{\ss}ner, Lattice simulations for light nuclei: Chiral effective field theory at leading order, \href{https://doi.org/10.1140/epja/i2006-10154-1}{Eur. Phys. J. A \textbf{31}, 105 (2007)}.

\bibitem{PhysRevLett.104.142501}
E. Epelbaum, H. Krebs, D. Lee, and Ulf-G. Mei{\ss}ner, Lattice effective field theory calculations for $A=3,4,6,12$ nuclei, \href{https://doi.org/10.1103/PhysRevLett.104.142501}{Phys. Rev. Lett. \textbf{104}, 142501 (2010)}.

\bibitem{EPJA45-335}
E. Epelbaum, H. Krebs, D. Lee, and Ulf-G. Mei{\ss}ner, Lattice calculations for $A=3,4,6,12$ nuclei using chiral effective field theory, \href{https://doi.org/10.1140/epja/i2010-11009-x}{Eur. Phys. J. A \textbf{45}, 335 (2010)}.

\bibitem{PLB732-110}
T. A. L{\"a}hde, E. Epelbaum, H. Krebs, D. Lee, Ulf-G. Mei{\ss}ner, and G. Rupak, Lattice effective field theory for medium-mass nuclei, \href{https://doi.org/10.1016/j.physletb.2014.03.023}{Phys. Lett. B \textbf{732}, 110 (2014)}.

\bibitem{Lu2019_PLB797-134863}
B.-N. Lu, N. Li, S. Elhatisari, D. Lee, E. Epelbaum, and Ulf-G. Mei{\ss}ner, Essential elements for nuclear binding, \href{https://doi.org/10.1016/j.physletb.2019.134863}{Phys. Lett. B \textbf{797}, 134863 (2019)}.

\bibitem{Lee:2005it}
D.~Lee and T.~Sch{\"a}fer,
Cold dilute neutron matter on the lattice. II. Results in the unitary limit,
\href{https://doi.org/10.1103/PhysRevC.73.015202}
{Phys. Rev. C \textbf{73}, 015202 (2006)}.

\bibitem{Lee:2005fk}
D.~Lee,
Ground state energy of spin-1/2 fermions in the unitary limit,
\href{https://doi.org/10.1103/PhysRevB.73.115112}
{Phys. Rev. B \textbf{73}, 115112 (2006)}.

\bibitem{Lee:2008xsa}
D.~Lee,
The Ground state energy at unitarity,
\href{https://doi.org/10.1103/PhysRevC.78.024001}
{Phys. Rev. C \textbf{78}, 024001 (2008)}.

\bibitem{Bour:2011xt}
S.~Bour, X.~Li, D.~Lee, U.~G.~Meissner and L.~Mitas,
Precision benchmark calculations for four particles at unitarity,
\href{https://doi.org/10.1103/PhysRevA.83.063619}
{Phys. Rev. A \textbf{83}, 063619 (2011)}.

\bibitem{Lee:2005nm}
D.~Lee,
Large-N droplets in two dimensions,
\href{https://doi.org/10.1103/PhysRevA.73.063204}
{Phys. Rev. A \textbf{73}, 063204 (2006)}.

\bibitem{PhysRevLett.112.102501}
E. Epelbaum, H. Krebs, T. A. L{\"a}hde, D. Lee, Ulf-G. Mei{\ss}ner, and G. Rupak, Ab initio calculation of the spectrum and structure of $^{16}$O, \href{https://doi.org/10.1103/PhysRevLett.112.102501}{Phys. Rev. Lett. \textbf{112}, 102501 (2014)}.

\bibitem{Nat.Comm.14-2777}
S. Shen, S. Elhatisari, T. A. L{\"a}hde, D. Lee, B.-N. Lu, and Ulf-G. Mei{\ss}ner, Emergent geometry and duality in the carbon nucleus, \href{https://doi.org/10.1038/s41467-023-38391-y}{Nat. Commun. \textbf{14}, 2777 (2023)}.

\bibitem{PhysRevLett.132.062501}
Ulf-G. Mei{\ss}ner, S. Shen, S. Elhatisari, and D. Lee, Ab initio calculation of the alpha-particle monopole transition form factor, \href{https://doi.org/10.1103/PhysRevLett.132.062501}{Phys. Rev. Lett. \textbf{132}, 062501 (2024)}.

\bibitem{PRL134-162503}
S. Shen, S. Elhatisari, D. Lee, Ulf-G. Mei{\ss}ner, and Z. Ren, Ab initio study of the beryllium isotopes $^7$Be to $^{12}$Be, \href{https://doi.org/10.1103/PhysRevLett.134.162503}{Phys. Rev. Lett. \textbf{134}, 162503 (2025)}.

\bibitem{PhysRevLett.106.192501}
E. Epelbaum, H. Krebs, D. Lee, and Ulf-G. Mei{\ss}ner, Ab initio calculation of the Hoyle state, \href{https://doi.org/10.1103/PhysRevLett.106.192501}{Phys. Rev. Lett. \textbf{106}, 192501 (2011)}.

\bibitem{PhysRevLett.109.252501}
E. Epelbaum, H. Krebs, T. A. L{\"a}hde, D. Lee, and Ulf-G. Mei{\ss}ner, Structure and rotations of the Hoyle state, \href{https://doi.org/10.1103/PhysRevLett.109.252501}{Phys. Rev. Lett. \textbf{109}, 252501 (2012)}.

\bibitem{PhysRevLett.110.112502}
E. Epelbaum, H. Krebs, T. A. L{\"a}hde, D. Lee, and Ulf-G. Mei{\ss}ner, Viability of carbon-based life as a function of the light quark mass, \href{https://doi.org/10.1103/PhysRevLett.110.112502}{Phys. Rev. Lett. \textbf{110}, 112502 (2013)}.

\bibitem{PRL119-222505}
S. Elhatisari et al., Ab initio calculations of the isotopic dependence of nuclear clustering, \href{https://doi.org/10.1103/PhysRevLett.119.222505}{Phys. Rev. Lett. \textbf{119}, 222505 (2017)}.

\bibitem{Zhang2025_PLB869-139839}
S. Zhang, S. Elhatisari, Ulf-G. Mei{\ss}ner, and S. Shen, Lattice simulation of nucleon distribution and shell closure in the proton-rich nucleus $^{22}$Si, \href{https://doi.org/10.1016/j.physletb.2025.139839}{Phys. Lett. B \textbf{869}, 139839 (2025)}.


\bibitem{PhysRevC.86.034003}
S. Bour, H.-W. Hammer, D. Lee, and Ulf-G. Mei{\ss}ner, Benchmark calculations for elastic fermion-dimer scattering, \href{https://doi.org/10.1103/PhysRevC.86.034003}{Phys. Rev. C \textbf{86}, 034003 (2012)}.

\bibitem{Nature528-111}
S. Elhatisari et al., Ab initio alpha-alpha scattering, \href{https://doi.org/10.1038/nature16067}{Nature \textbf{528}, 111 (2015)}.

\bibitem{Elhatisari2016_PRL117-132501}
S. Elhatisari et al., Nuclear binding near a quantum phase transition, \href{https://doi.org/10.1103/PhysRevLett.117.132501}{Phys. Rev. Lett. \textbf{117}, 132501 (2016)}.

\bibitem{PhysRevLett.125.192502}
B.-N. Lu, N. Li, S. Elhatisari, D. Lee, J. E. Drut, T. A. L{\"a}hde, E. Epelbaum, and Ulf-G. Mei{\ss}ner, Ab initio nuclear thermodynamics, \href{https://doi.org/10.1103/PhysRevLett.125.192502}{Phys. Rev. Lett. \textbf{125}, 192502 (2020)}.

\bibitem{PLB850-138463}
Z. Ren, S. Elhatisari, T. A. L{\"a}hde, D. Lee, and Ulf-G. Mei{\ss}ner, Ab initio study of nuclear clustering in hot dilute nuclear matter, \href{https://doi.org/10.1016/j.physletb.2024.138463}{Phys. Lett. B \textbf{850}, 138463 (2024)}.

\bibitem{Agar:2026nxr}
O.~Agar, Z.~Ren and S.~Elhatisari, From binding and saturation to criticality in nuclear matter with lattice effective field theory, \href{https://arxiv.org/abs/2604.09154}{arXiv:2604.09154 [nucl-th] (2026)}.

\bibitem{PhysRevLett.115.185301}
S. Bour, D. Lee, H.-W. Hammer, and Ulf-G. Mei{\ss}ner, Ab initio lattice results for Fermi polarons in two dimensions, \href{https://doi.org/10.1103/PhysRevLett.115.185301}{Phys. Rev. Lett. \textbf{115}, 185301 (2015)}.

\bibitem{Frame2020_EPJA56-24}
D. Frame, T. A. L{\"a}hde, D. Lee, and Ulf-G. Mei{\ss}ner, Impurity lattice Monte Carlo for hypernuclei, \href{https://doi.org/10.1140/epja/s10050-020-00257-y}{Eur. Phys. J. A \textbf{56}, 248 (2020)}.

\bibitem{EPJA60-215}
F. Hildenbrand, S. Elhatisari, Z. Ren, and Ulf-G. Mei{\ss}ner, Towards hypernuclei from nuclear lattice effective field theory, \href{https://doi.org/10.1140/epja/s10050-024-01427-y}{Eur. Phys. J. A \textbf{60}, 215 (2024)}.

\bibitem{Wigner1937_PhysRev51-106}
E. Wigner, On the consequences of the symmetry of the nuclear Hamiltonian on the spectroscopy of nuclei, \href{https://doi.org/10.1103/PhysRev.51.106}{Phys. Rev. \textbf{51}, 106 (1937)}.

\bibitem{Lee2021_PRL127-062501}
D. Lee, S. Bogner, B. A. Brown, S. Elhatisari, E. Epelbaum, H. Hergert, M. Hjorth-Jensen, H. Krebs, N. Li, B.-N. Lu, and Ulf-G. Mei{\ss}ner, Hidden spin-isospin exchange symmetry, \href{https://doi.org/10.1103/PhysRevLett.127.062501}{Phys. Rev. Lett. \textbf{127}, 062501 (2021)}.



\bibitem{Wang2025_arXiv2512.21942}
T. Wang, X. Feng, and B.-N. Lu, Multi-reference trial state for lattice quantum Monte Carlo simulations, \href{https://arxiv.org/abs/2512.21942}{arXiv:2512.21942 [nucl-th] (2025)}.

\bibitem{Niu2026_arXiv2603.25081}
Z.-W. Niu, S.-S. Zhang, and B.-N. Lu, Extracting resonance width from lattice quantum Monte Carlo simulations using analytical continuation method, \href{https://arxiv.org/abs/2603.25081}{arXiv:2603.25081 [nucl-th] (2026)}.

\bibitem{Liu2025EPJA61-85}
J. Liu, T. Wang, and B.-N. Lu, Perturbative quantum Monte Carlo calculation with high-fidelity nuclear forces, \href{https://doi.org/10.1140/epja/s10050-025-01568-8}{Eur. Phys. J. A \textbf{61}, 85 (2025)}.

\bibitem{Shi2026_PLB874-140303}
J.-A. Shi, C.-C. Wang, and B.-N. Lu, Observation of renormalization group invariance in symmetry-restored nuclear lattice effective field theory, \href{https://doi.org/10.1016/j.physletb.2026.140303}{Phys. Lett. B \textbf{874}, 140303 (2026)}.

\bibitem{Wang2026_arXiv2604.20681}
C.-C. Wang, J.-A. Shi, and B.-N. Lu, Cutoff-independent predictions from nuclear lattice effective field theory, \href{https://arxiv.org/abs/2604.20681}{arXiv:2604.20681 [nucl-th] (2026)}.



\bibitem{Ren2025_PRL135-152502}
Z. Ren, S. Elhatisari, and Ulf-G. Mei{\ss}ner, Ab initio study of the radii of oxygen isotopes, \href{https://doi.org/10.1103/y6s2-43ym}{Phys. Rev. Lett. \textbf{135}, 152502 (2025)}.

\bibitem{Song2026_PLB872-140086}
Y.-H. Song, M. Kim, Y. Kim, K. Cho, S. Elhatisari, D. Lee, Y.-Z. Ma, and Ulf-G. Mei{\ss}ner, Ab initio calculations of the carbon and oxygen isotopes: Energies, correlations, and superfluid pairing, \href{https://doi.org/10.1016/j.physletb.2025.140086}{Phys. Lett. B \textbf{872}, 140086 (2026)}.

\bibitem{Lahde:2019npb} T. A. L\"{a}hde and Ulf-G. Mei{\ss}ner, 
Nuclear Lattice Effective Field Theory: An Introduction, \href{https://doi.org/10.1007/978-3-030-14189-9}{Springer Cham, Lecture Notes in Physics (2019)}.


% T. A. Lähde and Ulf-G. Meißner. Nuclear Lattice Effec-
% tive Field Theory. Springer Cham (2019)

\bibitem{Forbes2011_PRL106-235303}
M. M. Forbes, S. Gandolfi, and A. Gezerlis, Resonantly Interacting Fermions in a Box, \href{https://doi.org/10.1103/PhysRevLett.106.235303}{Phys. Rev. Lett. \textbf{106}, 235303 (2011)}.

\bibitem{Carlson2011_PRA84-061602R}
J. Carlson, S. Gandolfi, K. E. Schmidt, and S. Zhang, Quantum Monte Carlo studies of unitary Fermi gases, \href{https://doi.org/10.1103/PhysRevA.84.061602}{Phys. Rev. A \textbf{84}, 061602(R) (2011)}.

\bibitem{Byers1961_PRL7-46}
N. Byers and C. N. Yang, Theoretical Considerations Concerning Quantized Magnetic Flux in Superconducting Cylinders, \href{https://doi.org/10.1103/PhysRevLett.7.46}{Phys. Rev. Lett. \textbf{7}, 46 (1961)}.

\bibitem{Loh1988_SyntheticMetals27-A499}
E. Y. Loh, Jr. and D. K. Campbell, Using twisted boundary conditions to find the thermodynamic limit of electron-phonon models, \href{https://doi.org/10.1016/0379-6779(88)90184-X}{Synth. Met. \textbf{27}, A499 (1988)}.

\bibitem{Valenti1991_PRB44-13203}
R. Valent\'i, C. Gros, P. J. Hirschfeld, and W. Stephan, Twisted boundary conditions and effective mass in the Hubbard model, \href{https://doi.org/10.1103/PhysRevB.44.13203}{Phys. Rev. B \textbf{44}, 13203 (1991)}.

\bibitem{Gros1992_ZPB86-359}
C. Gros, Evaluation of Fermi surface integrals by momentum-space discretization, \href{https://doi.org/10.1007/BF01309230}{Z. Phys. B Condens. Matter \textbf{86}, 359 (1992)}.

\bibitem{Gammel1993_SyntheticMetals55-4437}
B. M. Gammel, Thermodynamic limit of numerical simulations with twisted boundary conditions, \href{https://doi.org/10.1016/0379-6779(93)91068-F}{Synth. Met. \textbf{55}, 4437 (1993)}.

\bibitem{Gros1996_PRB53-6865}
C. Gros, Control of finite-size effects in Monte Carlo simulations, \href{https://doi.org/10.1103/PhysRevB.53.6865}{Phys. Rev. B \textbf{53}, 6865 (1996)}.

\bibitem{Lin2001PRE64-016702}
C. Lin, F. H. Zong, and D. M. Ceperley, Twist-averaged boundary conditions for quantum Monte Carlo simulations of solids, \href{https://doi.org/10.1103/PhysRevE.64.016702}{Phys. Rev. E \textbf{64}, 016702 (2001)}.

\bibitem{Bedaque2004_PLB593-82}
P. F. Bedaque, A new method to study scattering in a finite volume, \href{https://doi.org/10.1016/j.physletb.2004.04.045}{Phys. Lett. B \textbf{593}, 82 (2004)}.

\bibitem{Divitiis2004_PLB595-408}
G. M. de Divitiis, R. Petronzio, and N. Tantalo, On the properties of nucleons in a finite volume with twisted boundary conditions, \href{https://doi.org/10.1016/j.physletb.2004.06.035}{Phys. Lett. B \textbf{595}, 408 (2004)}.

\bibitem{Bedaque2005PLB616-208}
P. F. Bedaque and J.-W. Chen, Twisted boundary conditions and the determination of light hadron masses, \href{https://doi.org/10.1016/j.physletb.2005.04.053}{Phys. Lett. B \textbf{616}, 208 (2005)}.

\bibitem{Sachrajda2004_PLB609-73}
C. T. Sachrajda and G. Villadoro, Twisted boundary conditions in lattice simulations, \href{https://doi.org/10.1016/j.physletb.2005.01.033}{Phys. Lett. B \textbf{609}, 73 (2005)}.

\bibitem{Korber2016PRC93-054002}
C. K\"orber and T. Luu, Twisted boundary conditions for effective field theory in a finite volume, \href{https://doi.org/10.1103/PhysRevC.93.054002}{Phys. Rev. C \textbf{93}, 054002 (2016)}.







\bibitem{Ma2026}
Y.-Z. Ma, G. Palkanoglou, J. Carlson, S. Gandolfi, A. Gezerlis, G. Given, A. Hicks, D. Lee, K. E. Schmidt, and J. Yu, Evidence for Multimodal Superfluidity of Neutrons,
\href{https://arxiv.org/abs/2602.17611}
{arXiv:2602.17611 [nucl-th] (2026)}.

\bibitem{Guo2024_AtomDataNuclDataTab158-101661}
P. Guo et al., Nuclear mass table in deformed relativistic Hartree-Bogoliubov theory in continuum, II: Even-nuclei, 
\href{https://www.sciencedirect.com/science/article/abs/pii/S0092640X24000263}
{Atom. Data Nucl. Data Tables \textbf{158}, 101661 (2024)}.

\bibitem{Ring1980}
P. Ring and P. Schuck, The Nuclear Many-Body Problem, Springer-Verlag, New York (1980).

\bibitem{Burovski:2006ty}
E.~Burovski, N.~Prokof'ev, B.~Svistunov and M.~Troyer,
The Fermi-Hubbard model at unitarity,
\href{https://doi.org/10.1088/1367-2630/8/8/153}{New J. Phys. \textbf{8}, 153 (2006)}.

\end{thebibliography}
\end{document}